\documentclass[a4paper,10pt]{article}
\usepackage{publication}

\usepackage{natbib}
\usepackage{amsmath}
\usepackage{amsthm}
\usepackage{mathtools}
\usepackage{amsfonts}
\usepackage{physics}
\usepackage{mathtools}
\usepackage{lscape}
\usepackage{color}
\usepackage{xcolor}
\usepackage[pdftex]{graphicx}
\usepackage{accents}
\usepackage{dsfont}
\usepackage{etoolbox}
\usepackage{multirow}
\usepackage{multicol}
\usepackage{booktabs}

\newtheorem{theorem}{Theorem}

\newtheorem{corollary}{Corollary}

\newtheorem{definition}{Definition}

\DeclareMathOperator{\bmu}{\boldsymbol{\mu}}
\DeclareMathOperator{\bm}{\boldsymbol{m}}

\DeclareMathOperator{\btheta}{\boldsymbol{\theta}}
\DeclareMathOperator{\bth}{\boldsymbol{\theta}}

\DeclareMathOperator{\bbeta}{\boldsymbol{\eta}}

\DeclareMathOperator{\bomega}{\boldsymbol{\omega}}
\DeclareMathOperator{\bb}{\boldsymbol{b}}
\DeclareMathOperator{\bp}{\boldsymbol{p}}
\DeclareMathOperator{\bq}{\boldsymbol{q}}

\DeclareMathOperator{\by}{\boldsymbol{y}}
\DeclareMathOperator{\bY}{\boldsymbol{Y}}
\DeclareMathOperator{\bx}{\boldsymbol{x}}

\DeclareMathOperator{\0}{\boldsymbol{0}}
\DeclareMathOperator{\Ex}{\mathbb{E}}

\DeclareMathOperator{\dy}{\dd\boldsymbol{y}}

\newcommand{\Deta}{D\boldsymbol{\eta}}
\newcommand{\bWG}{\textsl{WG}}
\newcommand{\bLA}{\textsl{LA}}
\newcommand{\bKL}{\textsl{KL}}
\newcommand{\bSP}{\textsl{SP}}
\newcommand{\bHMC}{\textsl{HMCnuts}}
\newcommand{\olsi}[1]{{\accentset{\rule{0.95em}{0.05pt}}{#1}}}
\DeclarePairedDelimiterX{\inp}[2]{\langle}{\rangle}{#1, #2}

\allowdisplaybreaks

\title{Beyond Laplace: Closed-form wrapped Gaussian \\[0.15cm] posterior approximations on statistical manifolds}

\author[]{%
    Marcelo Hartmann$^{1,*}$ \\
    Luu Hoang Phuc Hau$^{1, 2}$ \\
    Anton Mallasto$^{3}$ \\
    Albert Kj{\o}ller Jacobsen$^{4}$ \\
    Georgios Arvanitidis$^{4}$ \\
    S{\o}ren Hauberg$^{4}$ \\
    H{\aa}vard Rue$^{5}$ \\
    Mark Girolami$^{6,7}$ \\[1em]
    {\small
    $^1$Department of Computer Science, University of Helsinki, Finland \\ 
    $^2$Division of Mathematical Sciences, Nanyang Technological University, Singapore \\
    $^3$Qutwo, Finland \\ 
    $^4$DTU Compute, Technical University of Denmark, Denmark \\
    $^5$Division of Computer, Electrical, Mathematical Science and Engineering, King Abdullah University of Science and Technology, Saudi Arabia \\
    $^6$Department of Engineering, University of Cambridge, United Kingdom \\
    $^7$The Alan Turing Institute, United Kingdom
    }
}
\contact{$^*$ marcelo.hartmann@helsinki.fi}


\date{\today}

\begin{document}

\maketitle

\begin{abstract}
\vspace{0.4cm} 
\noindent
In Bayesian statistics, the Laplace approximation provides a computationally efficient approximation to posterior distributions. However, its Gaussian form restricts it to elliptical shapes, limiting its ability to capture important posterior features such as skewness, heavy tails, and narrow high-probability regions. Recent work has addressed this limitation by exploiting Riemannian geometry to push forward Gaussian distributions from the tangent space to a chosen manifold, referred to wrapped Gaussians. While offering greater flexibility, they introduce substantial computational challenges. Sampling requires solving geodesic equations through the exponential map and density evaluation additionally depends on the logarithmic map and Jacobi fields, involving costly differential equation solvers alongside full matrices inverse operations, Christoffel symbols and curvature tensors. To overcome these limitations, we employ the theory of contrast functions to derive tractable approximations of the logarithmic and exponential maps on statistical manifolds endowed with the Fisher--Rao metric and the prior distribution geometry. The resulting methodology bypass the need to compute these geometric quantities and numerical solvers thereby removing the principal computational bottlenecks of existing wrapped Gaussian approaches. Empirical results across a range of models demonstrate that the proposed approximation captures complex posterior geometries while remaining orders of magnitude faster than current geometric state-of-the-art approximation.
\end{abstract}




\section{Introduction}

Since its introduction, Laplace’s approximation (\bLA)~\citep{laplace1812} has been a key method for fast Bayesian statistical inference~\citep[see, e.g.,][]{tierney1986accurate, kass1989validity, kass1990asymptotic, rue2009approximate}, as it offers a computationally efficient posterior approximation. The \bLA\, fits a Gaussian distribution to the posterior via a second-order Taylor expansion of the log-posterior about its mode. 
This yields a Gaussian approximation centered at the maximum a posteriori (MAP) point, with the covariance matrix given by the negative inverse Hessian of the log-posterior at the MAP. Although it brings fast inference, the elliptical shape of the Gaussian approximation fails to adequately capture posteriors with more involved distributional forms, such as skewness, heavy tails, narrow, or curved regions of high posterior mass.

A complementary line of work addresses some of these limitations by incorporating higher-order information into the Laplace approximation. In particular, \citet{durante:2024} develop a skew-modal approximation based on a third-order expansion of the log-posterior, thereby accounting for posterior asymmetry through higher-order derivatives. More recently, \citet{pozza:2026} considered a broader class of skew-symmetric perturbations of symmetric posterior approximations and showed that an optimal perturbation can be constructed without additional optimization costs. \citet{Dutta2026}, within the \emph{integrate nested Laplace approximation} (INLA) \citep[]{rue2009approximate}, propose a variational correction scheme that adjusts for the mean, marginal variance and marginal skewness of the \bLA\, via a skewed Gaussian copula. All these approaches primarily address posterior corrections through skewness only.


To jointly address posterior features that cannot be captured by an explicit skewness alone, another line of work leverages concepts from Riemannian geometry to construct more adaptable approximations. The central idea is to first encode the posterior geometric form into a Riemannian manifold, which means that the parameter space is endowed with a suitable choice of Riemannian metric. Second, a Gaussian distribution is defined at a chosen tangent space and pushed forward, through a transformation known as the \emph{exponential map}, onto the manifold. This is precisely the \emph{wrapped Gaussian} (\bWG) method proposed by \citet{chev:2022}. Motivated by \citet{hauberg:2018}, \citet{mallasto:2018}, and \citet{mallasto:19}, recent works of \citet{bergamin:2023}, \citet{yu:2024} and \citet{david:2026} used \bWG s to approximate posterior distributions, which they referred to as \emph{Riemann-Laplace} approximations. For the sake of consistency, we shall henceforth refer to it solely as \bWG.

The geometric construction of the \bWG, provides a fundamentally different correction to the \bLA. Rather than explicitly expanding the log-posterior to higher order to account for its asymmetry, the \bWG, inherits the intrinsic geometry of the chosen manifold. Since this geometry can be freely specified, it can be tailored to accommodate non-Gaussian posterior forms through an appropriate choice of manifold geometry.
This distinction is important: skew refinements such as those of \cite{durante:2024} and \cite{pozza:2026} primarily target posterior asymmetries, whereas the \bWG\, accounts for global geometric aspects induced by the curvature of the statistical manifold starting from a local Gaussian approximation. The two approaches are therefore complementary rather than competing corrections of the same type.

While offering greater flexibility, the geometric construction of the \bWG\, introduces substantial computational challenges. The computation of exponential maps (the push-forward) and their inverses, \emph{logarithmic maps}, involves solving a system of ordinary differential equations 
with no analytical solutions. Moreover, numerical solvers often require a large number of integration steps to either sample from the \bWG\, or to evaluate its density, consequently rendering the method unattainable in practice for even moderately sized parameter spaces.


In this work, we derive closed-form approximations of the logarithmic map and a low cost alternative to compute the exponential map. To achieve this, we use the theory of contrast functions from information geometry \citep{eguchi:1985, eguchi:1992, amarinaga:2000} and subsequently develop a tractable \bWG\, for posterior approximations on statistical manifolds. Specifically, we derive a closed-form approximation of the logarithmic map via local expansions of the Riemannian distance that is \emph{accurate to fourth order} and whose associated metric formally combines the Fisher–Rao and prior geometries within a unified contrast function framework not presented before. Given this approximation, the exponential map reduces to an \emph{ordinary least-squares problem}, solvable with standard gradient-based methods. Importantly, this construction does not require the explicit evaluation of third-order derivatives of the likelihood, in contrast to higher-order Laplace corrections based on third-order expansions \citep{durante:2024}.


Our findings yield two crucial computational advantages. First, sampling from the resulting \bWG\, approximation reduces to minimizing a least-squares objective, which any gradient-based optimizer can solve efficiently via automatic differentiation. Second, evaluating the approximate density is inexpensive, since the approximate logarithmic map is available in closed-form. To showcase the efficacy of our tractable \bWG\, approximations, we present experiments on generalized linear models, Gaussian processes, population growth models and neural networks, including comparisons with the recent skewness correction of \citet{pozza:2026}. Since the latter provide an optimal perturbation within their class of skew-symmetric corrections, this comparison allows us to assess whether the geometric correction provided by the \bWG\, yields additional improvements beyond the skewness refinement of \bLA.


Our contributions are as follows. \textbf{1)} We develop a closed-form approximation of the logarithmic map on statistical manifolds based on the theory of contrast functions, 
with direct applicability to approximate Bayesian inference 
\textbf{2)} We show that the resulting \bWG\, approximation defines a proper probability distribution and inherits the curvature induced by both the Fisher–Rao and prior geometries with the possibility to be extendable to other contrast functions. \textbf{3)} We show that sampling from the \bWG\, can be approach with a tractable formulation of the exponential map as a non-linear least-squares problem which means efficient sampling from the approximation via standard gradient-based optimization. \textbf{4)} We demonstrate empirically that the proposed \bWG\, approximation preserves the posterior distribution geometry, providing accurate posterior approximations at a reduced computational cost across a wide range of statistical models.

\subsection{Geometric perspective}

In its classical formulation, the \bLA\, can be viewed as constructed under the implicit assumption that the geometry of the parameter space is Euclidean. From a Riemannian perspective, however, Euclidean geometry is only one particular choice among more general geometries. This suggests a natural generalization. Instead of approximating the posterior by a Gaussian on the Euclidean parameter space, one can first introduce a Riemannian geometry that reflects the geometric shape of the posterior and then construct the approximation on this geometry.

\begin{figure}[!ht]
    \vskip 0.0in
        \begin{center}
            \centerline{
                \includegraphics[scale = 0.61]{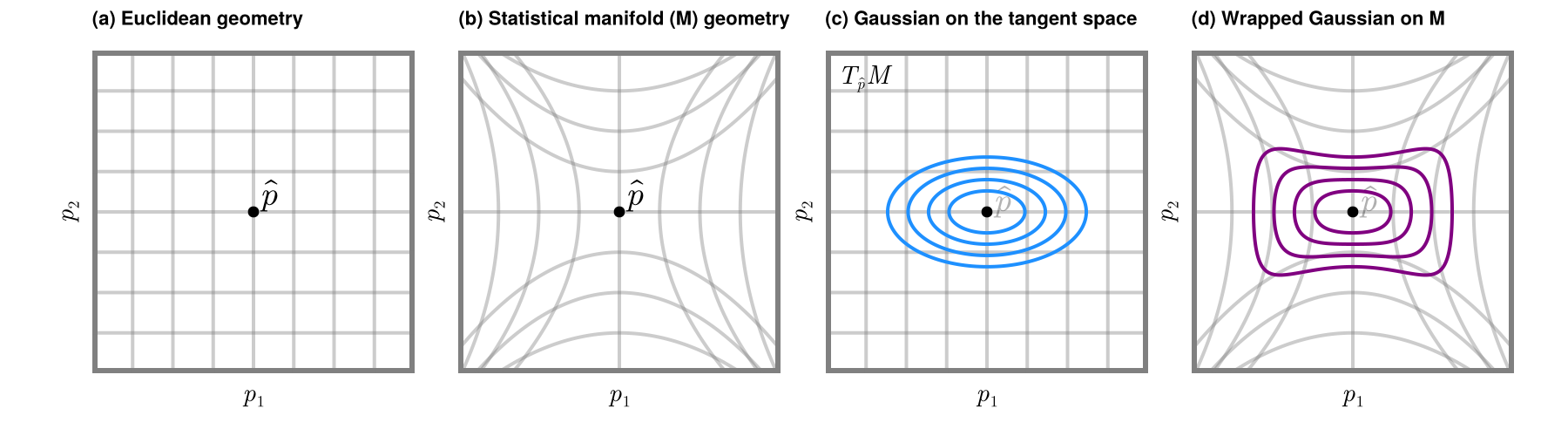}
            }
            \caption{
            Conceptual illustration of the effect of the underlying geometry on the form of a Gaussian distribution. In panel (a), the parameter space is endowed with the standard Euclidean geometry, represented by a flat grid. In panel (b), a non-Euclidean geometry is imposed on the same space, resulting in curved coordinate lines and illustrating the geometry of the statistical manifold $M$. Panel (c) shows a Gaussian distribution on the tangent space $T_{\hat\bp}M$ ({\color{blue} {\Large \textbf{-}}}, solid), which provides a local Euclidean approximation of $M$ around the base point $\hat\bp$. Finally, in panel (d), this Gaussian in (c) is mapped from the tangent space onto $M$ through the exponential map associated with the geometry in panel (b), transforming its Euclidean form into a \bWG\,  distribution on $M$ ({\color{purple} {\Large \textbf{-}}}, solid).
            }
            \label{fig:smvisuals}
        \end{center}
    \vskip -0.1in
\end{figure}

Let $M$ denote a statistical manifold associated with a Riemannian geometry, and let $\hat\bp \in M$ be a fixed basepoint, which can be regarded as the MAP estimate. The tangent space $T_{\hat\bp}M$ provides a local Euclidean representation of the manifold around $\hat\bp$. We can introduce a Gaussian distribution in this tangent space, set $V \sim \mathcal{N}(0, \Sigma)$, where $V$ represents random tangent vectors, and map $V$ through the exponential map. The resulting \bWG\, distribution is $\rho = \big(\mathrm{Exp}_{\hat\bp}\big)_{\#}(\mathcal{N}(\0, \Sigma))$, where $\big(\mathrm{Exp}_{\hat\bp}\big)_{\#}$ denotes the push forward by the exponential map. Since the exponential map is determined by the Riemannian geometry of $M$, the resulting \bWG\, therefore inherits such geometry and hence adapts to the distributional form of the posterior. Consequently, the Gaussian structure is retained only in the tangent space, while its image reflects the geometric features of $M$. In the special case of Euclidean geometry, $\mathrm{Exp}_{\hat\bp}(V) = \hat\bp + V$, and the push forward reduces exactly to the usual Gaussian approximation $\mathcal{N}(\hat\bp, \Sigma)$. See illustrative concepts in Figure \eqref{fig:smvisuals}.

In this sense, this construction provides a geometric mechanism for introducing a variety of deviations from the elliptical shape of the \bLA. The key advantage of \bWG s lies in its ability to capture global geometric aspects of the posterior beyond the immediate neighbourhood around $\hat\bp$. Rather than specifying a particular form of deviation from the Gaussian, such as skewness, the Riemannian geometry generalizes how the parameter space itself curves, squeezes, or stretches in different directions. The resulting \bWG, can, therefore, exhibit skewness, anisotropy, bending, and other non-elliptical features as consequences of its underlying geometry. Skewness is only one possible manifestation of these geometric corrections.

At the same time, this flexibility introduces two fundamental challenges. First, the Riemannian geometry must be chosen so that it meaningfully represents the posterior geometric form. The metric need not be regarded as a given primitive, rather, it can be reconstructed from the statistical inference problem whose geometry it is intended to describe. Second, the exponential map generally has no closed-form expression. Sampling consequently requires evaluating the exponential map, while density evaluation requires its inverse, the logarithmic map, together with its Jacobian. In general, these operations require numerical solutions of differential equations, formulated as an initial value problem (IVP) and a boundary value problem (BVP), respectively, making the geometric construction substantially more demanding than the classical \bLA. The aim of the present work is to retain this geometric flexibility while replacing these computationally expensive operations with tractable approximations.

In what follows, we develop the technical formalism required to construct the Riemannian geometry tailored to Bayesian statistical inference problems. We then introduce a tractable \bWG\, distribution on the resulting statistical manifold designed to preserve the geometric representation of the posterior form in the final proposed approximation method. The resulting approximation combines the geometric flexibility with computational efficiency required for practical Bayesian inference, making it suitable for applications across a broad range of models and experiments.

\section{Preliminaries and related works}\label{sec2}

We first review the essential concepts of Riemannian geometry needed to understand the main geometric tools used in this paper. We then introduce information geometry, which studies statistical manifolds endowed with a particular Riemannian structure. These concepts provide the foundation for constructing the \bWG\, distributions considered as posterior approximations. Readers may consult \cite{docarmo:1992}, \cite{kass::1997}, \citet{calin:2014}, \citet{chev:2022} and \citet{boumal:2023} for further details. Proofs of the new theoretical results and additional details on the main tools used here are provided in the supplementary material (appendix). Finally, we introduce the \bWG\, distribution, review related work, and discuss its main computational challenges.

\subsection{Riemannian manifolds and logarithmic maps}

A \emph{manifold} $M$ is a second-countable Hausdorff space that is locally homeomorphic to $\mathbb{R}^n$. Meaning that, every point $\bp \in M$ has an open neighbourhood that can be identified with an open subset of $\mathbb{R}^n$ via a homeomorphism, called a \emph{parametrisation} or chart. The countable collection of all such charts is called an \emph{atlas}, and $M$ is said to be $n$-dimensional. The Hausdorff condition rules out spaces where two distinct points $\bp \neq \bp'$ share every open neighbourhood, which would force any chart to assign them the same coordinates and violate the injectivity of the parametrisation. When all charts in the atlas are diffeomorphic on their overlapping images, $M$ is called a \emph{differentiable manifold}. Henceforth, we assume the existence of a global parametrisation (or global chart) and consider it. Given this parametrisation \( \bmu = (\mu_1, \ldots, \mu_n) \mapsto \bp(\bmu) \) we can define coordinate curves through $\bp$. Each coordinate curve has a tangent vector at $\bp$ and these vectors form the basis of the \emph{tangent space}. We denote these basis vectors as 
%
    $
    (\partial_1)_{\bp},\ldots,(\partial_n)_{\bp}
    $    
%
where the notation $(\partial_i)_{\bp} := \frac{\partial}{\partial \mu_i}\big|_{\bp}$ simultaneously means that they act as tangent vectors on $M$ and as directional derivatives on smooth function $f : M \to \mathbb{R}$ \citep{docarmo:1992}. 

Moreover, given the tangent vectors $U = \sum_i U^i\partial_i|_{\bp}$ and $V = \sum_i V^i\partial_i|_{\bp}$, we can define an inner product that is bilinear in its arguments on the tangent space at $\bp$ 
\begin{align} \label{eq:riemm}
    g_{\bp} : 
    (T_{\bp}M)^2 &\to \mathbb{R} \\
    (U, V) &\mapsto \sum_{i, j} U^i V^j 
    g_{\bp} \big( (\partial_i)_{\bp}, (\partial_j)_{\bp} \big).
    \nonumber
\end{align}
This is called metric tensor. For every $i, j$ we set $g_{\bp} \big( (\partial_i)_{\bp}, (\partial_j)_{\bp} \big) = G_{ij}(\bp)$, and $G(\bp)$ denotes a $n \times n$ matrix. Such a matrix collects the coefficients of the metric tensor. When $g_{\bp}$ is positive, or equivalently $G(\bp)$ is symmetric positive-definite matrix, the set $\{ g_{\bp}$, $\bp \in M \}$ is called Riemannian metric. When $M$ is endowed with such a metric it is called a \emph{Riemannian manifold}. 

The previous concepts allow us to generalize the notion of distance and differentiation on $M$. If $\gamma: [0, 1] \to M$ is a curve, we can calculate its length as 
%
    $
    L(\gamma) := \int_0^1 \sqrt{ g_{\gamma(t)} \big( 
    \dot{\gamma}(t),\dot{\gamma}(t) \big)} \, \dd t
    $
%
and the distance between two points $\bp$ and $\bq$ in $M$ is defined to be the infimum length of all curves starting at $\bp$ and ending at $\bq$. We call it Riemannian distance and denote it as $d(\bp, \bp')$. To differentiate on $M$ we also introduce the notion of \emph{connection}. Let $\mathcal{X}(M)$ denote the set of vector fields on $M$. The connection is a map $\nabla : \mathcal{X}(M)^2 \to \mathcal{X}(M)$, $(U,V) \mapsto \nabla_V U$, that is $\mathbb{R}$-linear in the first argument, $C^{\infty}(M)$-linear in the second, and satisfies the Leibniz rule \citep[Ch.~2]{docarmo:1992}. For tangent vectors $U$, $V$, the output $\nabla_V U = \sum_k \big(V(U^k) + \sum_{i, j} U^iV^j\Gamma_{ij}^k({\bp}) \big) \partial_k\big|_{\bp}$ and $\Gamma_{ij}^k({\bp})$, $i,j,k = 1,\ldots, n$, are the coefficients of the connection $\nabla$. In general, the metric $g_{\bp}$ and $\nabla$ are independent structures that can be chosen separately \citep{docarmo:1992}. However, given a metric $g_{\bp}$, there exists a unique 
connection
called the \emph{Levi-Civita} connection, which we denote $\nabla^{(0)}$. Its coefficients $\Gamma_{ij}^{k(0)}(\bp)$
are the called \emph{Christoffel symbols} of the second-kind. The Christoffel symbols of the first kind are $\Gamma_{ij, k}^{(0)}(\bp) = \sum_r G_{kr}(\bp) \Gamma_{ij}^{r(0)}(\bp)$ \cite[See][Ch. 2, Sec. 3]{docarmo:1992}. From now on, we will refer to both types of Christoffel symbols as coefficients of the connection, whose indexes will be lowered or raised according to the matrix coefficient $G$.

For a fixed metric $g$ and a smooth function $f : M \to \mathbb{R}$, the \emph{Riemannian gradient} is defined to be 
\begin{align} \label{eq:riemg}
    \mathrm{grad}\, f(\bp) = G^{-1}(\bp)\,\nabla f(\bp),
\end{align}
where $\nabla f$ is the Euclidean gradient. Additionally, fixed a connection $\nabla$ with coefficients $\Gamma_{ij}^k(\bp)$ and a direction $V$, the \emph{Riemannian Hessian} is given by the covariant derivative of the Riemannian gradient in the direction of $V$ as
\begin{align} \label{eq:riemh}
    \mathrm{Hess}\, f(\bp)[V] = G^{-1}(\bp)
    \Big(
    \nabla^2 f(\bp) - \sum_{k=1}^{n} \partial_k f(\bp)\,\Gamma^k(\bp)
    \Big) V,
\end{align}
where $\nabla^2 f$ is the Euclidean Hessian and $\Gamma^k(\bp)$ denotes the matrix with $(i, j)$ entries $\Gamma^k_{ij}(\bp)$ \citep[see][for details]{boumal:2023}. With the connection, we can compute the analogous of straight lines on $M$. A smooth curve $\gamma : (a, b) \to M$ is a \emph{geodesic} if $\nabla_{\dot\gamma} \dot\gamma = 0$, or equivalently, it satisfies the geodesic equations $\ddot{\gamma}^k(t) + \Gamma^k_{ij}(\gamma(t)) \dot{\gamma}^i(t)\dot{\gamma}^j(t) = 0$, $k = 1, \ldots, n$, a second-order system of ODEs \citep[Ch.~3]{docarmo:1992}. From geodesics we obtain two key maps central to this work. The \emph{exponential map} 
%
$
    \mathrm{Exp}_{\bp} :\, 
    T_{\bp}M \to M
$
with
$
    \mathrm{Exp}_{\bp}(V) 
    := \gamma_{\bp, V}(1)
$
%
where $\gamma_{\bp,V}$ is the unique geodesic with $\gamma_{\bp, V}(0) = \bp$ and $\dot\gamma_{\bp,V}(0) = V$, whenever this geodesic is defined up to time $1$. In particular, $\mathrm{Exp}_{\bp}$ is always defined on a sufficiently small neighbourhood of the origin in $T_{\bp}M$, and its restriction to such a neighbourhood is a diffeomorphism onto a neighbourhood of $\bp$. Its local inverse, the \emph{logarithmic map} 
\begin{align*}
    \mathrm{Log}_{\bp} :\,
    M 
    &
    \to T_{\bp}M
    \\
    \bp' &\mapsto \mathrm{Log}_{\bp}(\bp')
\end{align*}
sends a point $\bp'$ in such a neighbourhood to the unique vector $V$ for which the geodesic starting at $\bp$ in direction $V$ reaches $\bp'$ in unit time. When $M$ is geodesically complete, $\mathrm{Exp}_{\bp}$ is defined on all of $T_{\bp}M$. In this case, $\mathrm{Log}_{\bp}$ is single-valued and smooth
on $M\setminus \mathrm{Cut}(\bp)$, where $\mathrm{Cut}(\bp)$ denotes the cut
locus of $\bp$, namely the set of points at which geodesics starting from
$\bp$ stop being global distance minimizers. Under standard assumptions, the cut locus has zero Riemannian measure \citep[Lemma~2.3]{facca:2022}. Nevertheless, care is needed near the cut locus, where the
logarithmic map may fail to be smooth.

A useful fact is that the logarithmic map can be recovered from the squared Riemannian distance \citep[See][supp.\ materials]{pennec:2018} away from the cut locus. Let $d(\cdot, \bq) : M \to (0, \infty)$. Then
\begin{equation} \label{eq:logmap}
    \mathrm{Log}_{\bp}(\bq) = -\tfrac{1}{2}\,G^{-1}(\bp)
    \,
    \nabla d^2(\bp,\bq).
\end{equation}
where $\nabla d^2(\bp,\bq)$ is the Euclidean gradient of $d^2(\bp,\bq)$ w.r.t. $\bp$.

Following \citet[See Def. 2]{jump:2021}, we characterize an \emph{approximate} logarithmic map by its local behaviour. For each $\bp\in M$, we consider a smooth map $\olsi{\mathrm{Log}}_{\bp} : M \to T_{\bp} M$, $\bq \mapsto \olsi{\mathrm{Log}}_{\bp}(\bq)$ satisfying $\olsi{\mathrm{Log}}_{\bp}(\bp) = 0$ and derivative $D\olsi{\mathrm{Log}}_{\bp}(\bp) = I$ where $I$ is the identity map on the tangent space $T_{\bp}M$. These conditions ensure that $\olsi{\mathrm{Log}}_{\bp}$ agrees with the true logarithmic map to first order at $\bp$. We exploit these simple conditions for flexibility, allowing any map satisfying those to be a substitute for $\mathrm{Log}_{\bp}(\bq)$.

Throughout this work we mainly deal with Riemannian submanifolds $\bar{M} \subseteq M$ specified by an \emph{imbedding function} $h : \bar{M} \to M$. The metric of $M$ together with $h$ induces a metric on $\bar{M}$ called \emph{pullback}. This metric is given by $g_{\bar{\bp}} : (T_{\bar{\bp}} \bar{M})^2 \to \mathbb{R}$, $(V_1, V_2) \mapsto g(Dh(\bar{\bp})[V_1], Dh(\bar{\bp})[V_2])$ whose matrix coefficient is given by $G_{\bar{M}}(\bar{\bp}) = Dh(\bar{\bp})^\top G(h(\bar{\bp}))Dh(\bar{\bp})$ and where $Dh(\bar{\bp})$ denotes the Jacobian matrix of $h$. All the geometric theory presented above applies directly to this submanifold setting. The function $h$ plays a crucial practical role -- it encompasses a broad class of linear and non-linear regression functions -- frequently arising in Statistics, machine-learning (ML), and inverse problems, making this framework widely applicable \citep{kass::1997, arvanitidis:aistats:2021, arvanitidis:aistats:2022b}.

\subsection{Information geometry} \label{sec:ig}

Information geometry concerns the study of a particular class of Riemannian manifolds whose points are probability distributions. A \emph{statistical manifold} is a family of probability distributions
%
    $
    M = \big\{ \rho_{\bmu}(\by) : \Omega \to \mathbb{R}_{\geq 0} 
    \,\big|\, \bmu \in \mathrm{M} \subseteq \mathbb{R}^n
    \big\}
    $
%
where $\bmu$ is the parameter vector indexing each density. The map $\bmu \mapsto \rho_{\bmu}(\by)$ plays the role of a parametrisation provided it is injective -- that is, the family is \emph{identifiable} -- and any two parametrisations of $M$ (e.g. the mean and natural parameters of an exponential family) are diffeomorphic. From now on, we abuse notation and use $\bmu$ or $\bp$ interchangeably for a point of $M$, and similarly we will use $\btheta \in \Theta \subseteq \mathbb{R}^d$ and $\bar{\bp}$ for a point of $\bar{M}$. 

Here we make two assumptions that are still widely common in application areas. First, we assume the usual setting of independent distributions for each observation $y_i$, $i = 1, \ldots, n$. Second, we assume a generic function $h$ that plays the role of regression function, associating a covariate $\bx_i$ to an observation $y_i$, where $\bmu = h(\bth)$ and $h(\bth)_i := h(\bx_i, \bth)$ \citep{kass::1997}. Therefore, the submanifold defined by $h$ is $\bar{M} = \big\{ \rho_{\btheta}(\by) := \prod_{i = 1}^n\rho_{\mu_i}(y_i) \, \big| \, \mu_i = h(\bth)_i, i = 1,\ldots, n ,\, \bmu \in \mathrm{M} ,\, \bth \in \Theta \big\} $. Here, it should be understood that $\by$ is a random draw from $ \bY \sim \rho_{\btheta}$, and the set of covariates is fixed, they are not random variables.

To make $\bar{M}$ Riemannian, we can first specify a metric on $M$, and then pull it back to $\bar{M}$ with $h$. However, a natural way to do this operation directly in an intrinsic manner is through a \emph{contrast function} \citep[See][and supplementary materials for details]{eguchi:1985, eguchi:1992}, a smooth map $D : \bar{M}^2 \to \mathbb{R}_{\geq 0}$ whose Hessian on the diagonal is positive-definite everywhere and that it transforms covariantly, as a $(0, 2)$-tensor. Using short notation for $D_{\bar{M}}(\bth \| \bth') := D_{\bar{M}}(\rho_{\bth}(\by) \| \rho_{\bth'}(\by))$, $\partial_i := \partial/\partial\boldsymbol{\bth}_i$ and $\partial_{i'} := \partial/\partial \boldsymbol{\theta}'_i$, the Riemannian metric $g$ generated from the choice of $D_{\bar{M}}$ is then specified with the matrix coefficient
\begin{equation} \label{eq:div_metric}
    G_{ij}(\bth) := 
    \partial^2_{i'j'}
    D_{\bar{M}}(\bth \| \bth')
    \big|_{\bth' = \bth},
\end{equation}
see \citet[Sec. 11.3 for equivalent definitions]{calin:2014}. When $D_{\bar{M}}$ is the Kullback-Leibler contrast function, $G(\bth)$ is the \emph{Fisher information matrix} -- the most natural and studied choice on a statistical manifold -- owning its role in asymptotic inference \citep{lehmann:2003} and its interpretation of curvature as the sensitivity of the density to its parameters \citep{rao:1945, kass::1997, mark:2011}. Other common choices arise from Bregman, Chernoff, Jeffreys, Hellinger, and $f$-contrast functions \citep{amarinaga:2000, calin:2014}. The same contrast function also induces a pair of affine connections $\nabla, \nabla^*$ on $\bar{M}$. The coefficients of $\nabla$ are given by
\begin{align}
\label{eq:thirdordconst}
    \Gamma_{ij,k}(\bth) := 
    -
    \partial^3_{i, j, k'} D_{\bar{M}}(\bth \| \bth')
    \big|_{\bth' = \bth},
\end{align}
and applying the same construction to the \emph{dual contrast function} $D_{\bar{M}}^*(\bth \| \bth') := D_{\bar{M}}(\bth' \| \bth)$ gives the coefficients $\Gamma^*_{ij,k}(\bth)$ for the other connection $\nabla^*$ (defined similar to \eqref{eq:thirdordconst} but with $D^*$). They are called \emph{dual connections}, their partials commute and their average recovers the Levi-Civita connection $\nabla^{(0)}$ of the metric $g$ that is 
\begin{align}
    2\nabla^{(0)} = 
    \nabla + \nabla^*
\end{align}
generalising the Levi-Civita connection construction. This can be further generalized to a one-parameter family of compatible connections that plays a central role in information geometry \citep{pfan:1973, chentsov:1982, calin:2014}.

\subsection{Wrapped Gaussians}

Wrapped Gaussians distributions were first formalized in directional statistics as Gaussian-like distributions on circular domains \citep{mardia:2000} with applications in animal orientation and movement \citep{penswey:2008}. Subsequent work by \citet{hauberg:2018} studied \bWG s on spherical domains with applications on clustering and Kalman filtering for directional data. \cite{mallasto:2018} and \cite{mallasto:19} later developed similar constructions to more general Riemannian manifolds and to Gaussian processes that were applied to diffusion tensor imaging data. The basic idea underlying the \bWG\, is straightforward.
By placing a zero mean Gaussian measure on the tangent space $T_{\bar{\bp}}\bar{M}$, or equivalently $V \sim \mathcal{N}(\0, \Sigma)$, and pushing it forward to $\bar{M}$ via the exponential map $\mathrm{Exp}_{\bar{\bp}}$, we obtain
%
    $
    W \sim \big( 
        \mathrm{Exp}_{\bar{\bp}} \big)_{\#} \big( 
        V
    \big).
    $
%
The resulting distribution of $W$ is called a \emph{wrapped Gaussian}. Evaluating the resulting density on the manifold is difficult since it requires the change of variables through $\mathrm{Exp}_{\bar{\bp}}$, and hence the Jacobian determinant of its inverse \citep[See][Ch. 2]{chev:2022}. For the classical extrinsic constructions, with $\bar{M}$ as a subset of Euclidean space, e.g.\ the sphere $S^n \subset \mathbb{R}^{n+1}$, or the cone of $n\times n$ positive-definite matrices in $\mathbb{R}^{n(n + 1)/2}$, their Jacobians are available via a closed-form exponential map or logarithmic map, but these are rare cases. 

\citet{chev:2022} unified the extrinsic and intrinsic viewpoints in a single formulation based on local coordinates, clarifying that when the exponential map is given in a parametrisation $\bth$, denote $\mathrm{Exp}_{\bth^*}$ where $\bth^*$ is the basepoint, the probability density function with respect to the Lebesgue measure (E.\ Chevallier, personal communication, 7 January 2026) has the expression given by
\begin{align} \label{eq:riemwg}
    \rho_{\bWG}(\bth) =  
    \mathcal{N} \big( 
        \mathrm{Log}_{\bth^*}(\bth)|\0, \Sigma
    \big)
    \big|\det D\mathrm{Log}_{\bth^*}(\bth) \big|,
\end{align}
where $|\cdot|$ denotes the absolute value and $(\bth^*$, $\Sigma)$ can be seen as the parameters of the probability distribution $\rho_{\bWG}$. $D\mathrm{Log}_{\bth^*}(\bth)$ denotes the usual Jacobian of the logarithmic map in its respective parametrisation.

Inspired by \citet{hauberg:2018}, recent work of \citet{bergamin:2023} proposed the so-called Riemann-Laplace (\textsl{RL}), motivated by a second-order Taylor expansion on the manifold. If $\bth^*$ is the MAP and $f : \bar{M} \to \mathbb{R}$, where $f$ is the logarithm of a posterior distribution that is twice differentiable, using \eqref{eq:riemm}, \eqref{eq:riemg} and \eqref{eq:riemh} in a short computation of the Taylor-expansion \citep{boumal:2023} on the manifold around $\bth^*$ gives $f(\bth) \approx f(\bth^*)
    - \tfrac{1}{2}
    \big\|
        \mathrm{Log}_{\bth^*}(\bth)
    \big\|^2_{-\nabla^2 f(\bth^*)}
$, because the Euclidean gradient $\nabla f(\bth^*) = \0$ at the MAP. Then the posterior approximation reduces to 
\begin{align} \label{eq:riemlp}
    \rho_{\textsl{RL}}(\bth)
    &=
    \frac{e^{ -\frac{1}{2} \| \mathrm{Log}_{\bth^*}(\bth) \|^2_{-\nabla^2 f(\bth^*)}}}{Z}
\end{align}
with the normalizing constant given by $Z = \int_{\Theta} e^{ -\frac{1}{2} \| \mathrm{Log}_{\bth^*}(\bth) \|^2_{-\nabla^2 f(\bth^*)}} \dd\boldsymbol{\theta}$ (see Appendix for details). We note that these expressions were also already presented by \citet[Eq. 10 and 11]{hauberg:2018} and \citet[Eq. 2.4 or 2.5, and references therein]{chev:2022}, and differs from \eqref{eq:riemwg} by a Jacobian determinant term, a distinction that has not always been made clearly in the literature. 

The recipe of \cite{bergamin:2023} is, therefore, the same as that of \citet{chev:2022}; the novelty lies in the intrinsic formulation, in which the metric tensor is obtained as the pullback operation related to the log-posterior surface as the $d+1$ submanifold of the Euclidean space \citep[see][for priors works]{hartmann:22, hartmann:2023}, thereby inducing a Riemannian structure directly on the parameter space $\Theta$.

In either way, irrespective of the metric used to induce the geometric structure in \eqref{eq:riemwg} (or \eqref{eq:riemlp}), sampling from the density \eqref{eq:riemwg} means solving geodesic equations, which involves repeated inversion and differentiation of the metric tensor at each step of the numerical integration — a prohibitive computational burden. To evaluate this density, we further require the Jacobian determinant of the logarithmic map, computed via Jacobi field ODEs \citep[Ch.~4]{chev:2022}, which in turn involves the Riemannian curvatures (rank-4 tensor) and derivatives of the Christoffel symbols. Together, these bottlenecks make Equation \eqref{eq:riemwg} mostly representative, since it becomes practically impossible to perform such computations for models of moderate parameter sizes, even with automatic differentiation tricks \citep{baydin:2018}.


Such computational limitations are addressed in the next section, where we derive a closed-form approximation to the logarithmic map from a contrast function that captures the intrinsic geometry of the posterior distribution. We further show that the corresponding approximate exponential map reduces to a non-linear least-squares problem, bypassing the need for repeated inversion of the full matrix $G$, computation of Christoffel symbols, and evaluation of Riemannian curvature tensors. As a result, this now means that both sampling from and density evaluation of Equation \eqref{eq:riemwg} become computationally tractable, avoiding entirely the numerical ODE solvers required by previous \bWG\, constructions.

\section{Approximate logarithmic map}

This section presents the main theoretical results underlying our principal contribution: a closed-form approximation of the logarithmic map based on contrast functions. We begin with a new theorem establishing that the symmetrized contrast function locally approximates the squared Riemannian distance up to fourth-order terms. Building on this, we derive our main theorem on the approximate logarithmic map, and as a consequence, a corollary establishing the existence of a global inverse for a broad class of statistical models commonly used in Statistics and ML. We then define a \bWG\, distribution based on this approximate map, and show that both sampling and density evaluation become available at negligible computational cost. Partial proofs are given throughout, with full proofs deferred to the supplementary material (appendix).

The first result is an extension of Theorem 4.4.5 of \citet[p. 121]{calin:2014}.

\begin{theorem}[Approximate Riemannian distance] \label{theo:ard}
    Let $\bar{M}$ be a complete statistical manifold equipped with a Riemannian metric $g$ induced by a contrast function 
    $D_{\bar{M}}$.
    Let $C$ denote the symmetrized contrast function
    \[
        C(\bth \| \bth') = D_{\bar{M}}(\bth \| \bth') + D_{\bar{M}}^*(\bth \| \bth').
    \]
    Then it locally approximates the squared Riemannian distance $d : \bar{M}^2 \to \mathbb{R}_{\geq 0}$ associated with $g$ in the sense that
    \[
        C(\bth \| \bth') = d^2(\bth, \bth') + O \big( d^4(\bth, \bth') \big).
    \]
    \begin{proof}[Partial proof :]
    
        Let $\gamma :[0, 1] \rightarrow \bar{M}$ be a normalized minimizing geodesic curve such that $\gamma(0) = \bth$ and $\gamma(t) = \bth'$. Then $t = d(\bth, \bth')$ where $d$ is the Riemannian distance between $\bth$ and $\bth'$. A third-order Taylor expansion of the curve $\psi(t) := C(\bth \| \gamma(t))$ around $t = 0$ reveals the above relationship to the squared Riemannian distance due to the following reasons.

        As $C$ is also a contrast function due to $D_{\bar{M}}$, properties 1 and 2 (see supplementary material) makes the constant and linear coefficients of the curve $\psi$ vanish. The quadratic coefficient of $\psi$ involves the second derivatives of $D_{\bar{M}}$ and $D_{\bar{M}}^*$, which by construction define the matrix coefficient $G$ and, therefore, the metric $g$. As the geodesic has unit speed, $g(V,V) = \sum_{i, j} V^i V^j G_{ij}(\bth) = 1$ with $V = \dot\gamma(0)$, this contributes exactly $t^2$.
        
        The cubic coefficients of $\psi$ are more involved. Differentiating $\psi$ three times produces pure third-order derivative and mixed terms involving the geodesic acceleration $\ddot\gamma$. The geodesic equation relates this acceleration to the Levi-Civita connection $\ddot{\gamma}^k(t) = - \Gamma^{k(0)}_{ij}(\gamma(t)) \dot{\gamma}^i(t)\dot{\gamma}^j(t)$. Meanwhile, formulas due to \citet{eguchi:1985} relate the third derivatives of the dual contrast functions $D_{\bar{M}}$ and $D_{\bar{M}}^*$ to their respective dual connections $\nabla$ and $\nabla^*$. The key observation is that $\nabla^{(0)} = \tfrac{1}{2}(\nabla + \nabla^*)$. When the third derivatives of the forward contrast function 
        $\dd^3 / \dd t^3$
        $D_{\bar{M}}(\bth\|\gamma(t))$ are summed with that of the backward contrast function $\dd^3 / \dd t^3$
        $D_{\bar{M}}(\gamma(t)\|\bth)$ (the dual), the cubic terms associated with the connection $\nabla^{(0)}$ 
        exactly cancel with those third-order derivative terms that can be rewritten using Eguchi's formulas \citep[See][pg. 358]{eguchi:1985}. Thus, $\psi(t) = t^2 + O(t^4)$ and evaluating it at $t = d(\bth, \bth')$ gives the result.
    \end{proof}
\end{theorem}

It should be noted that the proof of the result above does not rely on any specific property of a particular contrast function beyond the general assumptions that it defines a Riemannian geometry. Consequently, the result extends naturally to any smooth contrast function considered in the literature.

We now turn to the paper's main theoretical result: a closed-form approximation of the logarithmic map that accounts for both the Fisher-Rao geometry and the prior distribution -- encoding the full posterior geometry -- and overcomes the computational difficulties that has limited \bWG s to low dimensional settings, making them practical as posterior approximations for Bayesian statistical inference in larger parameter spaces.
\begin{theorem}[Approximate logarithmic map] \label{theo:kla}
    Let $\bar{M}$ be the statistical manifold of probability distributions $\rho_{\btheta}$ and let $\pi$ be a prior distribution for $\bth$ with $\phi(\bth) := - \log \pi(\bth)$ and Hessian $\nabla^2\phi(\btheta) \succ 0$ (positive-definite). To encode the posterior geometry in a new contrast function, define $D_{\mathrm{post}}: \bar{M}^2 \to \mathbb{R}_+$ by
    \begin{align} \label{eq:klc}
        D_{\mathrm{post}}(\bth_1 \| \bth_2) 
        &= D_{\bar{M}}(\rho_{\bth_1} \| \rho_{\bth_2}) 
        +
        D_{\bar{M}}^*(\rho_{\bth_1} \| \rho_{\bth_2}) 
    \end{align}
    where $D_{\bar{M}}(\rho_{\bth_1}\|\rho_{\bth_2}) = \textsl{KL}(\rho_{\bth_1}\|\rho_{\bth_2}) + D_\phi(\bth_1\|\bth_2)$, $D^*_{\bar{M}}(\rho_{\bth_1}\|\rho_{\bth_2}) = \textsl{KL}(\rho_{\bth_2}\|\rho_{\bth_1}) + D_\phi(\bth_2\|\bth_1)$ the dual and $D_\phi$ the Bregman divergence of $\phi$. In $D_{\mathrm{post}}(\btheta_1 \| \btheta_2)$ fix $\btheta_2 = \btheta'$ and define the function $D'_{\textit{post}}(\cdot \| \btheta') : \bar{M} \to \mathbb{R}_+$. Then the function $\olsi{\mathrm{Log}}_{\bth} : \bar{M} \to T_{\bth} \bar{M}$ given by
    \begin{align} \label{eq:alm}
        \olsi{\mathrm{Log}}_{\bth}(\bth') 
        = -\tfrac{1}{2}\, G_{\Theta}^{-1}(\bth)\, \nabla_{\bth_1} D'_{\mathrm{post}}(\bth_1 \| \bth')\big|_{\bth_1 = \bth}
    \end{align}
    is an approximate logarithmic map that encodes the posterior geometry. The matrix $G_{\Theta}(\btheta) = \mathcal{I}_{\Theta}(\btheta) + \nabla^2 \phi(\btheta)$ is the matrix coefficient from the metric $g$ induced by the contrast function $D_{\bar{M}}$ (or its dual $D^*_{\bar{M}}$) and $\mathcal{I}_\Theta(\btheta)$ is the Fisher information matrix.
    \begin{proof}[Partial proof:]
        The function $D_{\mathrm{post}}$ defined as the sum of the symmetrized contrast function $D_{\bar{M}}$ defines a contrast function on $\bar{M}$. The Formula \eqref{eq:alm} follows from the idea of Theorem~\ref{theo:ard} and the recovery of logarithmic maps by \autoref{eq:logmap}. Since $\olsi{\mathrm{Log}}_{\bth}(\bth) = 0$ and $D\olsi{\mathrm{Log}}_{\bth}(\bth) = I$ it follows from \citet{jump:2021} that $\olsi{\mathrm{Log}}_{\bth}$ is an approximate logarithmic map. 
    \end{proof}
\end{theorem}

Theorem~\ref{theo:kla} eliminates the computational limitation that has restricted \bWG s to higher-dimensional parameter spaces. The approximate logarithmic map \eqref{eq:alm} bypasses the need for numerical integration of the geodesic equations and the Jacobi fields needed in previous implementations in its entirety. It requires the Hessian and the gradient of the divergence at $\bth^*$ once. Subsequent evaluation of the map itself is costless. These operations are readily and widely available through automatic differentiation tools. This shifts \bWG s from a niche method for small models to a more practical tool, increasing the scalability and accuracy of approximate Bayesian inference methods. We detail the range of applications in the next section; here we establish the geometric underpinnings.

Formally, we now work with the manifold $\bar{M}$ equipped with the metric $g$ whose matrix coefficient is $G_\Theta = \mathcal{I}_\Theta + \nabla^2\phi$. Following
\cite{miyamoto:2024}, all the manifolds we will work here are complete when endowed with the Fisher-Rao metric, i.e. $(\bar{M}, g_{\mathrm{fisher}})$. 
Thus, $(\bar{M}, g)$ is also complete as a metric space. By the Hopf--Rinow theorem \citep[Theorem~6.4.6]{klingenberg:1978}, the exponential map $\mathrm{Exp}_{\bth}$ is defined on the entirety of the tangent space $T_{\bth}\bar{M}$. This does not guarantee that geodesics minimize length, but it does ensure they are globally well-defined. Further note that the form of $G_\Theta$ coincides with the metric introduced by \citet{girolami:2011} in the context of Riemannian manifold Hamiltonian Monte Carlo, now formalized through the notion of contrast functions and dual connections. The expression \eqref{eq:alm} has also appeared before, as a scale version of the natural gradient of divergence measures observed by \citet{mallastos:2019}.

A practical question then arises concerning the inverse of $\olsi{\mathrm{Log}}_{\bth}$, as sampling would imply pushing forward through an approximate exponential map. The Inverse Function Theorem provides a local inverse on the neighborhood $U_{\bth} \subset \bar{M}$ where the Hessian matrix $\nabla^2_{\bth'} D_{\mathrm{post}}(\bth \| \bth')$ is nonsingular, $\olsi{\mathrm{Log}}_{\bth}$ is a diffeomorphism onto its image in $T_{\bth} \bar{M}$. A global inverse, however, is more intricate. It does not exist for overparameterised models such as neural networks, but within a large class of regression models in the exponential families -- which encompass various models used for downstream tasks in Statistics, including generalized linear models, fixed-effects, random-effects, mixed-effects, Gaussian latent models (Gaussian processes) and many others -- explicit conditions guaranteeing global inverse can be derived analytically. 
\begin{corollary}[Global inverse]\label{cor:dlmap}
    Let $\bar{M}$ be a full, regular and minimal exponential family of distributions, and let $\bbeta \in \Xi \subseteq \mathbb{R}^n$ the corresponding natural parametrisation (primal coordinates systems). Let $h : \Theta \to \Xi$, $\bth \mapsto h(\bth) = \bbeta$ be the imbedding function (or reparametrisation map) and let $A : \Xi \to \mathbb{R}$ denote the log-partition function that is strictly convex \citep[][]{banerjee:2005, kass::1997}. Assume that $\Theta$ is open, convex and the negative logarithm of the prior $\phi = -\log \pi$ has Hessian $\nabla^2 \phi(\bth) \succ 0$ in $\Theta$ while $\phi(\bth) \to \infty$ as $\bth \to \partial \Theta$ where $\partial \Theta$ 
    denotes the boundary of $\Theta$. If $h$ is an affine map with a full column rank Jacobian $Dh$, then the approximate logarithmic map $\olsi{\operatorname{Log}}_{\bth}$ admits a global inverse on its domain.
    \begin{proof}[Partial proof:]
        The gradients of the \bKL\, and Bregman divergences reduce to explicit formulas which are derivatives of the function $A$ and the negative log-prior $\phi$. Pre-multiplying it with $G_{\Theta}^{-1}$ yields a closed-form expression for $\olsi{\mathrm{Log}}_{\bth}$. To guarantee global invertibility, it suffices to show that for every tangent vector $V \in T_{\bth}\bar{M}$, the equation $V = \olsi{\mathrm{Log}}_{\bth}(\bth')$ has a unique solution $\bth'$. After a long algebraic rearrangement, the approximate logarithmic map is reformulated as the roots of the gradient of a strongly convex function $\bth' \mapsto u(\bbeta(\bth')) + v(\bth')$ dependent on $V$, and where $u$ and $v$ are strongly convex. From the optimization viewpoint, this is a first-order optimality condition for minimizing a function, and since the function is strongly convex and blows up at the boundary, the solution exists, and it is unique. Hence, the approximate logarithmic map $\bth' \mapsto \olsi{\mathrm{Log}}_{\bth}(\bth')$ has a global inverse, and it can be written as
        \begin{align} \label{eq:appexpmap}
            \olsi{\operatorname{Exp}}_{\bth}(V)
            = 
            \underset{\bth' \in \Theta}{\operatorname{argmin}} \,
            u(\bbeta(\bth')) + v(\bth').
        \end{align}
        Moreover, the determinant of the Jacobian of the approximate logarithmic map is non-vanishing everywhere; thus it is also a diffeomorphism.
    \end{proof}
\end{corollary}

The map \eqref{eq:alm} is now explicit. Once a basepoint $\bth^*$ is chosen, the metric inverse $G_\Theta(\bth^*)^{-1}$ is computed once and stored; evaluating $\olsi{\mathrm{Log}}_{\bth^*}$ requires only the gradient of the contrast function. Moreover, in this configuration, sampling from Equation \eqref{eq:riemwg} -- with the exact logarithmic map \eqref{eq:logmap} replaced with \eqref{eq:alm} -- reduces to the solution of a non-linear least-square problem, whose solution exists and is unique. Density evaluation is likewise efficient, as the Jacobian of Equation \eqref{eq:alm} can be obtained with readily available automatic differential tools. 
In the next section, we highlight these operations, alongside convergence guarantees in the common settings of statistical inference problems. 

\section{Wrapped Gaussian posterior approximations}

The next theorem shows that \bWG s \eqref{eq:riemwg} with approximate logarithmic map \eqref{eq:alm} can asymptotically recover the \bLA\, and, consequently, by the Bernstein-von Mises theorem, the true posterior distribution. Unlike the \bLA\,, however, \bWG s  incorporate geometric corrections induced by the manifold curvature, which, as the experiments will show, indicates strong improvements to the \bLA\,,
bringing the approximation closer to the posterior distribution in the finite-sample regime.
%
\begin{theorem}[Asymptotic equivalence to \bLA] \label{theo:asyn}
    Let $\bar{M} = \{\rho_{\bth} : {\bth} \in \Theta\}$ be a statistical manifold satisfying standard regularity conditions \citep[Sec. 7.3, 7.4]{mark:2011}, with true parameter $\bth_* \in \Theta$. Given a sample of size $n$ and a Gaussian prior, let $\hat\bth_n$ denote the MAP estimate and let $\bar\Sigma$ denote the inverse of the Hessian of the negative log-posterior at $\hat\bth_n$. Assume $\olsi{\mathrm{Log}}_{\hat\bth_n}$ has a global inverse (Corollary \eqref{cor:dlmap}). Then the \bWG,
    \begin{align*}
        \rho_{\bWG}(\bth) = \mathcal{N}\big(\olsi{\mathrm{Log}}_{\hat\bth_n}(\bth) \,\big|\, 0, \bar\Sigma\big) \,\big|\det D\olsi{\mathrm{Log}}_{\hat\bth_n}(\bth)\big|
    \end{align*}
    satisfies
    \begin{align*}
        \rho_{\bWG}(\bth) = \rho_{\textsl{LA}}(\bth) \, \big(1 + O(n^{-1}) 
        \big),
    \end{align*}
    where $\rho_{\textsl{LA}}$ is the \bLA.

    \begin{proof}[Partial proof:]
        By the central limit theorem, $\sqrt{n}(\hat\bth_n - \bth_*) \to \mathcal{N}(0, \mathcal{I}_\Theta(\bth_*)^{-1})$ in probability, and by the law of large numbers, $\bar\Sigma^{-1} \to n\mathcal{I}_\Theta(\bth_*)$ with the negative log-prior Hessian $\Sigma_\pi$ negligible at $O(1)$ compared to $n\mathcal{I}_\Theta$. Thus, typical posterior values $\bth$ lie at distance $\|\bth - \hat\bth_n\| = O(n^{-1/2})$ from the MAP. After Taylor expanding the gradient of the symmetrized contrast function $D_{\mathrm{post}}$ around $\hat\bth_n$, the third-order terms in the symmetrized contrast function cancel by Theorem~\ref{theo:ard}. The expansion simplifies to\looseness=-1
        \begin{align*}
            \olsi{\mathrm{Log}}_{\hat\bth_n}(\bth) 
            = 
            (\bth - \hat\bth_n) + H(\bth - \hat\bth_n) + O(n^{-2})
        \end{align*}
            where $H$ is a cubic correction involving Eguchi's $B$-tensor \citep[See][Prop. 2 and 3]{eguchi:1992}, which measures the manifold's curvature. Since $\|\bth - \hat\bth_n\| = O(n^{-1/2})$, the cubic term is $O(n^{-3/2})$. The Jacobian $D\olsi{\mathrm{Log}}_{\hat\bth_n}$ is then $I + O(n^{-1})$, with determinant $1 + O(n^{-1})$. As $\hat\bth_n \to \bth_*$ 
            when $n \to \infty$
            the \bWG\, matches the \bLA\, to $O(n^{-1})$.
    \end{proof}
\end{theorem}

This establishes \bWG\, as asymptotically well-founded and therefore theoretically justified.
When $\olsi{\mathrm{Exp}}_{\bth^*}$ is not guaranteed to have a global inverse, a \bWG\, density can still be defined. The basic construction puts a Gaussian $\mathcal{N}(\0, \Sigma)$ on $T_{\bth^*}\bar{M}$ and pushes forward through $\olsi{\mathrm{Exp}}_{\bth^*}$, which may be a many-to-one map. In this case, the \bWG\, admits a density representation w.r.t. the Lebesgue measure via the change of variable formula for non-injective maps \citep[See][Theorem 2.1]{james:2023}. Specifically, 
\begin{align} \label{eq:riemwg1}
    \rho_{\bWG}(\bth) = 
    \sum_{V \, \in\, \olsi{\mathrm{Log}}_{\bth^*}(\bth)}
    \mathcal{N}\big(V| \0, \Sigma\big)
    \Big|\det\big(D\,\olsi{\mathrm{Log}}_{\bth^*}(\bth)
    \big)
    \Big|
\end{align}
where the sum runs over all pre-images of $\bth$ under $\olsi{\mathrm{Exp}}_{\bth^*}$. As the density representation is available in closed-form, because $\olsi{\mathrm{Log}}_{\bth^*}$ is parametric, density evaluation requires only the Jacobian determinant of $\olsi{\mathrm{Log}}_{\bth^*}$ -- again, this can be computed via automatic differentiation, with no Jacobi field ODEs. Observe that, within the injectivity radius of $\bth^*$, or whenever $\olsi{\mathrm{Exp}}_{\bth^*}$ admits a global inverse, as in Corollary \eqref{cor:dlmap}, the above sum reduces to a single term with $V = \olsi{\mathrm{Log}}_{\bth^*}(\bth)$. 
Samples from $\eqref{eq:riemwg1}$
are obtained by solving the non-linear least-square problem, following the perspective of Corollary~\ref{cor:dlmap} (see appendix). That is, to sample from this \bWG\, first choose $(\bth^*$, $\Sigma)$ and follow the general steps:
\begin{align} \label{eq:nlsp}
    \textrm{draw} \, V &\sim \mathcal{N}(\0, \Sigma), \nonumber \\
    \textrm{set} \, \btheta_V &= 
    \underset{\btheta \in \Theta}{\operatorname{argmin}} \,
    \| V - \olsi{\mathrm{Log}}_{\bth^*}(\bth) \|^2.
\end{align}

Our approximations now open up new theoretical and practical directions: they induce entirely new classes of distributions — particularly for exponential families — that inherit the intrinsic Fisher-Rao and prior geometries, combining geometric structure with computational feasibility. Moreover, once Theorem~\ref{theo:ard} is general, it may be possible to create new classes of geometric induced distributions from various alternative contrast functions presented in the literature. 
Section~\ref{sec:experiments} demonstrates these advantages in concrete problems, where \bWG s can operate at a scale that was previously out of reach while maintaining good posterior approximation accuracy.

\section{Experiments}\label{sec:experiments}

This section presents several models that are widely used in various different areas of application. They comprise widely used generalized linear models, population growth models, Gaussian latent models (GLM, Gaussian processes) and non-linear regressions (in $\bth$) with neural networks in ML. In each example, we use the information geometry notation and denote each particular manifold by $\bar{M}$ from which the approximate logarithmic map is constructed. We also highlight whether each particular map is a diffeomorphism. It is important to note that we prefer this view because the elements of $\bar{M}$ are not likelihood functions, and we need to compute the respective \bKL\, divergence (contrast function) between two of its elements, which is also given in closed-form for the majority probabilistic models used in practice. For all the following examples, we will denote the working parametrisation as $\bth$ and its role will be made clear in the respective subsection. This is done to avoid cluttering up the notation. The basepoint of the approximate logarithm map is always set about the MAP estimate, i.e., $\bth^* = \hat{\bth}$ and the negative Hessian of the log-posterior distribution at the MAP denoted as $\bar{\Sigma}$. Related to the prior geometry, in all the following examples we choose Gaussian priors, this means that for $\pi(\bth) = \mathcal{N}(\bth|\0, \Sigma_\pi)$, the Hessian $\nabla^2 \phi(\bth) = \Sigma_\pi^{-1}$ with the choice of $\Sigma_\pi$ specified in the example. With a smaller modification to avoid extra matrix inversions, the approximate logarithmic maps are implemented in the following form,
\begin{small}
\begin{align} \label{eq:almE}
    \olsi{\mathrm{Log}}_{\hat{\bth}}(\bth) 
    &
    = 
    -\tfrac{1}{2} 
    \big(\Sigma_\pi \, \mathcal{I}_{\Theta}(\hat\bth) + I_D\big)^{-1}
    \Big(
    \Sigma_\pi \,
    \nabla_{\bth_1} 
    \big(
    \textsl{KL}(\rho_{\bth_1} \| \rho_{\bth}) 
    + 
    \textsl{KL}(\rho_{\bth} \| \rho_{\bth_1})
    \big)
    \big|_{\bth_1 = \hat{\bth}}
    +
    2(\hat{\bth} - \bth)
    \Big).
\end{align}
\end{small}
%

For computational implementations, note that the matrices $\overline{\Sigma}$, $\Sigma_\pi$ and $(\Sigma_\pi \, \mathcal{I}_{\Theta}(\hat{\btheta}) + I_D)^{-1}$ are all fixed, so that they can be stored in cache memory. The gradients of the \bKL\, divergences are evaluated at the MAP. This can also be pre-computed and stored only once, and the evaluation of the approximate logarithmic map alongside its derivatives w.r.t.\@ $\bth$ becomes very efficient. As sampling is now defined by a minimization problem, Equation \eqref{eq:nlsp}, it can be approached with any gradient-based optimization algorithm. 

Throughout the subsequent examples, we also benchmark against the symmetric skew approximation of \citet{pozza:2026} (\bSP), which perturbs \bLA\, by a 'skew' factor derived from the ratio of posterior densities evaluated at $\bth$ and at symmetric point $2\hat{\bth} - \bth$ around the MAP, requiring no additional optimization. We denote this approximate density as $\rho_{\bSP}$, with evaluation and sampling given, respectively, in Equation (5) and Algorithm 1 of \citet{pozza:2026} (See p. 9 and 11 respectively).

\subsection{Poisson and Bernoulli manifolds}

We start considering general formulations for two largely used cases, the Poisson and Bernoulli manifolds as follows
\begin{align*}
    \bar{M} = \left\lbrace 
        \rho_{\bth}(\by) = 
        \exp \big( \bbeta^\top T(\by) + B(\by) - A(\bbeta) \big)
        : 
        \eta_i = h(\bx_i, \bth), 
        \bth \in \Theta \subseteq \mathbb{R}^D, 
        \,
        \by \in \Omega 
    \right\rbrace
\end{align*}
where $\bbeta = (\eta_1, \ldots, \eta_n)$ and $T(\by) = (y_1, \ldots, y_n)$ in both model cases. For the Bernoulli case $\Omega = \{0, 1\}^n$, $B(\by) = 0$, and log-partition function $A(\bbeta) = \sum_{i = 1}^n \log(1 + e^{\eta_i})$. For the Poisson case we have $\Omega = \mathbb{N}^n$, $B(\by) = \sum_{i = 1}^n \log (1 / y_i!)$ and $A(\bbeta) = \sum_{i = 1}^n e^{\eta_i}$. 

\begin{figure}[!ht]
    \vskip 0.0in
        \begin{center}
            \centerline{
                \includegraphics[scale = 0.87]{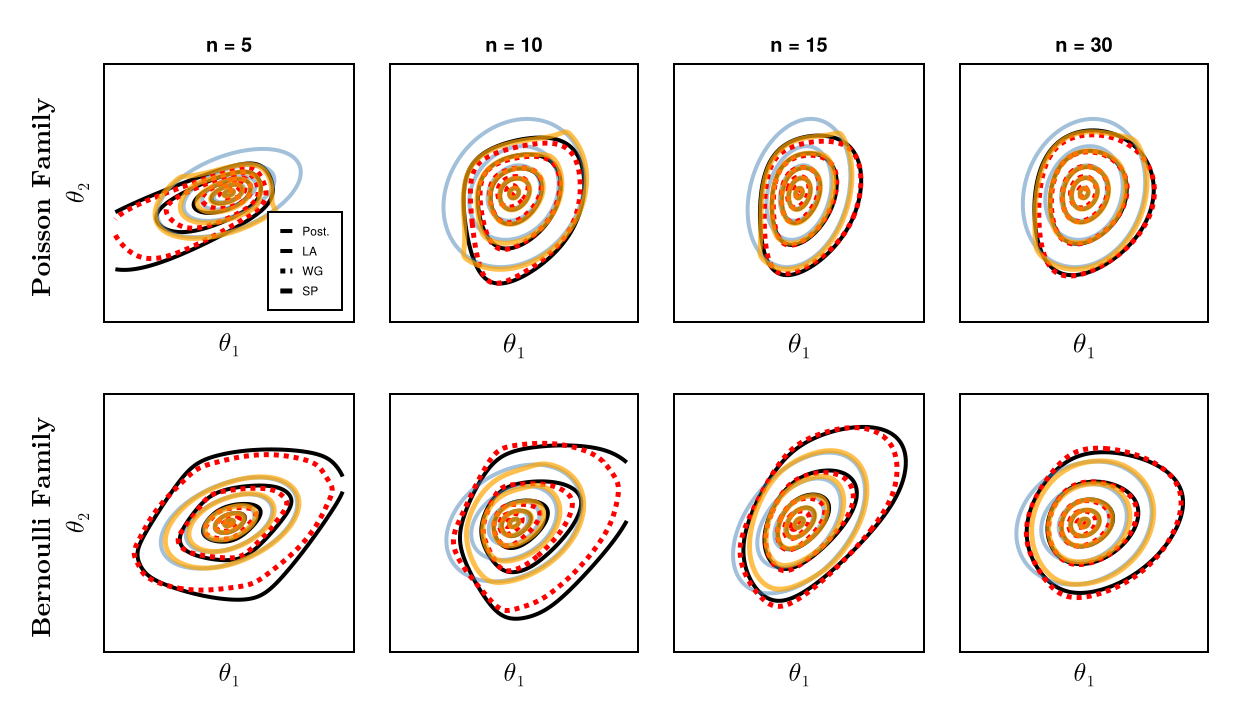}
            }
            \caption{For each number of data points $n \in \{5, 10, 15, 30\}$, we randomly pick values to form the constant matrix $\mathbf{X} \, (n \times 2)$. We also pick random covariance matrices for the prior distributions for each case of $n$, that is, the prior is given by $\btheta = (\theta_1, \theta_2) \sim \mathcal{N}(0, \Sigma)$ and the covariance is draw as $\Sigma \sim \mathrm{Wishart}(2, \mathrm{diag}(10, 10))$. True posterior is showed in ({\Large \textbf{-}}, solid), the \bWG\, in ({\color{red} {\Large \textbf{-}}}, dashed), \bLA\, in ({\color{cyan} {\Large \textbf{-}}}, solid) and \bSP\, in ({\color{orange} {\Large \textbf{-}}}, solid). In this case the \bWG\, approximation presents generally better fit compared to the \bLA\, and \bSP\, method. As the number of data points increase all approximations tend to a Gaussian shape. 
            }
            \label{fig:1}
        \end{center}
    \vskip -0.1in
\end{figure}

The majority of the statistical modelling approaches in real applications naturally account for the product manifold structure, once it is analogous to the independence assumption between data points. To further clarify this configuration, the following choices of imbedding function will make the underlying structure more transparent with the previous section (Subsection \eqref{sec:ig}), and perhaps surprisingly, by revealing how general it can be and the different types of models it accommodates. For the generic choice of $\bbeta = (\eta_1, \ldots, \eta_n) := h(\btheta) = \mathbf{X}\boldsymbol{\theta}$ where $\mathbf{X}$ is $n \times d$, $\btheta \in \mathbb{R}^d$ and $\operatorname{rank}(\mathbf{X}) = d$ it comprises three widely used cases. In particular, when $\mathbf{X} = \mathds{1}_n$ we get $\theta \in \mathbb{R}$ and this leads to 1-dimensional $\bar{M}$, still on the natural parametrisation. For arbitrary $\mathbf{X}$, this leads to the usual regression models, and for $\mathbf{X} = I_n$ we get $\bbeta = h(\btheta) = \btheta$, with $\btheta \in \mathbb{R}^n$, and this is associated with the form of a GLM. In Figure~\ref{fig:1}, we present visual examples for the Poisson and Bernoulli manifold in $d = 2$ dimensions.

An important special case arises when $\mathbf{X} = I_n$. In this case, we recover a GLM formulation for classification and counts, in other words, Gaussian process classification and Poisson count models with hierarchical Gaussian processes. Such models have been extensively studied in both the statistics and ML literature; see, for example, \citet{rue2009approximate, kuss:2005} and \citet{nickisch:2008}. Moreover, since $h$ is an affine transformation, the resulting approximate logarithmic maps are diffeomorphisms.

We also emphasize that, in the previous constructions, we worked directly with the natural parametrisation. However, even if the analysis were carried out in the expectation parametrisation, the approximate logarithmic map would still preserve convexity. This follows from the Legendre duality between the expectation and natural parametrisations in exponential families. This dual relationship ensures that convexity is preserved under the transformation induced by $h$ when it transforms between natural and expectation parametrisations, and vice-versa \citep[see][Section 4.2 and item 7, p. 1741]{barn:2014, banerjee:2005}. 

In Table \ref{tab:1} we present the result of an experiment comprising five different datasets. The experiment consists in measuring the Wasserstein-2 ($W_2$) distance and the Sliced-Wasserstein-2 ($SW_2$) distance between the approximate posteriors $\rho_{\textsl{WG}}$ \eqref{eq:riemwg1}, $\rho_{\textsl{NUTS}}$ given by the \bHMC\, samples \citep{hoffman:14}, $\rho_{\bSP}$ \citep{pozza:2026} and the traditional \bLA\, denoted as $\rho_{\bLA}$. The true posteriors are also obtained using different model constructions that are based on different imbedding choices. The first choice is the imbedding $h(\btheta) = \mathbf{X}\boldsymbol{\theta}$, which leads to standard regression models. The latter choice is $h(\btheta) = \btheta$, which leads to a GLM type of model. 
In the regression settings ("parametric models"), we collect only 10\% of the total number of data points for each dataset, so that the degrees of freedom are not so large that they would lead to posteriors that may be too close to a Gaussian, and so the effect of the approximations would disappear as we know that both posterior and approximations would converge to a Gaussian if the number of data points is large. The priors are chosen as $\btheta \stackrel{\mathrm{i.i.d}}{\sim} \mathcal{N}(0, \sigma^2 I_d)$ with $\sigma^2 = 100$. For the GLM type of models, we do the same task as before, but we choose $n = d = 100$ for all cases, except for the last microassay dataset, which has the total number of data points $n_{\mathrm{Total}} = 90$. In these later cases our priors are the $n$-dimensional multivariate Gaussian distribution given by $\btheta \sim \mathcal{N}(\0, K)$, where $K$ is a $n \times n$ covariance matrix constructed from the covariance function $k(\bx, \bx') = \sigma^2 \exp \big(- \tfrac{1}{2} \sum_{j = 1}^D (x_j - x'_j)^2 / \ell_j^2 \big)$ with the same variance $\sigma^2 = 100$ and length-scales $\ell_j = 1000 \ \forall j$, where $D$ is the covariate dimension. We make this choice so that the Gaussian prior creates the dependency on the true posterior. If the length-scales were to be small, it may be that the posterior is fully factorizable, thus independent, and the approximation for the full posterior distribution (joint) would not be necessary. For all models we drew $12 \times 10^3$ samples. 

\begin{table}[ht]
\centering
\footnotesize
\caption{Comparison of Wasserstein-2 ($W_2$) and Sliced Wasserstein-2 ($SW_2$) distances between the approximate posterior distributions $\rho_{\textsl{WG}}$, $\rho_{\textsl{LA}}$ and $\rho_{\textsl{SP}}$, and the reference posterior assumed to be from NUTS samples. Results are reported for Bernoulli and Poisson manifolds under regression and Gaussian latent models (GLM). Lower values indicate better agreement with the reference posterior. Across the 20 models combinations with five datasets, \bWG\, attains the lowest $W_2$ and $SW_2$ distances to the NUTS reference in nearly all cases, with close agreement in the cases of australian and microassay regressions models for the Bernoulli and Poisson manifolds respectively. In the appendix we present histograms and joint scatter plots for these two particular aforementioned cases.
} \label{tab:1}

    \begin{tabular}{llcccccccc}
    \toprule
    &
    & \multicolumn{4}{c}{Bernoulli manifold}
    & \multicolumn{4}{c}{Poisson manifold} \\
    \cmidrule(lr){3-6}
    \cmidrule(l){7-10}
    
    Dataset & Approx.
    & \multicolumn{2}{c}{Regression}
    & \multicolumn{2}{c}{GLM}
    & \multicolumn{2}{c}{Regression}
    & \multicolumn{2}{c}{GLM} \\
    
    &
    & \multicolumn{2}{c}{$n=0.1\,n_{\mathrm{Total}}$}
    & \multicolumn{2}{c}{$n=100$}
    & \multicolumn{2}{c}{$n=0.1\,n_{\mathrm{Total}}$}
    & \multicolumn{2}{c}{$n=100$} \\

    &
    & $W_2$ & $SW_2$
    & $W_2$ & $SW_2$
    & $W_2$ & $SW_2$
    & $W_2$ & $SW_2$ \\
    \midrule
    
    Australian
    & $\rho_{\textsl{WG}}$
    & $\mathbf{16.191}$  & $\mathbf{3.488}$
    & $\mathbf{66.204}$ & $\mathbf{2.106}$
    & $\underline{\mathbf{1.707}}$  & $\mathbf{0.224}$
    & $\mathbf{49.411}$ & $\mathbf{1.878}$ \\
    $n_{\mathrm{Total}}=690,\;D=15$
    
    & $\rho_{\textsl{LA}}$
    & $21.145$ & $4.726$
    & $71.970$ & $3.760$
    & $2.150$ & $0.389$
    & $55.696$ & $3.146$ \\
    
    & $\rho_{\textsl{SP}}$
    & $17.717$ & $3.866$
    & $69.580$ & $3.254$
    & $\underline{\mathbf{1.707}}$ & $0.244$
    & $54.372$ & $3.002$ \\
    \midrule
    
    German
    & $\rho_{\textsl{WG}}$
    & $\mathbf{9.102}$ & $\mathbf{1.083}$
    & $\mathbf{28.442}$ & $\mathbf{0.778}$
    & $\mathbf{1.676}$ & $\mathbf{0.073}$
    & $\mathbf{18.426}$ & $\mathbf{0.618}$ \\
    $n_{\mathrm{Total}}=1000,\;D=25$
    
    & $\rho_{\textsl{LA}}$
    & $11.580$ & $1.731$
    & $30.682$ & $1.362$
    & $1.747$ & $0.112$
    & $20.991$ & $1.099$ \\
    
    & $\rho_{\textsl{SP}}$
    & $9.608$ & $1.256$
    & $28.92$ & $0.891$
    & $1.718$ & $0.085$
    & $18.812$ & $0.756$ \\
    \midrule
    
    Heart
    & $\rho_{\textsl{WG}}$
    & $\mathbf{17.812}$ & $\mathbf{3.802}$
    & $\mathbf{37.812}$ & $\mathbf{0.888}$
    & $\mathbf{2.732}$ & $\mathbf{0.184}$
    & $\mathbf{24.532}$ & $\mathbf{0.791}$ \\
    $n_{\mathrm{Total}}=270,\;D=14$
    
    & $\rho_{\textsl{LA}}$
    & $22.410$ & $5.035$
    & $39.388$ & $1.438$
    & $2.899$ & $0.313$
    & $26.549$ & $1.362$ \\
    
    & $\rho_{\textsl{SP}}$
    & $19.604$ & $4.239$
    & $38.011$ & $1.365$
    & $2.747$ & $0.205$
    & $25.627$ & $1.090$ \\
    \midrule
    
    Pima
    & $\rho_{\textsl{WG}}$
    & $\textbf{1.687}$ & $\mathbf{0.326}$
    & $\mathbf{35.072}$ & $\mathbf{1.218}$
    & $\mathbf{8.585}$ & $\mathbf{2.757}$
    & $\mathbf{26.958}$ & $\mathbf{0.873}$ \\
    $n_{\mathrm{Total}}=532,\;D=8$
    
    & $\rho_{\textsl{LA}}$
    & $1.742$ & $0.717$
    & $38.877$ & $1.835$
    & $12.639$ & $4.303$
    & $30.676$ & $1.696$ \\
    
    & $\rho_{\textsl{SP}}$
    & $1.731$ & $0.530$
    & $35.851$ & $1.461$
    & $10.268$ & $3.291$
    & $28.449$ & $1.312$ \\
    \midrule
    
    
    
    
    Microassay 
    & $\rho_{\textsl{WG}}$
    & $\underline{\mathbf{6.914}}$ & $\mathbf{2.570}$
    & $\mathbf{3.866}$ & $\mathbf{0.033}$
    & $\mathbf{7.063}$ & $\mathbf{2.660}$
    & $\mathbf{3.684}$ & $\mathbf{0.040}$ \\
    $n_{\mathrm{Total}}=90,\;D=4$
    
    & $\rho_{\textsl{LA}}$
    & $9.454$ & $3.959$
    & $3.934$ & $0.057$
    & $9.490$ & $4.066$
    & $3.830$ & $0.087$ \\
    
    & $\rho_{\textsl{SP}}$
    & $\underline{\mathbf{6.915}}$ & $2.584$
    & $3.946$ & $0.055$
    & $7.080$ & $2.741$
    & $3.853$ & $0.087$ \\
    \bottomrule
    
    \end{tabular}
\end{table}

\subsection{Multinomial manifolds}
We now extend the Bernoulli construction to multivariate outcomes via the multinomial family. This manifold is denoted as follows,
\begin{align*}
    \bar{M} = \left\lbrace 
        \rho_{\bth}(\by) = 
        \prod_{i = 1}^n \prod_{c = 1}^C 
        (\mu_{c} \circ \bbeta_i)(\bth)^{y_{ic}}
        : 
        \eta_{ic} = h_c(\bx_{ic}; \bth_c),
        \bth \in \Theta \subseteq \mathbb{R}^D,
        \, \eta_{iC} = 0 \ \forall \, i 
    \right\rbrace
\end{align*}
where $\by \in \Omega = \{e_1, \ldots ,e_C \}^n$, $\mu_c \circ \bbeta_i = \exp(\eta_{ic})/\sum_{c = 1}^C \exp (\eta_{ic})$ and $\sum_{c = 1}^C \mu_c(\bbeta_i) = 1$, $\forall i$. The effective vector of natural parameters is denoted as $\bbeta_i = (\eta_{i, 1}, \ldots, \eta_{i, C-1})$ with $\bbeta =$ $(\bbeta_1$ $,\ldots,$ $\bbeta_n)$. 
The forms of the function $h$ are analogous to the previous cases, that is $\bbeta = h(\btheta) = \mathbf{X}\boldsymbol{\theta}$ where $\mathbf{X} = \mathrm{diag}(\mathbf{X}_1, \ldots, \mathbf{X}_{C -1})$ is a block matrix having dimension $n (C - 1) \times d (C-1)$. Each matrix $\mathbf{X}_c$ has dimension $n \times d$ with $d$ being the number of parameters associated with each class (they could also have a different number of regressors/parameters), and the full vector of parameters becomes $\btheta = (\theta_{1, 1}, \ldots, \theta_{1, d}, \ldots, \theta_{C-1, 1}, \ldots, \theta_{C-1, d}) \in \Theta = \mathbb{R}^{d(C-1)}$. If $\mathbf{X}_c = \mathds{1}_n$ then $\btheta = (\theta_1, \ldots, \theta_{C - 1})$. When $n = d$, we have $\mathbf{X} = I_{n(C-1)}$, and the model has the structure of a multi-class GLM (or multi-class Gaussian process). In all these cases, the approximate logarithmic maps are diffeomorphisms whenever $\mathrm{rank}(\mathbf{X}) = d(C - 1)$. The approximate logarithmic maps are diffeomorphisms since $h$ is an affine map. The latter case has also been extensively studied in the literature by \citet{Girolami:2006} and \citet{chai:2012}. It is also possible to consider a multivariate extension for the Poisson manifold, namely, the negative-multinomial distribution, see for example \citet{li:2020}.

\begin{figure}
    \begin{center}
    \hspace{-0.5cm}
    \begin{tabular*}{\columnwidth}{ ccc }
        \includegraphics[scale = 0.54]{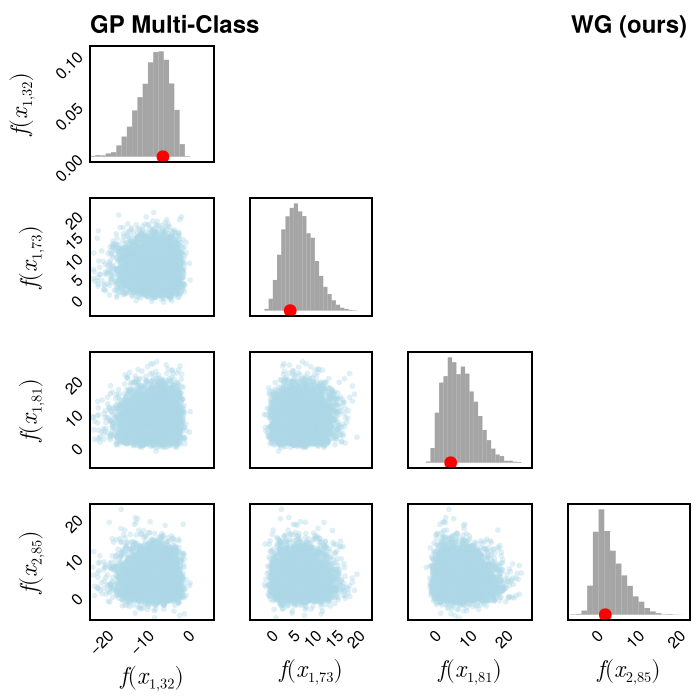}
        \hspace{-0.37cm}
        &
        \includegraphics[scale = 0.54]{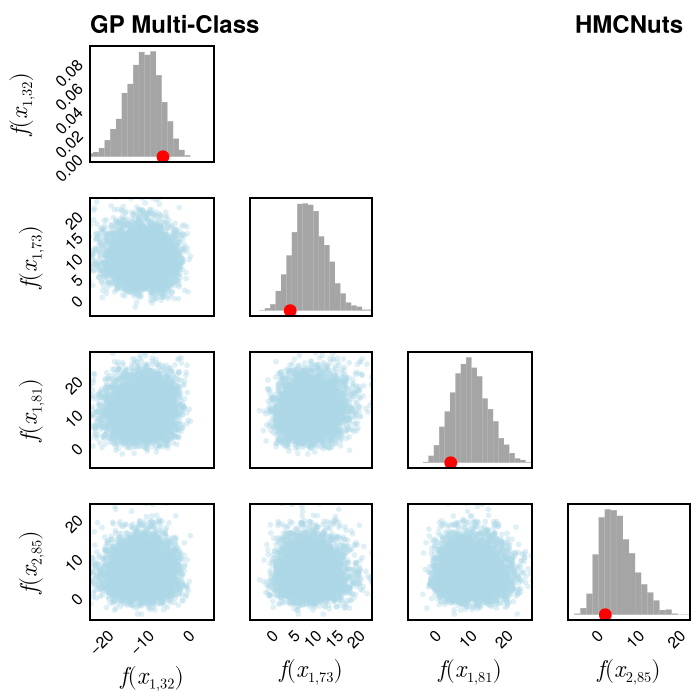}
        \hspace{-0.37cm}
        &
        \includegraphics[scale = 0.54]{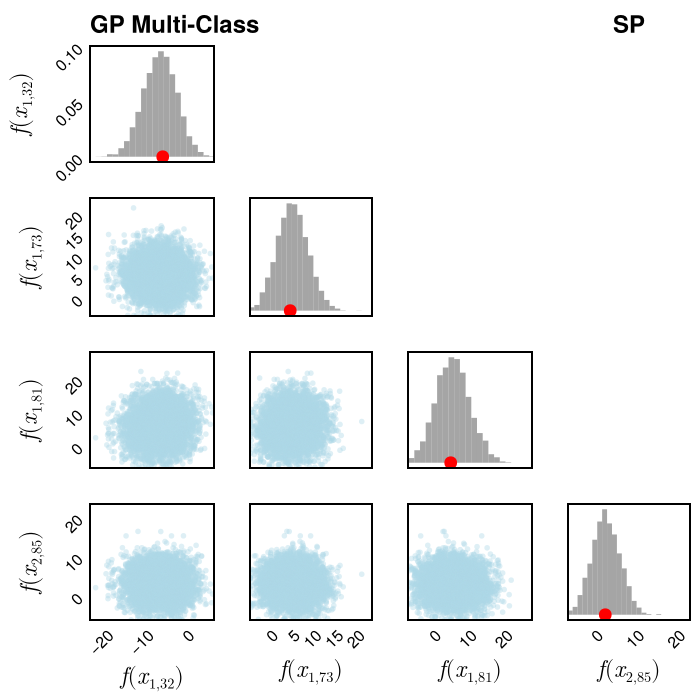}
    \end{tabular*}
        \caption{Posterior scatter plot comparison for particular function values in the multiclass manifold case. The left panel show the scatter plot of samples from $\bWG$. The middle panel shows the same plot for the \bHMC\, samples and the right panel shows the \bSP\, samples. The ({\color{red} $\bullet$}) dots depict the location of the MAP estimate. The $W_2$ distances between different approximation methods are $W_2(\rho_{\textsl{WG}}, \rho_{\textsl{NUTS}}) \approx 81.617$, $W_2(\rho_{\textsl{LA}}, \rho_{\textsl{NUTS}}) \approx 99.734$ and $W_2(\rho_{\textsl{SP}}, \rho_{\textsl{NUTS}}) \approx 95.052$. The $SW_2$ distances are $SW_2(\rho_{\textsl{WG}}, \rho_{\textsl{NUTS}}) \approx 2.367$, $SW_2(\rho_{\textsl{LA}}, \rho_{\textsl{NUTS}}) \approx 4.648$ and $SW_2(\rho_{\textsl{SP}}, \rho_{\textsl{NUTS}}) \approx 3.557$. Also, \bWG\, gives approximately $0.181$ relative improvement towards \bLA\,, while \bSP\, gives $0.046$ of relative improvement. This strongly indicates that the \bWG\, is closer to \bHMC\, than the \bLA\, or \bSP\, approximations.} 
        \label{fig:multiclass}
    \end{center}
\end{figure}

In this example, we use the US Postal Service (USPS) database of handwritten $0$-$9$ digits, which consists of 4699 segmented 16 $\times$ 16 greyscale images normalized so that the intensity of the pixels lies in $[-1, 1]$. We take the multinomial model with $C = 3$ and consider only the USPS $3$s vs $5$s vs $7$s data digits, which sums up to 1157 images. From this subset, we randomly select $100$ data points among the three classes. We approach this problem with the GLM and so the posterior dimension becomes $n(C - 1) = 200$. For the vector of latent variables we assume that $\btheta \sim \mathcal{N}(\0, K)$ where $K = \mathrm{diag}(K_1, K_2)$ with $K_1$ and $K_2$ also formed from the covariance function $k(\bx, \bx') = \sigma^2 \exp \big(- \tfrac{1}{2} \sum_{j = 1}^D (x_j - x'_j)^2 / \ell_j^2 \big)$ with $\sigma^2 = \ell_j = 100 \, \forall j$ where $D = 16^2$ is the covariate dimension. Particularly, the entries of both covariance matrices are given by $\{K_1\}_{i, j} = \{K_2\}_{i, j} = k(\bx_i, \bx_j)$ where $\bx_i$ is the $i^{\mathrm{th}}$ grayscale image in a vector representation. Observe that the parametrisation of $\bar{M}$ in consideration is different from that presented in \citet{rasm:2006} Section 3.5. In their work, the parametrisation is nonidentifiable as it considers an extra parameter for the last class $C$. We have avoided that, as it can bring extra difficulties for MCMC methods. We drew $5 \times 10^3$ sample from $\rho_{\bWG}$, $\rho_{\textsl{NUTS}}$, $\rho_{\bLA}$ and $\rho_{\bSP}$. Our method took approximately 17 minutes to complete a run, while the \bHMC\, took approximately 3 hours. We also use $W_2$ and $SW_2$ distances to evaluate the results of this experiment which are presented and discussed in Figure \ref{fig:multiclass}.

\subsection{Gaussian manifolds}
For the Gaussian class of manifolds, we begin by considering four cases. One is the usual Gaussian posterior in the case of linear models with Gaussian noise, and three other cases commonly discussed in the literature: the Squiggle, the Rosenbrock and Neal’s funnel densities. For each of these models, our methodology can be applied by making appropriate choices of the manifolds, the imbedding function, and the prior distributions. An exception appears in the case of Neal’s funnel, where the basepoint is not chosen as the MAP estimate, but rather as a null basepoint. We show that, with this particular choice of basepoint, the proposed methodology exactly recovers Neal’s funnel. We also consider settings involving non-affine imbedding functions. In particular, we study an example that arises from population growth models \citep{sharp:2022}, whose scenario is crafted to induce an ``unusual'' geometry of the posterior distribution in a low-dimensional regime. We finalise with an example of a non-affine imbedding function using an overparameterised neural-network as in \citet{bergamin:2023}.

\subsubsection{Classical cases in the literature}

\begin{figure}[!ht]
    \vskip 0.0in
        \begin{center}
            \begin{tabular*}{\columnwidth}{ cc }
                \includegraphics[scale = 0.56]{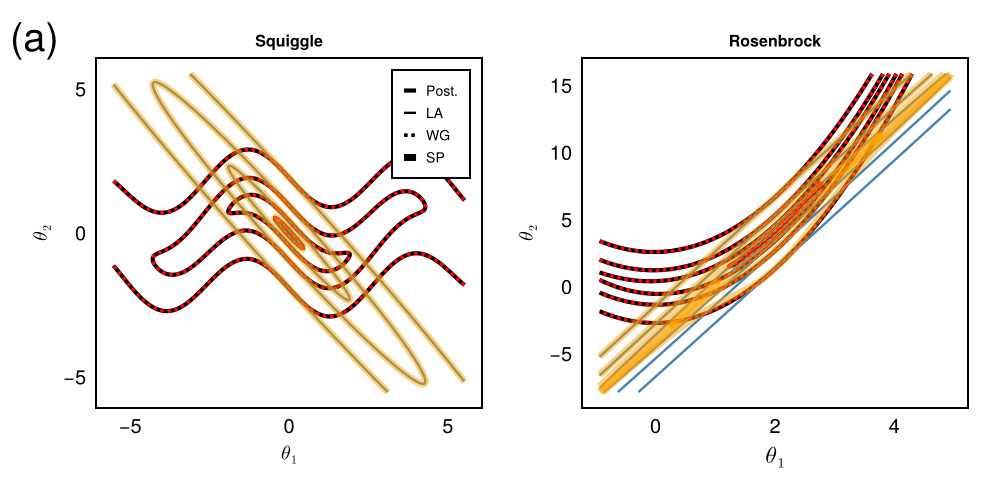}
                \hspace{-0.3cm}
                &
                \includegraphics[scale = 0.56]{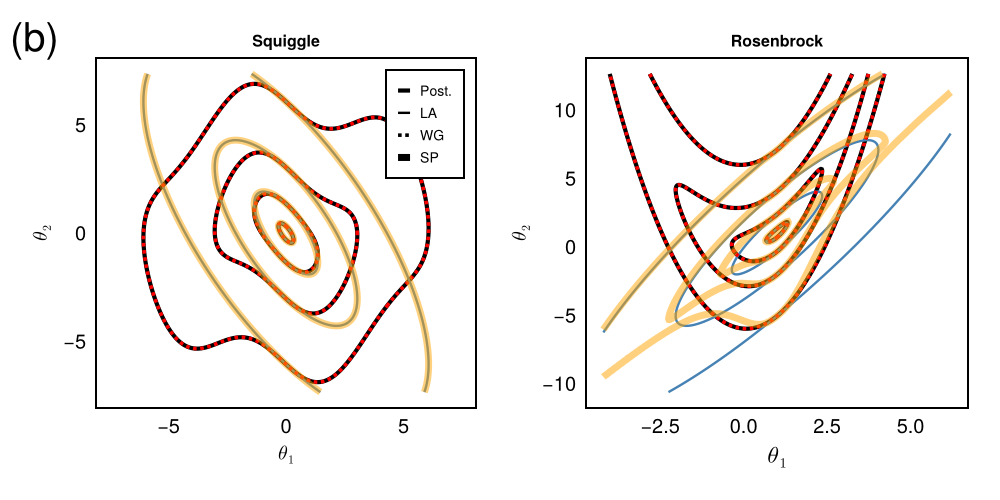}
                \hspace{-0.3cm}
            \end{tabular*}
            \caption{
                True target in ({\color{black} {\Large \textbf{-}}}, solid), \bWG\, in ({\color{red} {\Large \textbf{-}}}, dashed), \bLA\, in ({\color{blue} {\Large \textbf{-}}}, solid) and \bSP\, in ({\color{orange} {\Large \textbf{-}}}, solid). Both target distributions, $\rho_1$ (squiggle) and $\rho_2$ (Rosenbrock), exhibit strongly curved, non-elliptical geometries that a Gaussian approximation cannot capture. \bWG\, closely tracks the true target in both cases, correctly recovering the oscillatory and banana-shaped contours. \bSP\, offers minimal improvement over \bLA\, in panels (a) left-side and (b) right-side. Since it only redistributes mass between mirrored points about the basepoint rather than reshaping the approximating density, it cannot correct for the underlying strong curvature, and its gains remain none or small relative to the \bWG.
            }
            \label{fig:classics}
        \end{center}
    \vskip -0.1in
\end{figure}

For the Gaussian linear case and Neal's funnel, we recover the exact posterior; the proofs are left in the supplementary materials (appendix). For the squiggle, the density function is given $\rho_1(\theta_1, \theta_2) = \mathcal{N}\big(\theta_1, \theta_2  + \sin(a \theta_1)|(0, 0), \Sigma_{\by} \big)$. We exemplify two cases, (a) using $(a, \Sigma_{\by}) = (1.5, \bigl[\begin{smallmatrix} 2 & 0 \\ 0 & 0.1\end{smallmatrix}\bigr])$ and (b) $(a, \Sigma_{\by}) = (1, I_2)$. For the Rosenbrock case the density is $\rho_2(\theta_1, \theta_2) = \mathcal{N}\big(a - \theta_1, b (\theta_2 -  \theta_1^2) |(0, 0), \Sigma_{\by} \big)b$ and the cases we show are (a) $(a, b, \Sigma_{\by}) = (2, 5, \bigl[\begin{smallmatrix} 6 & -1 \\-1 & 5\end{smallmatrix}\bigr])$ and (b) $(a, b, \Sigma_{\by}) = (1, 1, I_2)$. We then generically define the statistical manifold as 
%
    $
    \bar{M} = 
    \big\{
    \rho_{\bth}(\by) = 
    \mathcal{N}(\by|\bmu, \Sigma_{\by}) : 
    \bmu = h(\bth),
    \,
    \by \in \mathbb{R}^2,
    \,
    \Sigma_{\by} 
    \, \mathrm{fixed} \,
    \big\}.
    $
%
For $\rho_1$ we set the imbedding function as $\bmu = h_1(\btheta) = \big(\theta_1, \theta_2  + \sin(a \theta_1)\big)$. For $\rho_2$ we have $\bmu = h_2(\btheta) = \big(a - \theta_1, b (\theta_2 -  \theta_1^2) \big)$.  Both cases consider $\bth \in \mathbb{R}^2$ and $\Sigma_\pi = \mathrm{diag}(\infty) $. Results are presented and explained in Figure \eqref{fig:classics}.

\subsubsection{Non-affine imbeddings}
For the next two examples, we consider the logistic population growth model and a feed-forward neural network. To study both cases, we define the statistical manifold as 
%
    $
    \bar{M} = 
    \big\{
    \rho_{\bth}(\by) = 
    \mathcal{N}(\by|\bmu, \sigma^2 I_{n})
    : (\bmu, \sigma^2) = h(\bth), 
    \bth \in \Theta = \mathbb{R}^d,
    \,
    \by \in \mathbb{R}^{n}
    \big\}
    $
%
where the imbedding function $h$ models both mean and variance of the Gaussian distribution. From the contrast function theory, we also equip $\bar{M}$ with the metric $G_{\Theta}(\bth) = Dh(\bth)^\top\mathrm{diag}\big( \big(\tfrac{1}{\sigma^2}, \ldots, \tfrac{1}{\sigma^2}, \tfrac{n}{2(\sigma^2)^2}\big)(\bth)\big) Dh(\bth)$ + $\Sigma_\pi^{-1}$, which additionally takes into account the variance parameters $\sigma^2$. \\

\begin{figure}
    \begin{center}
    \hspace{-0.5cm}
    \begin{tabular*}{\columnwidth}{ ccc }
        \includegraphics[scale = 0.53]{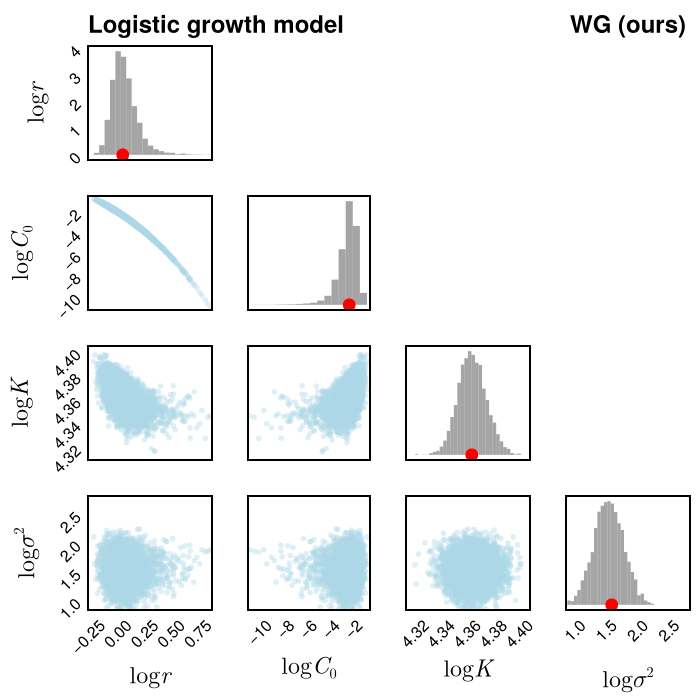}
        \hspace{-0.3cm}
        &
        \includegraphics[scale = 0.53]{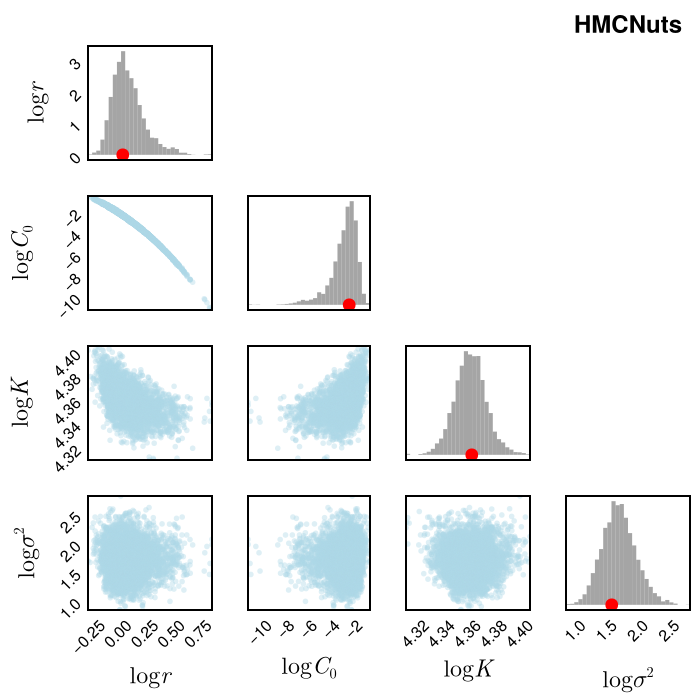}
        \hspace{-0.3cm}
        &
        \includegraphics[scale = 0.53]{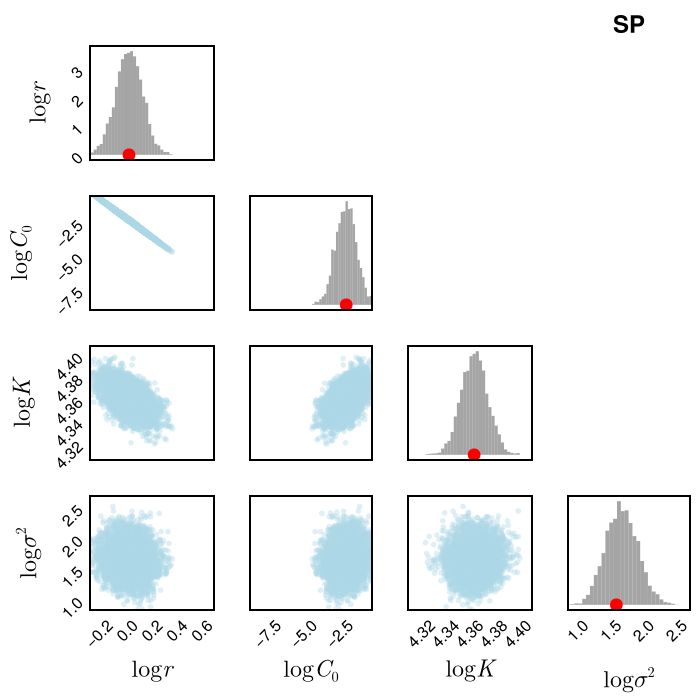}
    \end{tabular*}
        \caption{Posterior scatter plot comparison by sampling from the different approximation methods, \bWG\,, \bHMC\, and \bSP. The ({\color{red} $\bullet$}) dots depict the location of the MAP estimate. On the left panel, we show the scatter plot of samples from \bWG. The middle panel show the scatter plot from \bHMC\, and the right panel shows the sample from the \bSP\, method. Clearly, the samples from \bWG\, are in closer agreement with the samples from \bHMC\, in comparison with the samples from \bLA\, and \bHMC.      
        The $W_2$ distances for the approximation methods are $W_2(\rho_{\textsl{WG}}, \rho_{\textsl{NUTS}}) \approx 0.516$, $W_2(\rho_{\textsl{LA}}, \rho_{\textsl{NUTS}}) \approx 0.583$ and $W_2(\rho_{\textsl{SP}}, \rho_{\textsl{NUTS}}) \approx 0.560$. The $SW_2$ distances are $SW_2(\rho_{\textsl{WG}}, \rho_{\textsl{NUTS}}) \approx 0.245$, $SW_2(\rho_{\textsl{LA}}, \rho_{\textsl{NUTS}}) \approx 0.363$ and $SW_2(\rho_{\textsl{SP}}, \rho_{\textsl{NUTS}}) \approx 0.352$. This again shows that the \bSP\, method is only able to account for skewness and not able to properly account for the curved shape of the posterior.} 
        \label{fig:ode}
    \end{center}
\end{figure}

\noindent
\textbf{Population growth model}: \ For the first case, we take the logistic population growth model in systems biology as presented in \citet{sharp:2022}. This model has a differential equation and a solution respectively given by
%
    $
    \dd_t C(t) = r C(t) \left(
        1 - C(t) / K
    \right) 
    $
    and
    $
    C(t) = \tfrac{C(0)K}{C(0) + (K - C(0)) \exp(-r t)},
    $
%
where $C(t)$ denotes the number of individuals in the population at time $t$. The parameter $K > 0$ is the carrying capacity, describing the upper limit to which the population can grow and $r > 0$ is the intrinsic growth rate. We also treat the initial population size $C_0 =: C(0) > 0$ and the noise level $\sigma^2$ as parameters of the model for which we aim to do inference. The full parameter vector is therefore $\bomega = (r, C_0, K, \sigma^2)$. For the working parametrisation, we reparametrise $\bomega$ via $r = \exp(\theta_1)$, $C_0 = \exp(\theta_2)$, $K = \exp(\theta_3)$, and $\sigma^2 = \exp(\theta_4)$, where $\btheta = (\theta_1, \ldots, \theta_4) \in \Theta = \mathbb{R}^4$. For a given set of time points $t_1, \ldots, t_n$, we denote the observed data as $y_{t_i} = C(t_i) + E_i$, where $E_i \sim \mathcal{N}(0, \sigma^2)$, $i = 1, \ldots, n$. The imbedding function is defined as $(\bmu,\, \sigma^2) = h(\btheta) = \big(C\left(t_1;\, \btheta_{1:3}\right), \ldots, C\left(t_n;\, \btheta_{1:3}\right),\, \exp(\theta_4)\big)$ with priors $\theta_j \stackrel{\mathrm{i.i.d.}}{\sim} \mathcal{N}(0,\, 10)$ for $j = 1, \ldots, 4$. The results are presented and described in Figure \eqref{fig:ode}. The data considered can be found in \citet[Figure~9]{sharp:2022}.
\\ 

\noindent
\textbf{Neural networks}. \ We consider a feed-forward neural network with $L$
layers and coordinate-wise activation function $\phi$, defining the map
$f_{\btheta} : \mathbb{R}^{d^i} \rightarrow \mathbb{R}^{d^o}$ as
%
    $
    f_{\btheta_L}(\bx) =
    \phi_L \big(\cdots \phi_1(\bx^\top W^0 + \bb_0)
    W^1 + \bb_1\big) \cdots \big) W^L + \bb_L ,
    $
%
where $\btheta_L \in \Theta_L \subseteq \mathbb{R}^{d_L}$, with $d_L = (d^i + 1)\,k_1 + \sum_{i=1}^{L-1}(k_i + 1)\,k_{i+1} + (k_L + 1)\,d^o$, and $\Theta_L = \mathbb{R}^{(d^i + 1) \times k_1} \times \prod_{i=1}^{L-1} \mathbb{R}^{(k_i + 1) \times k_{i+1}} \times\mathbb{R}^{(k_L + 1) \times d^o}$. For a given data set of size $n$, any imbedding function $h$ based on this neural network can generically be written as $(\bmu,\, \sigma^2) = h(\btheta) = (f_{\btheta_L}(\bx_1), \ldots, f_{\btheta_L}(\bx_n),\,\exp(\theta_{d_L+1}))$, with $\btheta = (\btheta_L,\, \theta_{d_L + 1})$. For this case, we revisit the example presented by \citet{bergamin:2023} that considers the non-linear one-dimensional regression problem of \citet{snelson:2005}; see Figure~1. We set $\phi_1 = \phi_2 = \tanh$, $L = 2$, $d^i = d^o = 1$, and $k_1 = k_2 = 10$, giving $d_L + 1 = d = 142$. The dataset consists of $n = 163$ observations. Our example differs from \citet{bergamin:2023} since we consider the parameter $\sigma^2$ in the full inference procedure, not pre-fixing it to a certain value. Following common practice in the ML community, the priors are set to $\theta_j \stackrel{\mathrm{i.i.d}}{\sim} \mathcal{N}(0,\, 1)$ for $j = 1, \ldots, d+1$. Note that $h$ is heavily overparameterised for identifying the $\bmu$-space (the image of $f_{\btheta_L}$), which can render all the approximations (including \bHMC) ``ill-conditioned'' and unstable even when the degrees of freedom $n - d$ is positive with strong informative priors. 

\begin{figure}[!ht]
    \vskip 0.0in
        \begin{center}
            \centerline{
                \includegraphics[scale = 0.72]{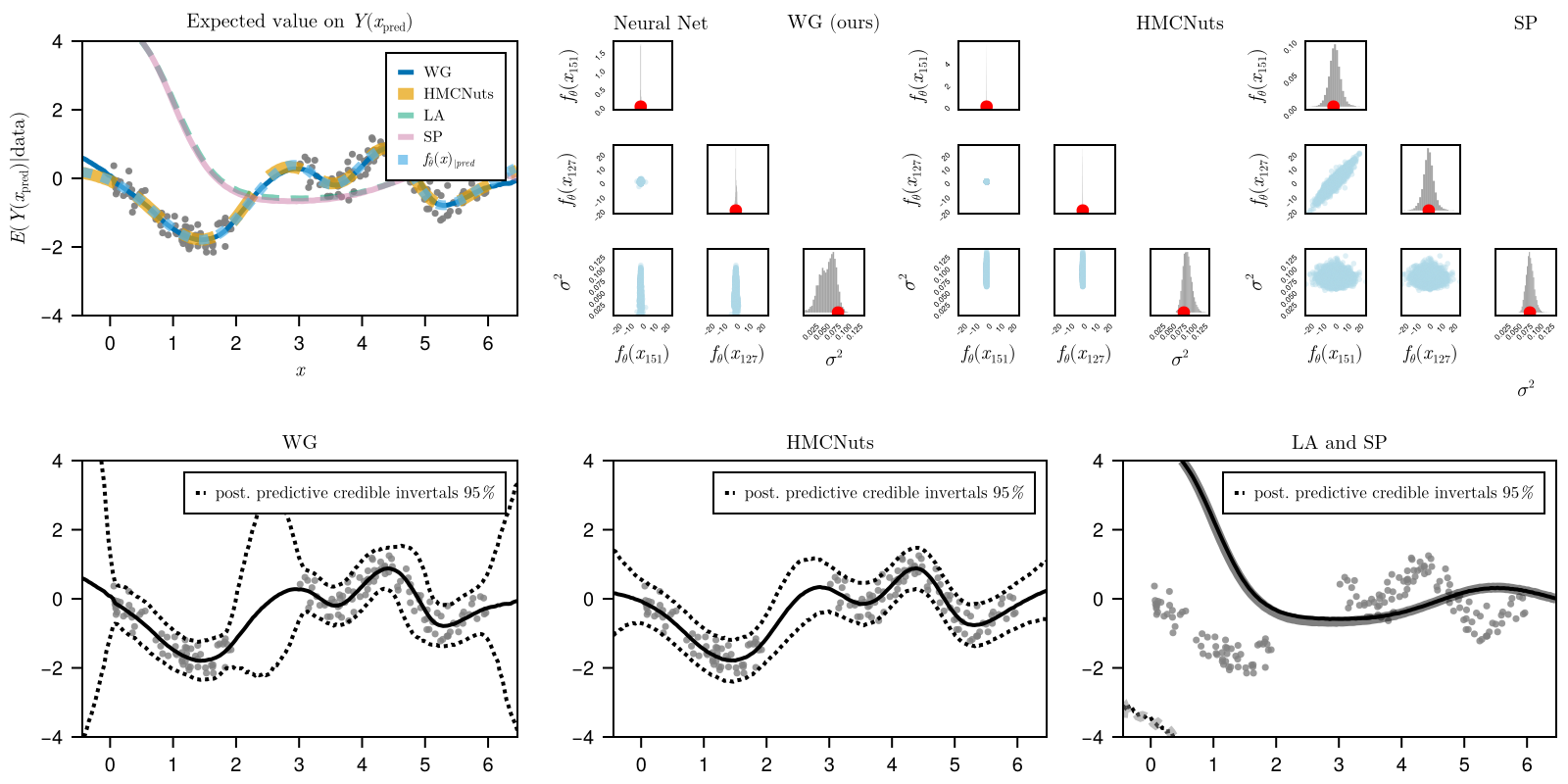}
            }
            \caption{The upper-left panel depicts the estimate of $\Ex(Y(\bx_{\mathrm{pred}}))$ along the $x$-axis for four different methods : \bWG, \bHMC\, \bLA\, and \bSP. Notably, \bWG, \bHMC\, and the image of $f_{\boldsymbol{\hat\theta}}(\cdot)$ show all good fits, while the \bLA\, and \bSP\, are essentially uncontrolled. On the upper-right panels, we show the scatter plot of the predictive distributions of $f_{\btheta}(\cdot)$ at particular inputs (or covariates) and different approximation methods. \bWG\, and \bHMC\, are in closer agreement compared to the \bLA\, and \bSP. The bottom panels depict the predictive distributions for the different methods along the $x$-axis and their $95\%$ credible intervals. The bottom-left panel shows visually similar predictive performance compared to \citet[see Appendix, Figure D3]{bergamin:2023}.
            The bottom-middle panel shows the predictive performance for \bHMC. Its credible intervals seem conservative in the regions where no data is present ($x \notin [2, 3])$. The \bLA\, and \bSP\, do not work properly in the predictive task, as shown in the bottom-right panel in ({\color{black} {\Large \textbf{-}}}, solid) and ({\color{gray} {\Large \textbf{-}}}, solid) respectively.} 
            \label{fig:nn-all}
        \end{center}
    \vskip -0.1in
\end{figure}

We used the BFGS algorithm to compute the MAP estimate (taking approximately 2.5 seconds),
we then computed the Hessian matrix $\overline{\Sigma}$ without imposing any numerical stabilization and post-processing of it with a fixed $\sigma^2$ value, as is commonly done \cite[see][Section~5]{bergamin:2023}, for example, in the Python or Julia package {\tt LaplaceRedux.jl}. 
We again used BFGS to solve the associated least-squares problem to draw samples from \eqref{eq:riemwg1}. For a sample size of $6 \times 10^3$, our method required approximately 4 hours. For the \bHMC\, sampler, we used the package {\tt advancedHMC.jl}, which allows real-time monitoring of diagnostic quantities assessing sample quality. During preliminary runs, we observed that the tree-depth consistently reached its default maximum value ({\tt max\_depth} = 10). This suggests that the sampler was unable to identify a natural stopping point (i.e., no U-turn was detected), indicating stronger dependence between samples and, consequently, a lower mixing and effective sample size. To mitigate this, we increased the maximum tree depth to {\tt max\_depth} = 13, allowing trajectories to reach lengths of up to $2^{13} = 8192$ steps, hence a better exploration of the posterior distribution. For the same sample size, the \bHMC\, sampler required approximately 19 hours to complete a run, still frequently reaching the maximum tree-depth. The corresponding results are shown Figure~\ref{fig:nn-all} and Table~\ref{tab:prednn}. It clearly shows that both \bHMC\, and \bWG\, perform better than the classical \bLA\, and \bSP\, to predict the data. Both $W_2$ and $SW_2$ distances,
for the push-forward distributions by $f_{\btheta}(\bx)$ using the \bWG\, approximation and \bHMC\, as reference, are smaller compared to the other approximations. 
\begin{table}[h]
    \begin{center}
    \caption{$W_2$ and $SW_2$ distances between the push-forward distribution by $f_{\btheta}(\bx)$ induced by the different approximate posteriors (approximate predictive distributions) and the NUTS posterior as reference. The $W_2$ and $SW_2$ distance between using the \bWG\, approximation and \bHMC\, is shown in the first row. The second and third rows shows the same distance measures for \bLA\, and \bSP\, method. The performance of the \bWG s is clearly better compared to the other approximations.
    }
    \label{tab:prednn}
    \begin{tabular}{lcc}
        \toprule
        Approximation method & $W_2$ & $SW_2$ \\
        \midrule
        $f_{\btheta}(\bx)_{\#}\big(\rho_{\textsl{WG}}\big)$ & $\mathbf{2.932}$ & $\mathbf{0.239}$  \\[0.2cm]
        $f_{\btheta}(\bx)_{\#}\big(\rho_{\textsl{LA}}\big)$ & $56.283$ & $5.025$ \\[0.2cm]
        $f_{\btheta}(\bx)_{\#}\big(\rho_{\textsl{SP}}\big)$ & $54.204$ & $4.929$ \\       \bottomrule
    \end{tabular}
    \end{center}
\end{table}
%

Finally, as neural networks are overparameterised, we run an additional experiment to assess the sensitivity of the predictive distribution to the choice of MAP estimate. Since the primary goal in ML is prediction - that is, inference on $f_{\btheta}(\bx)$ rather than on $\btheta$ itself - we examine whether the push-forward predictive  distributions obtained from different MAP estimates, found by restarting the optimisation from different initialisations, remain consistent with one another. For this extra experiment, we did not consider the \bSP\, approximation, as previous results showed almost no improvement compared to \bLA. After running this experiment, results suggest that, although different initialisations yield different MAP estimates and hence different Gaussian distributions on the tangent space, the resulting predictive distributions of $f_{\btheta}(\bx)$ are remarkably similar across different MAPs estimates. Full details are given in the supplementary material (appendix).

\section{Concluding remarks and future work} \label{sec:crft}

This work addresses a longstanding challenge in Bayesian statistics: how to improve upon the classical \bLA\, while preserving the computational efficiency that has made it one of the most widely used tools for approximate inference. Through  differential-geometric methods, information geometry and, particularly, the theory of contrast functions \citep{eguchi:1985}, we derived a closed-form approximation to the logarithmic map on statistical manifolds and used it to construct a new class of geometric motivated probability distributions. These were then used as closed-form \bWG\, approximations of posterior distributions due to the particular choice Fisher-Rao and prior geometries that naturally capture the shape of the posterior.

These new results further reveal several potential new research directions. One specifically important avenue concerns improving fast approximate inference schemes in GLM and Gaussian processes \citep{ohagan:1978, rasm:2006}, as they comprise many models with different names such as: generalized linear latent variable, random-effects, fixed-effects, mixed-effects, factor-models, Gaussian processes with monotonicity information \citep{riihimaki:10}, logistic Gaussian process density estimation \citep{riihimaki:2014}, heteroscedastic regression models \citep{julian:2004, hartmann:2019l} and possibly many new others.

Our methodology can be used directly as a plug-in method into the widely used approximate inference framework INLA \citep{rue2009approximate, simpson:2016}, potentially improving posterior approximations, marginal likelihoods, type-II marginal likelihood estimation, as well as approximations for marginals and conditional densities. Another natural direction is Bayesian model comparison and model selection. Since marginal likelihood approximations rely on Laplace expansions, replacing these with \bWG\, approximations may yield more accurate evidence estimates and improved model selection procedures, particularly in finite-sample settings where posterior curvature may play an important role.

The proposed construction is also well suited to variational inference and expectation propagation methods \citep{opper:2001, minka:2001}. The availability of a tractable density evaluation together with efficient sampling makes \bWG\, distributions attractive candidates for defining novel variational families. In this setting, it is also possible to modify the intrinsic geometry of the \bWG\, distribution proposed here. The basepoint $\bth_*$ and the covariance matrix $\Sigma$ would then constitute the parameters of the variational family alongside the choice of metric $g$. In this particular direction \citet{frank:2021} presents a closed related work, their method aims find a transformation $g$, on the original parameter space so that the pullback of its inverse approximates Euclidean metric only locally. Our method has global geometric aspect compared to theirs, as it find a mathematically valid approximation of the exponential map and its inverse. 

From the perspective of information geometry, the proposed approximate logarithmic map with $h$ as the identity function (or in dual/moment parametrisation) in the exponential family straightforwardly provides a new computationally efficient surrogate for Riemannian distances on statistical manifolds endowed with the Fisher-Rao metric. The case of the multivariate Gaussian family is especially important, as its closed-form Riemannian distance has been a long standing problem with a wide research interest in many application areas, see for example \citet{pinele:2020} and references therein.

Many challenges remain unresolved. From the perspective of statistics and machine-learning communities, large-scale applications may still be constrained by the storage and manipulation of the matrix coefficient $G$. The development of low-rank or sparse approximations of it represents an important direction for future work in modern deep-learning architectures with millions of parameters. As these architectures are overparameterised, understanding the extent to which \bWG\, approximations centred at different basepoints can yield invariant predictive distributions on the output of the neural network is an active area of research \citep{roy:2024}. Preliminary experiments with the neural network considered in this paper provide empirical evidence of such an invariance property. If this phenomenon holds generally, it could eliminate the need to compute eigenvalues and eigenvectors of $G$ to identify directions in parameter space that leave the network outputs essentially unchanged. In this sense, the methodology proposed in this paper would automatically account for such directions through its geometric construction.


\section{Competing interests}

We have no competing interests to disclose.

\section{Author contributions statement}

M.H. independently conceived all ideas presented in the paper, designed and conducted the experiments, analyzed the results, and presented the formal proofs. L.H. verified the technical derivations, proposed alternative directions, consistency, and made improvements to the exposition. A.J. and G.A. provided the extra experiments proposed by M.H., with neural-networks and discussed it. A.M., S.H., H.R. and M.G. contributed with suggestions and reviewed the manuscript, which improved its clarity and the discussion of related work.

\section{Acknowledgments}

M.H. thanks Professor Nihat Ay (TU Hamburg, Germany) for drawing attention to the theory of contrast functions following a talk given in the Department of Computer Science at the invitation of Assistant Professor Pierre-Alexandre Murena. M.H. also thanks professor Luiz Hartmann (Federal University of São Carlos, Brazil) and professor Fabiana Coswosck (Federal University of Espírito Santo, Brazil) for the overview and discussions related to the foundational concepts in this work. The authors thank Professor Shinto Eguchi (Institute of Statistical Mathematics, Tokyo, Japan) for reading the manuscript and providing valuable comments and suggestions. The authors also thank Professor Theo Damoulas (University of Warwick) and MSc. Mengqi Chen for reading early versions of the manuscript and providing constructive feedback.This work was supported by the Research Council of Finland grant number 369502.

\section{Code and data availability}
To appear

\bibliographystyle{abbrvnat}
\bibliography{reference.bib}

\begin{appendices}

\section{Appendix: Contrast functions and dual connections}

A central result in information geometry is that Riemannian metrics and connections can be generated from suitable divergence functions, such as for example the Kullback-Leibler divergence \citep{kl:1951}. This viewpoint was introduced in a series of seminal works by \citet{eguchi:1983, eguchi:1985, eguchi:1992}, in which he showed that a contrast function induces a Riemannian metric through its second-order derivatives and a pair of dual connections through its third-order derivatives.

\begin{definition}[Contrast function]
A \textit{contrast function} on $M$ denoted as $D_M : M^2 \to \mathbb{R}_+$ is a smooth function that satisfies the following properties
\begin{center}
    \begin{enumerate}
    \item \, Positive : $D_M(\bp||\bp') \geq 0 \ 
    \forall \ \bp, \bp' \in M$,
    
    \item \, Non-degenerate : $D_M(\bp||\bp') = 0 
    \Leftrightarrow \bp = \bp'$.
    
    \item \, The first variation along the diagonal vanishes : 
    $$ 
    \partial_j D_M(\bp||\bp')|_{\bp = \bp'} 
    = \partial_{j'} D_M(\bp||\bp')|_{\bp' = \bp} = 0 \ 
    \forall j \in \{1, \ldots, d \}.
    $$ 
    \vspace{-0.7cm}
    
    \item \, The Hessian along the diagonal,
    $$ 
    G_{ij}(\bp) = \partial^2_{i', j'} D_M(\bp||\bp')|_{\bp' = \bp} 
    $$
    is strictly positive-definite and smooth $\forall \bp \in M$ \citep[See equivalent definitions in][Sec. 11.2, formulas 11.3.12-11.3.15]{calin:2014} .
    \end{enumerate}
\end{center}
\end{definition}

When a divergence function $D_M$ satisfies the aforementioned properties, it is also possible to show that reparametrizations on $M$ transform $G$ covariantly or accordingly to a $(0, 2)$-tensor \citep[see][Section 11.2]{calin:2014}, which means that it can be used as a metric tensor in a manifold turning it into a Riemannian manifold. Beyond the Riemannian metric, contrast functions also induce additional geometric structure on $M$. In particular, they generate a pair of affine connections that play a central role in information geometry.
\begin{definition}[Dual contrast function and dual connections] 
Given a contrast function on $M$, $D_M : M^2 \to \mathbb{R}$, we define its dual contrast function $D_M^* : M^2 \to \mathbb{R}$ by swapping the arguments,
\begin{align*}
    D_M^*(\bp \| \bp') = D_M(\bp' \| \bp).
\end{align*}
The function $D_M$ induces an affine connection $\nabla$ whose coefficients are
\begin{align*}
    \Gamma_{ij,k}(\bp)
    =
    - \frac{\partial^3}{\partial {\bp}_i \partial {\bp}_j \partial {\bp}'_k}
    D_M(\bp \| \bp')\Big|_{\bp'=\bp}.
\end{align*}
Similarly, the contrast function $D_M^*$ induces another connection $\nabla^*$ with coefficients
\begin{align*}
    \Gamma^*_{ij,k}(\bp)
    =
    - \frac{\partial^3}{\partial {\bp}'_i \partial {\bp}'_j \partial {\bp}_k}
    D_M(\bp \| \bp')\Big|_{\bp = \bp'}.
\end{align*}
The pair $(\nabla,\nabla^*)$ is called a pair of \emph{dual connections}. Both connections are torsion-free because their coefficients are symmetric in the lower indices. Moreover, they are related to the Levi--Civita connection $\nabla^{(0)}$ of the metric $g$ (or $G$) by
\begin{align*}
    2\nabla^{(0)} = \nabla + \nabla^*,
\end{align*}
see \citep[][for technical details and derivations]{pfan:1973, chentsov:1982, calin:2014}.
\end{definition}
The pair of dual connections plays a fundamental role in information geometry. Many important results are naturally expressed in terms of this dualistic structure. For instance, exponential families are characterized as manifolds that are flat with respect to the exponential connection $\nabla$, while mixture families are flat with respect to the dual connection $\nabla^*$ \citep{amarinaga:2000}. Moreover, the generalized Pythagorean theorem for contrast functions, which underlies projection methods and statistical estimation, is formulated in terms of orthogonality defined by these dual connections \citep{amarinaga:2000}. This structure also leads to the family of $\alpha$-connections, which interpolates between $\nabla$ and $\nabla^*$ and provides a unified geometric framework for statistical inference \citep[See][Sec. 1.11]{calin:2014}

\section{Appendix: Proofs}
\textbf{Theorem 1}
    %
    \begin{proof}: 
        Let $(M, g)$ be a Riemannian manifold where $g$ is induced by a contrast function $D_M: M^2 \rightarrow \mathbb{R}$, and whose Levi-Civitta connection is denoted as $\nabla^{(0)}$. Let $\gamma$ be a geodesic curve minimizing the length on $M$ that joins the points $\bp$ and $\bp'$ where $\gamma(0) = \bp$ and $\gamma(t) = \bp'$. Since the arc-length along the geodesic is the Riemannian distance $d$, let $t = d(\bp, \bp')$. Consider the functions $\psi(t) = C(\bp \| \gamma(t)):=\psi_F(t) + \psi_B(t)$ with $\psi_F(t) = D_M(\bp||\gamma(t))$ and $\psi_B(t) = D_M(\gamma(t) || \bp)$, and the index notation with Einstein summation. A third-order Taylor expansion of $\psi_F$ at $t = 0$ gives,
        \begin{align*}
            \psi_F(t) = \psi_F(0) + t \psi_F'(0)   
            + \frac{t^2}{2} \psi''_F(0)   
            + \frac{t^3}{6} \psi'''_F(0)
            + \mathcal{O}(t^4)
        \end{align*}
        with coefficients given by,
        \begin{align*}
            \psi_F(0) &= D_M(\gamma(0)||\gamma(0)) = 0,
            \\
            \psi'_F(0) &= 
            \partial_{\gamma^i} D_M(\bp||\gamma(0)) 
            \dot\gamma^i(0)
            = 0,
            \\
            \psi''_F(0) &= 
            \partial_{\gamma^i, \gamma^j} D_M(\bp||\gamma(0)) 
            \dot\gamma^i(0) \dot\gamma^j(0) 
            + 
            \partial_{\gamma^i} D_M(\bp||\gamma(0)) 
            \ddot\gamma^i(0)
            = G_{ij}(\bp) \dot\gamma^i(0) \dot\gamma^j(0),
            \\
            \psi'''_F(0) &= 
            \partial_{\gamma^i, \gamma^j, \gamma^k} D_M(\bp||\gamma(0)) 
            \dot\gamma^i(0) \dot\gamma^j(0) \dot\gamma^k(0) 
            +
            2 \partial_{\gamma^i, \gamma^j} D_M(\bp||\gamma(0)) 
            \ddot\gamma^i(0) \dot\gamma^j(0) 
            \\
            &+
            \partial_{\gamma^i, \gamma^k} D_M(\bp||\gamma(0))
            \dot\gamma^k(0) \ddot\gamma^i(0)
            +
            \partial_{\gamma^i} D_M(\bp||\gamma(0)) 
            \dddot\gamma^i(0).
        \end{align*}
        Similarly for $\psi_B$,
        \begin{align*}
            \psi_B(t) = \psi_B(0) + t \psi_B'(0)   
            + \frac{t^2}{2} \psi''_B(0)   
            + \frac{t^3}{6} \psi'''_B(0)
            + \mathcal{O}(t^4)
        \end{align*}
        with coefficients
        \begin{align*}
            \psi_B(0) &= D_M(\gamma(0)||\gamma(0)) = 0,
            \\
            \psi'_B(0) &= 
            \partial_{\gamma^i} D_M(\gamma(0)||\bp) 
            \dot \gamma^i(0),
            = 0
            \\
            \psi''_B(0) &= 
            \partial_{\gamma^i, \gamma^j} D_M(\gamma(0)||\bp) 
            \dot \gamma^i(0) \dot \gamma^j(0) + 
            \partial_{\gamma_i} D_M(\gamma(0)||\bp) 
            \ddot \gamma^i(0)
            = G_{ij}(\bp) \dot\gamma^i(0) \dot\gamma^j(0),
            \\
            \psi'''_B(0) &= 
            \partial_{\gamma^i, \gamma^j, \gamma^k} D_M(\gamma(0)||\bp) 
            \dot \gamma^i(0) \dot \gamma^j(0) \dot \gamma^k(0)
            +
            2 \partial_{\gamma^i, \gamma^j} D_M(\gamma(0)||\bp) 
            \ddot\gamma^i(0) \dot\gamma^j(0) 
            \\
            & +
            \partial_{\gamma^i} \partial_{\gamma^k} D_M(\gamma(0)||\bp) 
            \ddot\gamma^i(0) \dot\gamma^k(0) 
            +
            \partial_{\gamma^i} D_M(\gamma(0)||\bp) 
            \dddot\gamma^i(0).
        \end{align*}
        Since $\gamma$ is a geodesic, it holds that $\ddot \gamma^k(t) = 
        - \sum_{ij} \Gamma^{k(0)}_{ij}(\gamma(t))\dot\gamma^i(t) \dot \gamma^j(t)$, where $\Gamma^{k(0)}_{ij}$ are the coefficients of Levi-Civita connection $\nabla^{(0)}$. The second and third terms of the third derivatives can be jointly rewritten as
        \begin{align*}
            G_{ij}(\bp) \ddot\gamma^i(0) \dot\gamma^j(0) 
            &=
            - 
            G_{ij}(\bp) \Gamma^{i(0)}_{rs}(\bp) \dot\gamma^r(0) \dot \gamma^s(0) \dot \gamma^j(0)
            \\
            &=
            - \Gamma^{(0)}_{rs, j}(\bp) \dot\gamma^r(0) \dot \gamma^s(0) \dot \gamma^j(0)
        \end{align*}
        since $\Gamma^{(0)}_{rs, j} := G_{ij} \Gamma^{i(0)}_{rs}$. Summing these coefficients from the third-order terms of $\psi_F$ and $\psi_B$ we have, 
        \begin{align} \label{eq:coeff1}
            6 \, G_{ij}(\bp) \ddot\gamma^i(0) \dot\gamma^j(0) 
            &=
            - 6 \,
            \Gamma^{(0)}_{rs, j}(\bp) \dot\gamma^r(0) \dot \gamma^s(0) \dot \gamma^j(0).
        \end{align}
        From \cite{eguchi:1985}, page 358, Equations 3.5 and 3.6, we have that the third-order derivatives of the $D_M(\cdot||\bp)$ and its dual (reverse) $D_M(\bp||\cdot)$ hold the following relations
        \begin{align*}
            \partial_{\gamma^i, \gamma^j, \gamma^k} D_M(\gamma(0)||\bp) 
            &=
            \Gamma_{ij, k}(\bp) + \Gamma^{*}_{jk, i}(\bp) + \Gamma_{ki, j}(\bp)
            \\
            \partial_{\gamma^i, \gamma^j, \gamma^k} D_M(\bp||\gamma(0)) 
            &=
            \Gamma^{*}_{ij, k}(\bp) + \Gamma^{*}_{jk, i}(\bp) + \Gamma_{ki, j}(\bp),
        \end{align*}
        where $\Gamma_{ij, k}(\bp)$ and $\Gamma^*_{ij, k}(\bp)$ are the coefficients of the pair of the dual connections $\nabla$ and $\nabla^*$ induced by the contrast function $D_M$ respectively. Since the $D_M(\cdot \| \cdot)$ is smooth in both entries, we can swap the order of third-derivatives w.r.t. $i, j, k$. Swapping $i \leftrightarrow j$ and using the fact that the dual connections are symmetric, we get
        \begin{align*}
            \partial_{\gamma^j, \gamma^i, \gamma^k} D_M(\gamma(0)||\bp) 
            &=
            \Gamma_{ij, k}(\bp) + \Gamma^{*}_{ki, j}(\bp) + \Gamma_{jk, i}(\bp)
            \\
            \partial_{\gamma^i, \gamma^j, \gamma^k} D_M(\bp||\gamma(0)) 
            &=
            \Gamma^{*}_{ij, k}(\bp) + \Gamma^{*}_{jk, i}(\bp) + \Gamma_{ki, j}(\bp)
        \end{align*}
        Summing these third-order terms from $\psi_F$ and $\psi_B$ and using the relation between the pair of dual connections and the Levi-Civita connection, we have
        \begin{align} \label{eq:coeff2}
            \Big(\partial_{\gamma^i, \gamma^j, \gamma^k} D_M(\gamma(0)||\bp)
            +
            \partial_{\gamma^i, \gamma^j, \gamma^k} & D_M(\bp||\gamma(0))
            \Big) \dot \gamma^i(0) \dot \gamma^j(0) \dot \gamma^k(0)
            \\ \nonumber
            &= 
            \big(
            2 \Gamma_{ij, k}^{(0)}(\bp)
            +
            2 \Gamma_{ki, j}^{(0)}(\bp)
            +
            2 \Gamma_{kj, i}^{(0)}(\bp)
            \big) \dot \gamma^i(0) \dot \gamma^j(0) \dot \gamma^k(0)
            \\ \nonumber
            &=
            6 \, \Gamma_{ij, k}^{(0)}(\bp) \dot \gamma^i(0) \dot \gamma^j(0) \dot \gamma^k(0).
        \end{align}
        Finally, when summing the curves $\psi_F$ and $\psi_B$ the third-order coefficients \eqref{eq:coeff1} and \eqref{eq:coeff2} cancel out. As geodesics parametrized by the arc-length are unit-speed curves, we also have $g(\dot\gamma(0), \dot\gamma(0)) = 1$. Therefore, the third-order approximation of $\psi$ is given by
        \begin{align*}
            \psi(t) &= \psi_F(t) + \psi_B(t) = 
            t^2 G_{ij}(\bp) \dot\gamma^i(0) \dot\gamma^j(0)
            + \mathcal{O}(t^4) 
            = t^2 g(\dot\gamma(0), \dot\gamma(0)) + \mathcal{O}(t^4) 
            \\
            C(\bp \| \bp')
            &= d(\bp, \bp')^2 + \mathcal{O}(d(\bp, \bp')^4).
        \end{align*}
        This proves the theorem.
    \end{proof}
    %
%
\noindent 
\textbf{Theorem 2}
%
\begin{proof}
    From properties 3 and 4 of the definition of contrast function we have,
    \begin{align*}
        \olsi{\mathrm{Log}}_{\bth}(\bth) 
        &= -\tfrac{1}{2}\, G_{\Theta}^{-1}(\bth) \, \nabla_{\bth_1} D'_{\mathrm{post}}(\bth_1 \| \bth)\Big|_{\bth_1 = \bth} = 
        -\tfrac{1}{2}\, G_{\Theta}^{-1}(\bth) \0 = \0.
    \end{align*}
    Evaluating the derivative of $\olsi{\mathrm{Log}}_{\bth}$ at $\bth$ yields
    \begin{align*}
        D\olsi{\mathrm{Log}}_{\bth}(\bth) 
        &= -\tfrac{1}{2}\, G_{\Theta}^{-1}(\bth)\, \nabla^2_{\bth_2, \bth_1}
        D'_{\mathrm{post}}(\bth_1 \| \bth_2) \Big|_{\bth_1 = \bth, \bth_2 = \bth} 
        \\
        &=
        -\tfrac{1}{2}\, G_{\Theta}^{-1}(\bth)\, \nabla^2_{\bth_2, \bth_1}
        \big(
            D_{\bar{M}}(\bth_1 \| \bth_2)
            +
            D_{\bar{M}}^*(\bth_1 \| \bth_2) 
        \big)\Big|_{\bth_1 = \bth, \bth_2 = \bth}
        \\
        &= 
        -\tfrac{1}{2}\, G_{\Theta}^{-1}(\bth)\,
        \big( - G_{\Theta}(\bth) - G_{\Theta}(\bth) \big)
        = I.
    \end{align*}
    Hence, it satisfies the definition of approximate logarithmic map in \cite{jump:2021}.
\end{proof}
\noindent
\textbf{Corollary 1}
    %
    \begin{proof}:
        Let us start writing a member of the exponential family in $\btheta$ through $\bbeta$, denoting $\bbeta(\btheta) := h(\btheta)$. Let's denote the Jacobian of $h$ as $D h(\btheta) := \Deta(\btheta)$ and the chain rule derivative w.r.t. $\btheta$ as $\nabla_{\btheta} = \Deta(\btheta)^\top \nabla_{\bbeta}$. We write a member of the exponential family as,
        \begin{align*}
            \rho_{\btheta}(\by) =
            \exp \big(
                \bbeta(\btheta)^\top T(\by)
                + B(\by) - A(\bbeta(\btheta))
            \big)
        \end{align*}
        where $T(\by)$ is the vector of sufficient statistics, $B(\by)$ the base measure and $A(\bbeta(\btheta))$ the log-partition function in $\btheta$. Besides, 
        \begin{align*}
            \nabla_{\bth}(1) &=
            \nabla_{\bth} \int_{\Omega} \rho_{\btheta}(\by) \dy \\
            \0 &=
            \int_{\Omega}
            \nabla_{\bth}
            \exp \big(
                \bbeta(\btheta)^\top T(\by)
                + B(\by) - A(\bbeta(\btheta))
            \big)
            \dy \\
            &=
            \Deta(\btheta)^\top 
            \int_{\Omega} T(\by) \rho_{\btheta}(\by) \dy
            -
            \Deta(\btheta)^\top \nabla_{\bbeta} A(\bbeta)|_{\bbeta = \bbeta(\bth)}
            \int_{\Omega} \rho_{\btheta}(\by) \dy 
            \\
            &=
            \Deta(\btheta)^\top 
            \mathbb{E}_{\rho_{\bth}}(T(\bY))
            -
            \Deta(\btheta)^\top \nabla_{\bbeta} A(\bbeta)|_{\bbeta = \bbeta(\bth)},
        \end{align*}
        hence $\mathbb{E}_{\rho_{\bth}}(T(\bY)) = \nabla_{\bbeta} A(\bbeta)|_{\bbeta = \bbeta(\bth)}$. Note that the integral changes to a summation when $\bY$ is discrete random vector. 
        
        The forward $\textsl{KL}$ divergence then reads
        \begin{align*}
            \textsl{KL}(\rho_{\btheta_1}(\by) || \rho_{\btheta_2}(\by)) &=
            \int_\Omega
            \log \frac{\rho_{\btheta_1}(\by)}{\rho_{\btheta_2}(\by)}
            \rho_{\btheta_1}(\by) \dy
            \\
            &=
            \int_\Omega  
            \left( \log \frac{
                \exp \Big(
                \bbeta(\btheta_1)^\top T(\by)
                + B(\by) - A(\bbeta(\btheta_1))
                \Big)}{
                \exp \Big(
                \bbeta(\btheta_2)^\top T(\by)
                + B(\by) - A(\bbeta(\btheta_2))
                \Big)}
            \right)
            \rho_{\btheta_1}(\by) \dy 
            \\
            &=
            \int_\Omega \left(
                (\bbeta(\btheta_1) - \bbeta(\btheta_2))^\top T(\by)
                - (A(\bbeta(\btheta_1)) - A(\bbeta(\btheta_2)))
            \right)
            \rho_{\btheta_1}(\by) \dy
            \\
            &=
            \big( \bbeta(\btheta_1) - \bbeta(\btheta_2) \big)^\top 
            \mathbb{E}_{\rho_{\btheta_1}}(T(\bY)) - 
            \big( A(\bbeta(\btheta_1)) - A(\bbeta(\btheta_2)) \big) 
            \\
            &=
            \big( \bbeta(\btheta_1) - \bbeta(\btheta_2) \big)^\top 
            \nabla_{\bbeta} A(\bbeta)|_{\bbeta = \bbeta(\btheta_1)} 
            - 
            \big(A(\bbeta(\btheta_1)) - A(\bbeta(\btheta_2)) \big),
        \end{align*}
        where we have used that $\mathbb{E}_{\rho_{\btheta_1}}(T(\bY)) = \nabla A_{\bbeta}(\bbeta)|_{\bbeta = \bbeta(\btheta_1)}$. For the reverse $\textsl{KL}$ we just swap the parameters $\btheta_1 \leftrightarrow \btheta_2$. The gradient of the forward $\textsl{KL}$ divergence is respectively given by, 
        %
        \begin{align*}
            \nabla_{\btheta_1}
            \textsl{KL}(\rho_{\btheta_1}(\by) || \rho_{\btheta_2}(\by)) 
            &
            =
            \Deta(\btheta_1)^\top \nabla_{\bbeta}
            \Big(
                (\bbeta(\btheta_1) 
                - \bbeta(\btheta_2))^\top 
                \nabla_{\bbeta} A(\bbeta)|_{\bbeta = \bbeta(\btheta_1)} 
                \\
                & \hspace{3cm} 
                - (A(\bbeta(\btheta_1)) - A(\bbeta(\btheta_2)))
            \Big) 
            \\ 
            &=
            \Deta(\btheta_1)^\top
            \Big(
                \nabla^2_{\bbeta} A(\bbeta)|_{\bbeta = \bbeta(\btheta_1)}
                (\bbeta(\btheta_1) - \bbeta(\btheta_2))
                \\
                & \hspace{3cm} 
                - 
                \nabla_{\bbeta} A(\bbeta)|_{\bbeta = \bbeta(\btheta_1)}
                + 
                \nabla_{\bbeta} A(\bbeta)|_{\bbeta = \bbeta(\btheta_1)}
            \Big),
        \end{align*}
        analogous for the reverse $\textsl{KL}$
        %
        \begin{align*}
            \nabla_{\btheta_1}
            \textsl{KL}(\rho_{\btheta_2}(\by) || \rho_{\btheta_1}(\by))
            &=
            \Deta(\btheta_1)^\top \nabla_{\bbeta}
            \Big(
                \big(\bbeta(\btheta_2) - \bbeta(\btheta_1) \big)^\top 
                \nabla_{\bbeta} A(\bbeta)|_{\bbeta = \bbeta(\btheta_2)} 
                \\
                & 
                \hspace{3cm}
                - \big(A(\bbeta(\btheta_2)) - A(\bbeta(\btheta_1))\big)
            \Big) 
            \\ 
            &=
            \Deta(\btheta_1)^\top
            \Big(
                - \nabla_{\bbeta} A(\bbeta)|_{\bbeta = \bbeta(\btheta_2)} 
                + \nabla_{\bbeta} A(\bbeta)|_{\bbeta = \bbeta(\btheta_1)} 
            \Big).
        \end{align*}
        %
        Denote the Bregman divergence as 
        \begin{align*}
            D_\phi(\btheta_1 \| \btheta_2) = 
            \phi(\btheta_1) - \phi(\btheta_2) 
            - 
            \inp{\nabla_{\btheta_2}\phi(\btheta_2)}{\btheta_1 - \btheta_2}.
        \end{align*}
        The gradient of the forward divergence is given by, 
        \begin{align*}
            \nabla_{\btheta_1} 
            D_\phi(\btheta_1 \| \btheta_2)
            =
            \nabla_{\btheta_1} \phi(\btheta_1)
            -
            \nabla_{\btheta_2} \phi(\btheta_2)
        \end{align*}
        and for the reverse divergence, 
        \begin{align*}
            \nabla_{\btheta_1} 
            D_\phi(\btheta_2 \| \btheta_1)
            =
            \nabla^2_{\btheta_1} \phi(\btheta_1)(\btheta_1 - \btheta_2).
        \end{align*}
        Summing all gradients, we obtain
        %
        \begin{align*}
            \nabla_{\btheta_1}
            \textsl{KL}( & 
            \rho_{\btheta_1}(\by) || \rho_{\btheta_2}(\by)) 
            +
            \nabla_{\btheta_1}
            \textsl{KL}(\rho_{\btheta_2}(\by) || \rho_{\btheta_1}(\by))
            +
            \nabla_{\btheta_1} 
            D_\phi(\btheta_1 \| \btheta_2)
            +
            \nabla_{\btheta_1} 
            D_\phi(\btheta_2 \| \btheta_1)
            =
            \\
            & 
            \Deta(\btheta_1)^\top 
            \Big(
                \nabla^2_{\bbeta} A(\bbeta)|_{\bbeta = \bbeta(\btheta_1)}
                (\bbeta(\btheta_1) - \bbeta(\btheta_2))
                + \nabla_{\bbeta} A(\bbeta)|_{\bbeta = \bbeta(\btheta_1)}
                - \nabla_{\bbeta} A(\bbeta)|_{\bbeta = \bbeta(\btheta_2)} 
            \Big)
            \\
            &
            +
            \nabla_{\btheta_1} \phi(\btheta_1)
            -
            \nabla_{\btheta_2} \phi(\btheta_2)
            +
            \nabla^2_{\btheta_1} \phi(\btheta_1)(\btheta_1 - \btheta_2).
        \end{align*}
        %
        Then the approximate logarithmic map $\btheta_2 \mapsto \olsi{\mathrm{Log}}_{\btheta_1}(\btheta_2)$ is given by
        \begin{align*} 
            \olsi{\mathrm{Log}}_{\btheta_1}(\btheta_2) 
            &
            = 
            -\tfrac{1}{2} 
            G_{\Theta}(\btheta_1)^{-1}
            \Big[
            \Deta(\btheta_1)^\top 
            \Big(
                \nabla^2_{\bbeta} A(\bbeta)
                |_{\bbeta = \bbeta(\btheta_1)}
                (\bbeta(\btheta_1) - \bbeta(\btheta_2))
                \\
                &
                + 
                \nabla_{\bbeta} A(\bbeta)|_{\bbeta = \bbeta(\btheta_1)}
                - 
                \nabla_{\bbeta} A(\bbeta)|_{\bbeta = \bbeta(\btheta_2)} 
                \Big)
                \\
                &
                +
                \nabla_{\btheta_1} \phi(\btheta_1)
                -
                \nabla_{\btheta_2} \phi(\btheta_2)
                +
                \nabla^2_{\btheta_1} \phi(\btheta_1)(\btheta_1 - \btheta_2)
            \Big],
        \end{align*}
        where $G_\Theta(\bth)$ $=$ $\Deta(\bth)^\top \mathcal{I}_\Xi(\bbeta(\bth)) \Deta(\bth) + \nabla^2\phi(\bth)$ is the matrix coefficient obtained from the contrast function formed by the sum of the \bKL\, and Bregman divergence in the parameterisation $\bth$. The matrix $\mathcal{I}_{\Xi}$ is the Fisher information matrix in the natural parametrisation $\bbeta$. Let $V \in T_{\btheta_1} \bar{M}$ where $\btheta_1$ is the fixed basepoint of the approximate logarithmic map. To analyse the bijection of this map, we want to show that, if for any given $V$ there exists a unique $\btheta_2 \in \Theta$ such that when $\olsi{\mathrm{Log}}_{\btheta_1}$ is evaluated at $\btheta_2$, it maps back to $V$. To do so, first consider the short notation 
        $\bbeta_1 = \bbeta(\btheta_1)$,
        $\bbeta_2 = \bbeta(\btheta_2)$,
        $\Deta_1 := \Deta(\btheta_1)$,
        $\Deta_2 := \Deta(\btheta_2)$,
        $ \nabla^2_{\bbeta} A(\bbeta_1) := \nabla^2_{\bbeta} A(\bbeta)|_{\bbeta = \bbeta(\btheta_1)}$, 
        $\nabla_{\bbeta} A(\bbeta_1) := \nabla_{\bbeta} A(\bbeta)|_{\bbeta = \bbeta(\btheta_1)}$ 
        and 
        $\nabla_{\bbeta} A(\bbeta_2) := \nabla_{\bbeta} A(\bbeta)|_{\bbeta = \bbeta(\btheta_2)}$. 
        Then we rewrite the approximate map $V = \olsi{\mathrm{Log}}_{\btheta_1}(\btheta_2)$, up to multiplication by $-1$, as follows,
        \begin{align*}
            0 
            &= 
            -\tfrac{1}{2} G_{\Theta}(\btheta_1)^{-1}
            \bigg[
            \Deta_1^\top 
            \Big(
                \nabla^2_{\bbeta} A(\bbeta_1)
                (\bbeta_1 - \bbeta_2)
                + \nabla_{\bbeta} A(\bbeta_1)
                - \nabla_{\bbeta} A(\bbeta_2)
            \Big)
            \\ & \ \phantom{=}
                +
                \nabla_{\btheta_1} \phi(\btheta_1)
                - \nabla_{\btheta_2} \phi(\btheta_2)
                + \nabla^2_{\btheta_1} \phi(\btheta_1)(\btheta_1 - \btheta_2)
            \bigg]
            - V 
            \\
            &=
            \Deta_1^\top 
            \Big(
                \nabla^2_{\bbeta} A(\bbeta_1)
                (\bbeta_1 - \bbeta_2)
                + \nabla_{\bbeta} A(\bbeta_1)
                - \nabla_{\bbeta} A(\bbeta_2)
            \Big) 
            \\ & \ \phantom{=}
                +
                \nabla_{\btheta_1} \phi(\btheta_1)
                - \nabla_{\btheta_2} \phi(\btheta_2)
                + \nabla^2_{\btheta_1} \phi(\btheta_1)(\btheta_1 - \btheta_2)
            + 2 G_\Theta(\btheta_1) V 
            \\
            &= 
            \Deta_1^\top 
            \Big(
                - \nabla^2_{\bbeta} A(\bbeta_1) \bbeta_2
                + \nabla_{\bbeta} A(\bbeta_1) 
                - \nabla_{\bbeta} A(\bbeta_2) 
                + \nabla^2_{\bbeta} A(\bbeta_1) \bbeta_1 
            \Big) 
            \\ & \ \phantom{=}
                +
                \nabla_{\btheta_1} \phi(\btheta_1)
                - \nabla_{\btheta_2} \phi(\btheta_2)
                + \nabla^2_{\btheta_1} \phi(\btheta_1)(\btheta_1 - \btheta_2)
                + 2 G_\Theta(\btheta_1) V \\
            &= 
            \Deta_1^\top 
            \Big(
                -
                \nabla_{\bbeta} 
                \Big(
                \tfrac{1}{2}
                \norm{\bbeta}^2_{\nabla^2_{\bbeta} A(\bbeta_1)} + A(\bbeta) 
                \Big)
                \Big|_{\bbeta = \bbeta(\btheta_2)}
                + \nabla_{\bbeta} A(\bbeta_1) 
                + \nabla^2_{\bbeta} A(\bbeta_1) \bbeta_1
            \Big) 
            \\ & \ \phantom{=}
                -
                \nabla_{\btheta_2}
                \Big(
                \tfrac{1}{2}\norm{\btheta_1 - \btheta_2}^2_{\nabla^2_{\btheta_1} \phi(\btheta_1)} 
                +
                \phi(\btheta_2)
                \Big) + \nabla_{\btheta_1} \phi(\btheta_1)
                + 2 G_\Theta(\btheta_1) V
            \\
            &=
            \Deta_1^\top 
            \Big(
                \nabla_{\bbeta} 
                \Big(
                \tfrac{1}{2}
                \norm{\bbeta}^2_{\nabla^2_{\bbeta} A(\bbeta_1)} + A(\bbeta) 
                - \inp{\bbeta}{U_*}
                \Big)
                \Big|_{\bbeta = \bbeta(\btheta_2)}
            \Big) 
            \\ & \ \phantom{=} 
            + \nabla_{\btheta_2} 
            \Big(
                \tfrac{1}{2}\norm{\btheta_1 - \btheta_2}^2_{\nabla^2_{\btheta_1} \phi(\btheta_1)} 
                + \phi(\btheta_2)
                - 
                \inp{\btheta_2}{V_*}
            \Big)
            \\
            &=
            \Deta_1^\top 
            \Big(
                \nabla_{\bbeta} u(\bbeta) \Big|_{\bbeta = \bbeta(\btheta_2)}
            \Big)
            +
            \nabla_{\btheta_2} v(\btheta_2).
        \end{align*}
        Where we denote the functions 
        $$
        u(\bbeta) = \tfrac{1}{2}
        \norm{\bbeta}^2_{\nabla^2_{\bbeta} A(\bbeta_1)} + A(\bbeta) 
        - \inp{\bbeta}{U_*}
        $$ 
        with $U_* = \nabla_{\bbeta} A(\bbeta_1) + \nabla^2_{\bbeta} A(\bbeta_1) \bbeta_1$
        and 
        $$ v(\btheta_2) = 
        \tfrac{1}{2}\norm{\btheta_1 - \btheta_2}^2_{\nabla^2_{\bth_1} \phi(\bth_1)} 
        + \phi(\btheta_2) - \inp{\btheta_2}{V_*}
        $$ 
        with 
        $$
        V_* = \nabla_{\btheta_1} \phi(\btheta_1)
        + 2 G_\Theta(\btheta_1) V.
        $$ 
        Note that the function $u$, in $\bbeta$, is strongly convex because of the quadratic term in $\bbeta$ with constant positive-definite matrix $\nabla^2_{\bbeta} A(\bbeta_1) \succ 0$ (in practice we rarely would evaluate it on the boundary).
        As the map $h$ is affine with Jacobian $\Deta_1 = \Deta_2 = F$ (constant w.r.t $\bth_2$) full column rank, it is also strongly convex in $\bth_2$. Hence, the first term in the last passage is the gradient of the strongly convex function $u$. The function $v$ is also a strongly convex function in $\bth_2$, due to the quadratic form in $\bth_2$ and the fixed positive-definite matrix
        $\nabla^2_{\btheta_1} \phi(\btheta_1) \succ 0$. Therefore, the last expression is the sum of the gradients of strongly convex functions with a convex domain $\Theta$. Furthermore, $u \circ h+v$ also blows up at the boundary of $\Theta$ (since $\phi$ does) or when $\norm{\bth_2} \to \infty$. Therefore, for the given $V$, $u \circ h + v$ has a unique minimizer, and $\btheta_2$ is the unique minimizer. Hence, for each element $V \in T_{\btheta_1} \bar{M}$ there exists a unique $\btheta_2 \in \bar{M}$, and the approximate logarithmic map is, therefore, bijective.
        
        Moreover, the Jacobian of the approximate logarithmic map is computed as, 
        \begin{align*} 
            D\olsi{\mathrm{Log}}_{\btheta_1}(\btheta_2) 
            &= 
            \tfrac{1}{2} G_{\Theta}(\btheta_1)^{-1}
            \Big[
            F
            ^\top 
            \Big( 
                \nabla^2_{\bbeta} A(\bbeta_1) 
                +
                \nabla^2_{\bbeta} A(\bbeta_2)
            \Big)
            F
                + \nabla^2_{\btheta_2} \phi(\btheta_2)
                + \nabla^2_{\btheta_1} \phi(\btheta_1)
            \Big].
        \end{align*}
        As the determinant of this Jacobian is the determinant product of two symmetric positive-definite matrices, it is non-vanishing everywhere. Hence, the approximate logarithm map is also a diffeomorphism. 
        
        Note also that, because $\bth_2 \mapsto (u \circ h + v)(\bth_2)$ is strongly convex, the gradient of the function $g(\bth_2) = \norm{\nabla_{\bth_2} (u \circ h + v)(\bth_2)}^2 $ satisfies the global Polyak-{\L}ojasiewicz condition \citep{kamiri:2016}, hence every stationary point of $\nabla_{\bth_2} g$, say $\bth_2^*$, is the unique minimizer of $u \circ h + v$.
    \end{proof}
     %
%
\noindent
\textbf{Theorem 3}
%
    \begin{proof}:
        By the Central Limit Theorem (CLT), $\sqrt{n}(\hat{\btheta}_n - \btheta_*) \rightarrow \mathcal{N}(\0, \mathcal{I}_{\Theta}(\btheta_*)^{-1})$ 
        converges in probability, and $\bar{\Sigma}^{-1} \rightarrow n \mathcal{I}_\Theta(\btheta_*) $ by the Law of Large Numbers, with $\Sigma^{-1}_{\pi} = O(1)$ negligible compared to $n\mathcal{I}_\Theta(\btheta_*)$.
        Typical posterior values lie around $\boldsymbol{\delta} := \| \bth - \hat\bth_n\| = O(n^{-\frac{1}{2}})$, so it suffices to evaluate $\olsi{\mathrm{Log}}_{\hat\bth_n}$ around the $\hat\bth_n$.
        
        We Taylor-expand the gradient of $D := D_{\mathrm{post}}(\bth \| \bth')$ in the first argument around $\bth = \hat\bth_n$ and the second argument around $\bth' = \hat\bth_n$, in the direction $\boldsymbol{\delta} = \bth - \hat\bth_n$. Write $\partial_i$ for derivatives in the first argument and $\partial_{i'}$ for derivatives in the second. Using the Einstein summation and evaluating it at $(\hat\bth_n, \hat\bth_n)$, the gradient satisfies
        \begin{align*}
            \partial_{m} D
            &= 
            \partial_m \, \partial_{i'}D
            \boldsymbol{\delta}^{i'}
            +
            \tfrac{1}{2}
            \partial_m \, \partial_{i'}\partial_{j'} D
            \boldsymbol{\delta}^{i'}\boldsymbol{\delta}^{j'}
            +
            \tfrac{1}{6} \partial_m \, \partial_{i'}\partial_{j'}\partial_{k'} D \, 
            \boldsymbol{\delta}^{i'}\boldsymbol{\delta}^{j'}\boldsymbol{\delta}^{k'}
            +
            O(\| \boldsymbol{\delta} \|^4) 
        \end{align*}
        On the diagonal $(\hat\bth_n, \hat\bth_n)$, the first derivative of $D$ vanishes, and its mixed second derivative gives $\partial_m \partial_{i'} D = -2G_{mi'}$ where $G = G_\Theta(\hat\bth_n)$ is the matrix coefficient. By Theorem 1, the third-order terms in the approximation of the symmetrized contrast function cancel, so $\partial_m \partial_{i'} \partial_{j'} D = 0$, thus
        \begin{align*}
            \partial_{m} D = 
            -2 G_{mi'}
            \boldsymbol{\delta}^{i'}
            +
            \tfrac{1}{6} 
            \partial_{m}\partial_{i'}\partial_{j'}\partial_{k'} D
            \boldsymbol{\delta}^{i'}\boldsymbol{\delta}^{j'}\boldsymbol{\delta}^{k'}
            +
            O(\| \boldsymbol{\delta} \|^4)
        \end{align*}
        where $D_{mi'j'k'} = \partial_{m}\partial_{i'}\partial_{j'}\partial_{k'} D$.
        Plugging into the approximate logarithmic map formula
        \begin{align*}
            \olsi{\mathrm{Log}}_{\hat\bth_n}(\bth)_i
            &=
            -\tfrac{1}{2} (G^{-1})^{im} \partial_m D
            \\
            &= -\tfrac{1}{2} (G^{-1})^{im}
            \big(
            - 2 G_{mi'} \boldsymbol{\delta}^{i'}
            + \tfrac{1}{6} D_{m i'j'k'}
            \boldsymbol{\delta}^{i'}\boldsymbol{\delta}^{j'}\boldsymbol{\delta}^{k'}
            +
            O(\| \boldsymbol{\delta} \|^4)
            \big) 
            \\
            &= 
            \boldsymbol{\delta}^{i}
            -
            \tfrac{1}{12} (G^{-1})^{im} D_{m i'j'k'}
            \boldsymbol{\delta}^{i'}\boldsymbol{\delta}^{j'}\boldsymbol{\delta}^{k'}
            +
            O(\| \boldsymbol{\delta} \|^4).
        \end{align*}
        From \citet[see Prop. 2 and 3]{eguchi:1992} the second term in the approximate logarithmic map directly relates to Eguchi's curvature-like $B$-tensor as follows. In our symmetrized $D$, Eguchi's fourth-order differential operator $D^E$ has coefficients $\bar{g}(D^E(\partial_{i'}, \partial_{j'}) \partial_{k'}, \partial_m) = - D_{m i' j' k'}$ where $\bar{g}$ is the metric-tensor with coefficients $2G$ induce by $D$. When raising its indices we get $D^{Em}_{i'j'k'} = - \tfrac{1}{2} (G^{-1})^{ms} D_{s i' j' k'}$.
        The $B$-tensor has coefficients 
        $
        \bar{g}(B(\partial_{i'}, \partial_{j'})\partial_{k'}, \partial_m) = 
        -
        D_{mi'j'k'} 
        $
        $
        - 
        2G_{ms}
        \big(
        \partial_{i'}(\Gamma_{j'k'}^{s(0)})^D
        +
        $
        $
        (\Gamma_{i'r'}^{s(0)})^D
        (\Gamma_{j'k'}^{r'(0)})^D
        \big)
        $ 
        where $(\Gamma_{j'k'}^{s(0)})^D$ are the Christoffel symbols of the second-kind associated with the metric $2G$. Raising its indices, we have
        $B^m_{i'j'k'} = D^{Em}_{i'j'k'} - 
        \big(
        \partial_{i'}(\Gamma_{j'k'}^{m(0)})^C
        +
        (\Gamma_{i'r'}^{m(0)})^C
        (\Gamma_{j'k'}^{r'(0)})^C
        \big)$. The approximate logarithmic map can be rewritten as
        \begin{align*}
            \olsi{\mathrm{Log}}_{\hat\bth_n}(\bth)
            &=
            \boldsymbol{\delta}
            + 
            H(\boldsymbol{\delta}, \boldsymbol{\delta}, \boldsymbol{\delta})
            +
            O(\norm{\boldsymbol{\delta}}^4) 
        \end{align*}
        where the vector $H(\boldsymbol{\delta}, \boldsymbol{\delta}, \boldsymbol{\delta})$ $=$ $\tfrac{1}{6}\big( \big( B^1_{i'j'k'} + \partial_{i'}(\Gamma_{j'k'}^{1(0)})^D + (\Gamma_{i'r'}^{1(0)})^D (\Gamma_{j'k'}^{r'(0)})^D \big) 
        \boldsymbol{\delta}^{i'}\boldsymbol{\delta}^{j'}\boldsymbol{\delta}^{k'},$
        $
        \ldots,
        $
        $
        \big( B^{d_\Theta}_{i'j'k'} + \partial_{i'}(\Gamma_{j'k'}^{d_\Theta(0)})^D + (\Gamma_{i'r'}^{d_\Theta(0)})^D (\Gamma_{j'k'}^{r'(0)})^D \big)
        \boldsymbol{\delta}^{i'}\boldsymbol{\delta}^{j'}\boldsymbol{\delta}^{k'}
        \big)$ and $d_\Theta = \dim(\bar{\Theta})$. This expression now makes explicit the role of the Riemannian curvature notion in the local corrections of the Laplace approximation which is based on the $B$-tensor and Christoffel symbols coefficients.
        
        The Jacobian follows by differentiation so that $D\olsi{\mathrm{Log}}_{\hat{\btheta}_n}(\btheta)
        = I + O(\| \boldsymbol{\delta} \|^2) = I + O(n^{-1}) $ and similarly its determinant as
        $\det D\olsi{\mathrm{Log}}_{\hat{\btheta}_n}(\btheta)
        = 1 + O(\| \boldsymbol{\delta} \|^2) = 1 + O(n^{-1})$. Since $\hat\btheta_n \rightarrow \btheta_*$ 
        in probability, 
        we conclude
        \begin{align*}
            \rho_{\textsl{WG}}(\btheta) = 
            \rho_{\textsl{LA}}(\btheta)(1 + O(n^{-1}))
        \end{align*}
    \end{proof}
    %
%
\section{Appendix: Exact approximations}

\subsection{Exactness of the Gaussian posterior case for linear models} 

Consider the following Gaussian linear regression case $(Y_1, \ldots, Y_n)|\btheta \sim \mathcal{N}(\bmu, \Sigma)$ with $\bmu = \mathbf{X}\boldsymbol{\theta}$ where $\mathbf{X}$ is a $n \times d_\Theta$ matrix, $d_\Theta$ the parameters space dimension and the prior distribution $\btheta \sim \mathcal{N}(\bm, Q)$. The posterior distribution is given in closed-form as another Gaussian $\rho_{\by}(\btheta) = \mathcal{N}(\btheta| \bmu_{\mathrm{post}}, \Sigma_{\mathrm{post}})$ whose parameters follow the Kalman's equations $\bmu_{\mathrm{post}} = \bm + Q\mathbf{X}^\top(\Sigma + \mathbf{X} Q \mathbf{X}^\top)^{-1} (\by - \textbf{X}\boldsymbol{m})$ and $\Sigma_{\mathrm{post}} = Q - Q \mathbf{X}^\top (\Sigma + \mathbf{X} Q \mathbf{X}^\top)^{-1}\mathbf{X} Q$. This is a Gaussian where the MAP is $\hat{\btheta} = \bmu_{\mathrm{post}}$ and the negative inverse Hessian of the log-posterior at the MAP is $- \nabla^2 \log \rho_{\by}(\hat{\btheta})^{-1} = \Sigma_{\mathrm{post}}$.
Define the manifold
\begin{align*}
    \bar{M} = 
    \Big\{
    \rho_{\bth}(\by) = 
    \mathcal{N}(\by|\bmu(\bth), \Sigma) : 
    \bmu = h(\bth) = \mathbf{X}\boldsymbol{\theta},
    \,
    \by \in \mathbb{R}^n,
    \Sigma \, \mathrm{fixed}
    \Big\}
\end{align*}
In this case, consider the contrast function $D_{\mathrm{post}}$ constructed with respect to the $\bKL$-divergence
%
    $
    \textsl{KL}(\rho_{\bth_1} || \rho_{\bth_2}) 
    =
    \tfrac{1}{2} \norm{\bmu(\bth_2) - \bmu(\bth_1)}^2_{\Sigma^{-1}}
    $
%
and its dual. By including the prior geometry through the Bregman divergence, the sum of the gradients of the dual divergences w.r.t. $\bth_1$ gives
\begin{align*}
    \nabla_{\btheta_1} D_{\mathrm{post}}(\bth_1 \| \bth_2)
    &=
    \nabla_{\btheta_1} 
    \Big( 
    \textsl{KL} (\rho_{\btheta_1} || \rho_{\btheta_2}) 
    +
    \textsl{KL} (\rho_{\btheta_2} || \rho_{\btheta_1}) 
    +
    D_\phi(\btheta_1 \| \btheta_2)
    +
    D_\phi(\btheta_2 \| \btheta_1)
    \Big) \\
    &= 
    2 \big( \mathbf{X}^\top \Sigma^{-1}\mathbf{X} 
    + 
    Q^{-1}\big) \big(\btheta_1 - \btheta_2\big).
\end{align*}
From the contrast function, the final matrix coefficient of the metric is given as
\begin{align*}
    \nabla^2_{\btheta_1} 
    \big( \textsl{KL} (\rho_{\bth_1} || \rho_{\bth}) 
    +
    D_\phi(\btheta_1 \| \bth) 
    \big)
    \big|_{\bth_1 = \bth} 
    =
    G_{\Theta}(\btheta)
    = 
    \mathcal{I}_{\Theta}(\btheta) + \Sigma_\pi^{-1}
    = 
    \mathbf{X}^\top \Sigma^{-1} \mathbf{X} + Q^{-1}.
\end{align*}
Take as the basepoint $\btheta_1 = \hat{\btheta} = \bmu_{\mathrm{post}}$ and $\Sigma = - \nabla^2 \log \rho_{\by}(\hat{\btheta})^{-1} = \Sigma_{\mathrm{post}}$. The approximate logarithmic map and its Jacobian become
\begin{align*}
    \olsi{\mathrm{Log}}_{\bmu_{\mathrm{post}}}(\btheta) 
    &= 
    -\tfrac{1}{2} 
    \big( \mathbf{X}^\top \Sigma^{-1} \mathbf{X} + Q^{-1} \big)^{-1}
    2 \big( \mathbf{X}^\top \Sigma^{-1}\mathbf{X} 
    + 
    Q^{-1}\big) \big(\bmu_{\mathrm{post}} - \btheta \big) 
    \\
    &
    = 
    \btheta - \bmu_{\mathrm{post}} \\
    D\olsi{\mathrm{Log}}_{\bmu_{\mathrm{post}}}(\btheta) &= I.
\end{align*}
The wrapped-Gaussian distribution simplifies to
\begin{align*}
    \rho_{\textsl{WG}}(\btheta)
    &=
    \mathcal{N}
    \big(
    \olsi{\mathrm{Log}}_{\bmu_{\mathrm{post}}}(\btheta)|
    \0,
    \Sigma_{\mathrm{post}}
    \big)
    |D\olsi{\mathrm{Log}}_{\bmu_{\mathrm{post}}}(\btheta)| 
    \\
    &
    = 
    \mathcal{N}
    \big( \btheta | \bmu_{\mathrm{post}}, \Sigma_{\mathrm{post}} \big)
\end{align*}
and the approximation is exact.

\subsection{The Neal's funnel}

The Neal's funnel probabilistic model is given by, 
\begin{align*}  
    \rho(\btheta) =
    \prod_{d = 1}^{D - 1} 
    \mathcal{N}\big(\theta_d|0, \exp(\tfrac{1}{2}\theta_D)\big)   
    \mathcal{N}(\theta_D|0, \sigma), \ \ \btheta \in \Theta = \mathbb{R}^D.
\end{align*}
Define the manifold 
\begin{align*}
    \bar{M} = 
    \Big\{
    \rho_{\bth}(\by) = 
    \mathcal{N}(\by|\bmu(\bth), I_n) : 
    \bmu(\bth) = h(\bth),
    \by \in \mathbb{R}^D
    \Big\}
\end{align*}
where 
\begin{align*}
    h : \Theta 
    &
    \to \mathbb{R}^D 
    \\
    \btheta 
    &
    \mapsto
    \big(
    \theta_1 \exp(-\tfrac{1}{2} \theta_D), 
    \cdots, 
    \theta_{D - 1}
    \exp(-\tfrac{1}{2} \theta_D),
    \ \tfrac{\theta_D}{\sigma}
    \big)
\end{align*}
is a diffeomorphism \citep[See][for details]{Stan:2023}. The Jacobian of $h$ is
\begin{align*}
    Dh(\btheta) &=
    \begin{bmatrix}
        \exp(-\tfrac{1}{2} \theta_D) & 0
        & \ldots & -\frac{\theta_1}{2} \exp(-\tfrac{1}{2} \theta_D)
        \\
        \vdots & \ddots & \ddots & \vdots 
        \\
        0 & 0 & \ldots & -\frac{\theta_{D - 1}}{2} \exp(-\tfrac{1}{2} \theta_D)
        \\
        0 & 0 & \cdots & \frac{1}{\sigma}
    \end{bmatrix}.
\end{align*}
Take the symmetrized contrast function 
$$
C(\bth_1 \| \bth_2) = \tfrac{1}{2} \norm{\bmu(\bth_2) - \bmu(\bth_1)}^2 + \tfrac{1}{2} \norm{\bmu(\bth_1) - \bmu(\bth_2)}^2.
$$ 
These are Gaussian \bKL\, divergences assuming covariances as identity matrix. Note also that this would be the same as additionally considering the symmetrized Bregman divergence in $C$ and then setting the prior variance to infinity. The gradient of $C$ is
%
    $
    \nabla_{\btheta_1}C(\bth_1 \| \bth_2)
    = 
    2Dh(\btheta_1)^\top \big( h(\btheta_1) - h(\btheta_2) \big).
    $
%
From the first term of $C$, the Hessian along the diagonal gives the matrix coefficient $G_{\Theta}(\btheta) = Dh(\btheta)^\top Dh(\btheta)$. 
Hence, the approximate logarithmic map can be simplified to
\begin{align*}
    \olsi{\mathrm{Log}}_{\btheta_1}(\btheta_2) 
    &=
    -\tfrac{1}{2} G_\Theta(\btheta_1)^{-1}
    \nabla_{\btheta_1} 
        C(\bth_1, \bth_2)
    \\
    &=
    Dh(\btheta_1)^{-1} 
    Dh(\btheta_1)^{-\top}
    Dh(\btheta_1)^\top 
    \big( h(\btheta_2) - h(\btheta_1) \big)
    \\
    &=
    Dh(\btheta_1)^{-1} 
    \big( h(\btheta_2) - h(\btheta_1) \big)
\end{align*}
Let $\btheta_2 = \btheta$ and set the basepoint $\btheta_1 = \0$. We get the Jacobian $Dh(\btheta_1) = \mathrm{diag}(1, \ldots, \tfrac{1}{\sigma})$ and $h(\btheta_1) = \0$. After some length calculations, the negative inverse Hessian of the logarithm of Neal's funnel at the origin gives $\Sigma = -\nabla^2 \log \rho(\bth_1)^{-1} = \mathrm{diag} (1, \ldots, \sigma^2)$. The approximate logarithmic map and the determinant of its Jacobian give,
\begin{align*}
    \olsi{\mathrm{Log}}_{\btheta_1}(\btheta) 
    &=
    \mathrm{diag}(1, \ldots, \sigma) h(\btheta) 
    \\
    |D\olsi{\mathrm{Log}}_{\btheta_1}(\btheta)| 
    &= 
    |\mathrm{diag}(1, \ldots, \sigma)
    Dh(\btheta)| 
    = 
    \underbrace{\exp(-\tfrac{1}{2}\theta_D) \ldots \exp(-\tfrac{1}{2}\theta_D)}_{D-1 \, \mathrm{times}}
\end{align*}
The wrapped Gaussian is defined, 
\begin{align*}
    \rho_{\textsl{WG}}(\btheta)
    &=
    \mathcal{N}
    \big(
    \olsi{\mathrm{Log}}_{\btheta_1}(\btheta)|
    \0,
    \Sigma
    \big)
    |D\olsi{\mathrm{Log}}_{\btheta_1}(\btheta)|
    \\
    &= 
    (2\pi)^{-\frac{D}{2}}|\Sigma|^{-\frac{1}{2}}
    \exp\big(
    -\tfrac{1}{2}
    \norm{ 
    \olsi{\mathrm{Log}}_{\btheta_1}(\btheta)
    }_{\Sigma^{-1}}^2
    \big)
    |D\olsi{\mathrm{Log}}_{\btheta_1}(\btheta)| 
    \\
    &= 
    (2\pi)^{-\frac{D}{2}}|\Sigma|^{-\frac{1}{2}}
    \exp\left(
    -\tfrac{1}{2}
    \norm{ 
    h(\btheta)^\top Dh(\btheta)^\top\Sigma^{-1}Dh(\btheta)h(\btheta)
    }_{\Sigma^{-1}}^2
    \right)
    |D\olsi{\mathrm{Log}}_{\btheta_1}(\btheta)| 
    \\
    &=
    (2\pi)^{-\frac{D}{2}}\sigma^{-1}
    \exp\big(
    -\tfrac{1}{2}\norm{h(\btheta)}^2
    \big)
    \exp(-\tfrac{1}{2}\theta_D) \ldots \exp(-\tfrac{1}{2}\theta_D)
    \\
    &=
    (2\pi)^{-\frac{D}{2}}\sigma^{-1}
    \exp\left(
    -\frac{1}{2}
    \sum_{d = 1}^{D - 1}
    \left(
    \frac{\theta_d}{\exp(\tfrac{1}{2}\theta_D)}
    \right)^2
    - \frac{1}{2}
    \left(
    \frac{\theta_D}{\sigma}
    \right)^2
    \right)
    \exp(-\tfrac{1}{2}\theta_D) \ldots \exp(-\tfrac{1}{2}\theta_D)
    \\
    &=
    \prod_{d = 1}^{D - 1} 
    (2\pi)^{-\frac{1}{2}}
    \exp(-\tfrac{1}{2}\theta_D)
    \exp\left(
    -\frac{1}{2}
    \left(
    \frac{\theta_d}{\exp(\tfrac{1}{2}\theta_D)}
    \right)^2
    \right)
    (2\pi)^{-\frac{1}{2}} \sigma^{-1}
    \exp\left(
    -\frac{1}{2}
    \left(
    \frac{\theta_D}{\sigma}
    \right)^2
    \right)
    \\
    &=
    \prod_{d = 1}^{D - 1} 
    \mathcal{N}\big(\theta_d|0, \exp(\tfrac{1}{2}\theta_D)\big)   
    \mathcal{N}(\theta_D|0, \sigma)
    \\ 
    &
    = \rho(\btheta).
\end{align*}
This shows the approximation is exact.

\section{Appendix : Riemann-Laplace approximation}

Let $\bar{M}$ be a manifold and $f : \bar{M} \to \mathbb{R}$ the logarithm of a posterior distribution. Let $\btheta$ be a global parametrisation of $\bar{M}$ and $\bth^*$ the basepoint (MAP estimate). Let $\mathrm{Exp}_{\bth^*}(V)$ denote the exponential map and $\mathrm{Log}_{\bth^*}(\bth)$ the logarithmic map (the inverse map of the exponential map).
The second-order Taylor-expansion \citep{boumal:2023} on the manifold is given by 
\begin{align*} 
    f(\mathrm{Exp}_{\bth^*}(V)) &\approx
    f(\bth^*) + 
    g_{\bth^*}\big(\mathrm{grad}\, f(\bth^*), V\big) 
    + 
    \tfrac{1}{2}\, g_{\bth^*}\big(\mathrm{Hess}\, f({\bth^*})[V], V\big)
    +
    O(g_{\bth^*}(V, V)^3).
\end{align*}
Plugging equations (1), (2) and (3) (main paper) in the equation above, and ignoring the remainder term, we obtain
\begin{align*}
    f(\mathrm{Exp}_{\bth^*}(\mathrm{Log}_{\bth^*}(\bth)))
    &\approx
    f(\bth^*) + 
    g_{\bth^*}\big(\mathrm{grad}\, f(\bth^*), \mathrm{Log}_{\bth^*}(\bth)\big) 
    \\ 
    &+ 
    \tfrac{1}{2}\, g_{\bth^*}\big(\mathrm{Hess}\, f({\bth^*})[\mathrm{Log}_{\bth^*}(\bth)], \mathrm{Log}_{\bth^*}(\bth) \big)
    \\
    f(\bth)
    &\approx 
    f(\bth^*) +
    \Big\langle 
        G(\bth^*)^{-1} \nabla f(\bth^*), \, 
        \mathrm{Log}_{\bth^*}(\bth) 
    \Big\rangle_{G(\bth^*)} 
    \\
    &
    +
    \tfrac{1}{2}
    \Big\langle
    G(\bth^*)^{-1} \Big(
        \nabla^2 f(\bth^*) 
        - 
        \sum_k \Gamma^k(\bth^*) \partial_k f(\bth^*)
    \Big), \,
    \mathrm{Log}_{\bth^*}(\bth)
    \Big\rangle_{G(\bth^*)} 
    \\
    &\approx 
    f(\bth^*) 
    + 
    \big\langle 
        \nabla f(\bth^*),\, \mathrm{Log}_{\bth^*}(\bth) 
    \big\rangle 
    \nonumber \\
    &
    - \tfrac{1}{2}
    \bigg( 
        \big\|
            \mathrm{Log}_{\bth^*}(\bth)
        \big\|^2_{-\nabla^2 f(\bth^*)} 
    +
    \big\|
        \mathrm{Log}_{\bth^*}(\bth)
    \big\|^2_{\sum_{k=1}^{d}\Gamma^k(\bth^*)\,\partial_k f(\bth^*)}
    \bigg). 
\end{align*}
Since at $\bth^*$ we have the gradient $\nabla f(\bth^*) = \0$, the last expression simplifies to
\begin{align*}
    f(\bth)
    &\approx 
    f(\bth^*)
    - \tfrac{1}{2}
    \big\|
        \mathrm{Log}_{\bth^*}(\bth)
    \big\|^2_{-\nabla^2 f(\bth^*)}.
\end{align*}
Following \cite{chev:2022}, as the logarithmic map is given in a parametrisation, we can find a density expression using the Lebesgue measure, as in the classical Laplace approximation. That is, define the Riemann-Laplace approximation as
\begin{align*}
    \rho_{\textsl{RL}}(\bth) 
    &= \frac{e^{    
    - \tfrac{1}{2}
    \big\|
        \mathrm{Log}_{\bth^*}(\bth)
    \big\|^2_{-\nabla^2 f(\bth^*)}
    }
    }{
    \displaystyle \int_{\Theta}
        e^{    
        - \tfrac{1}{2}
        \big\|
            \mathrm{Log}_{\bth^*}(\bth)
        \big\|^2_{-\nabla^2 f(\bth^*)}
        }
        \dd\boldsymbol{\bth}
    }.
\end{align*}
The extra difficult in this approach, besides the computation of the true logarithmic map, resides solving the integral characterizing the normalizing constant in the denominator. Observe that when the geometry is Euclidean and the parametrization in Cartesian, we recover the classical Laplace approximation because $\mathrm{Log}_{\bth^*}(\bth) = \bth - \bth^*$. 

\section{Appendix : Extra scatter plots for Poisson and Bernoulli manifolds}

\begin{figure}[!h]
    \begin{center}
    \hspace{-0.8cm}
    \begin{tabular*}{\columnwidth}{ cc }
        \includegraphics[scale = 0.56]{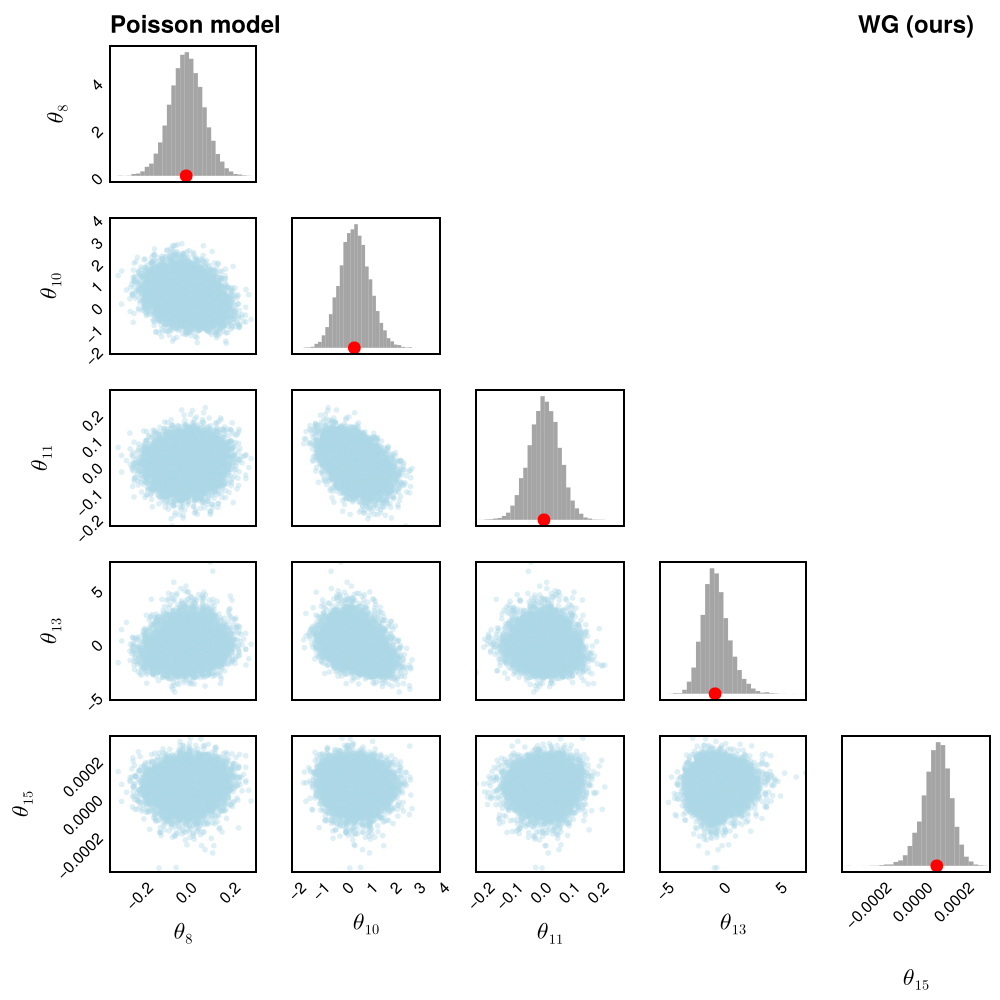}
        \hspace{-0.37cm}
        &
        \includegraphics[scale = 0.56]{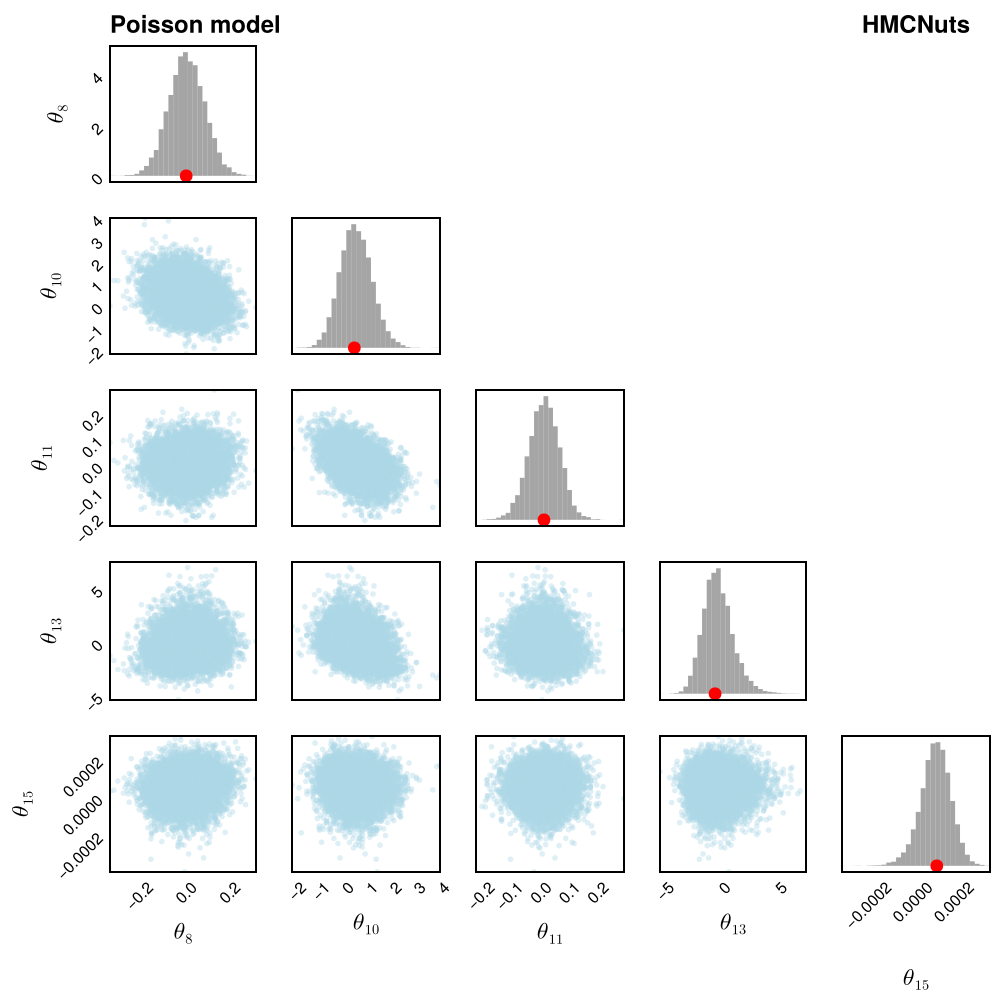}
        \\
        \includegraphics[scale = 0.56]{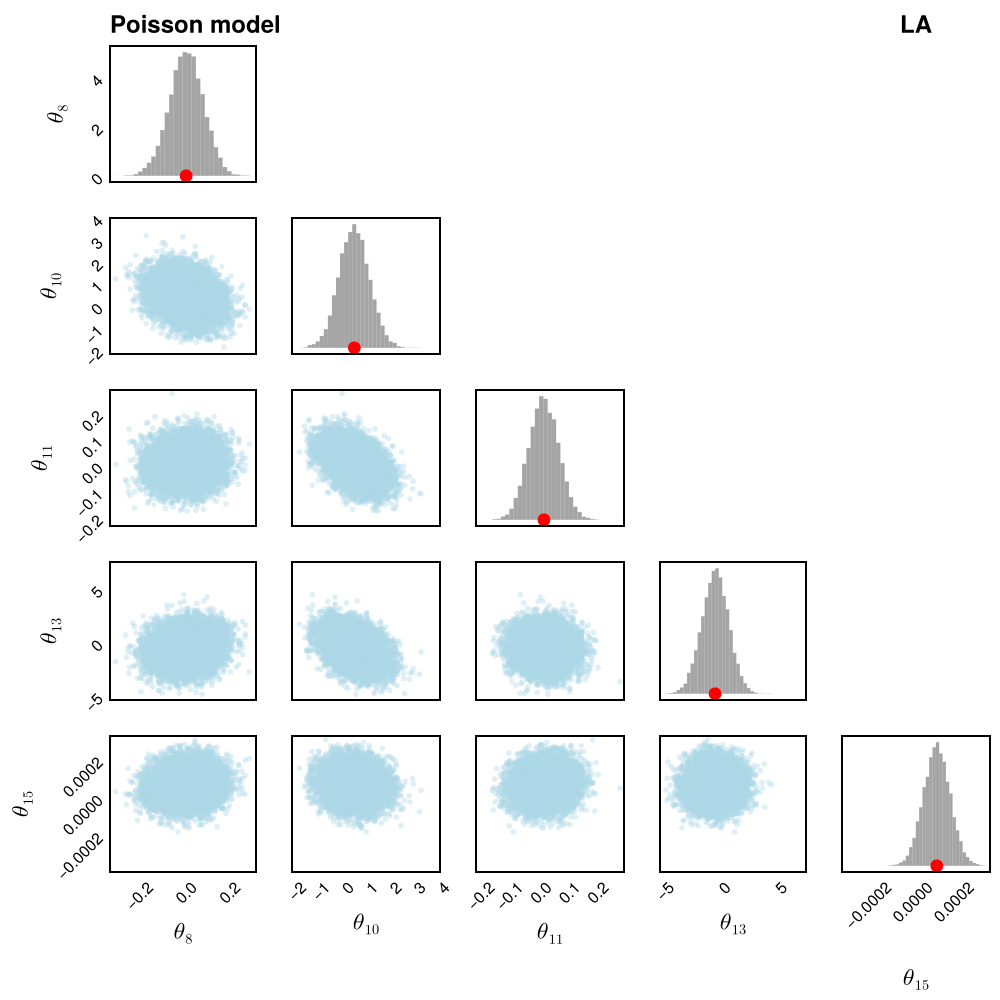}
        \hspace{-0.37cm}
        &
        \includegraphics[scale = 0.56]{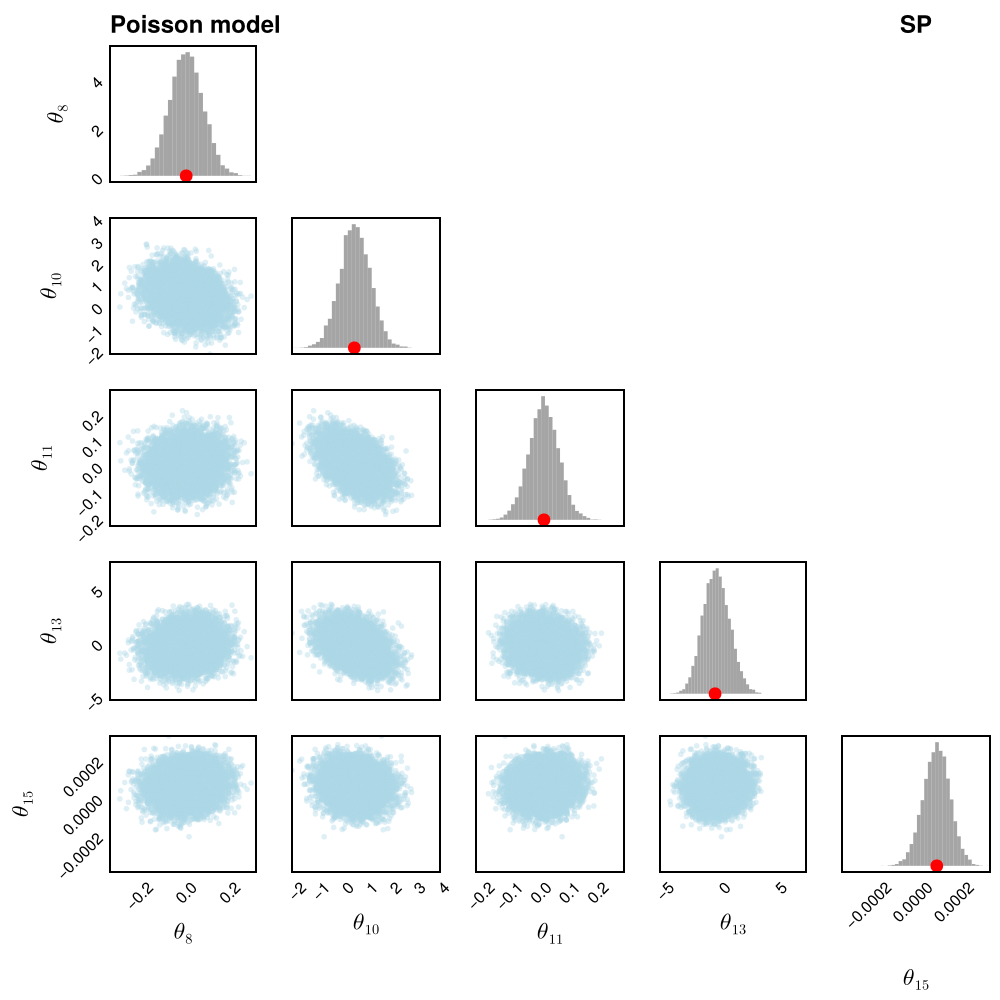}
    \end{tabular*}
        \caption{
            Illustration of the resulting scatter plot for the approximate methods \bWG\,, \bHMC\, (reference), \bLA\, and \bSP\, in the australian dataset case with Poisson regression model. We had randomly picked to parameters $\bth_8$, $\bth_{10}$, $\bth_{11}$, $\bth_{13}$ and $\bth_{15}$ with all 12000 samples for visualization. 
            In the top-left figure we show the scatter plots and histograms as estimators of the joint and marginal densities of the top-right distributions. On the bottom-left and bottom-right we respectively reproduce the same figures for the \bLA\, and \bSP\, cases.
        } 
        \label{fig:reg-poisson-1}
    \end{center}
\end{figure}

\begin{figure}[!hb]
    \begin{center}
    \begin{tabular*}{\columnwidth}{ cc }
        \includegraphics[scale = 0.5]{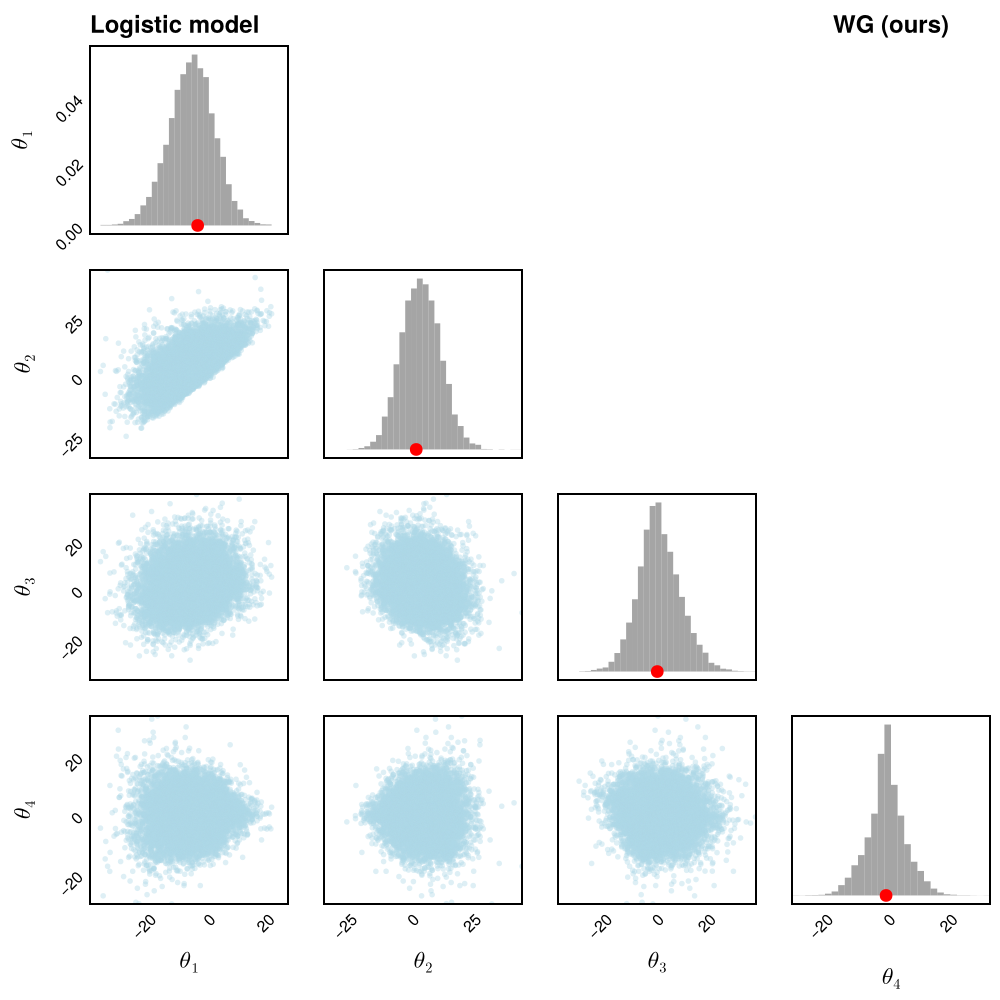}
        \hspace{-0.37cm}
        &
        \includegraphics[scale = 0.5]{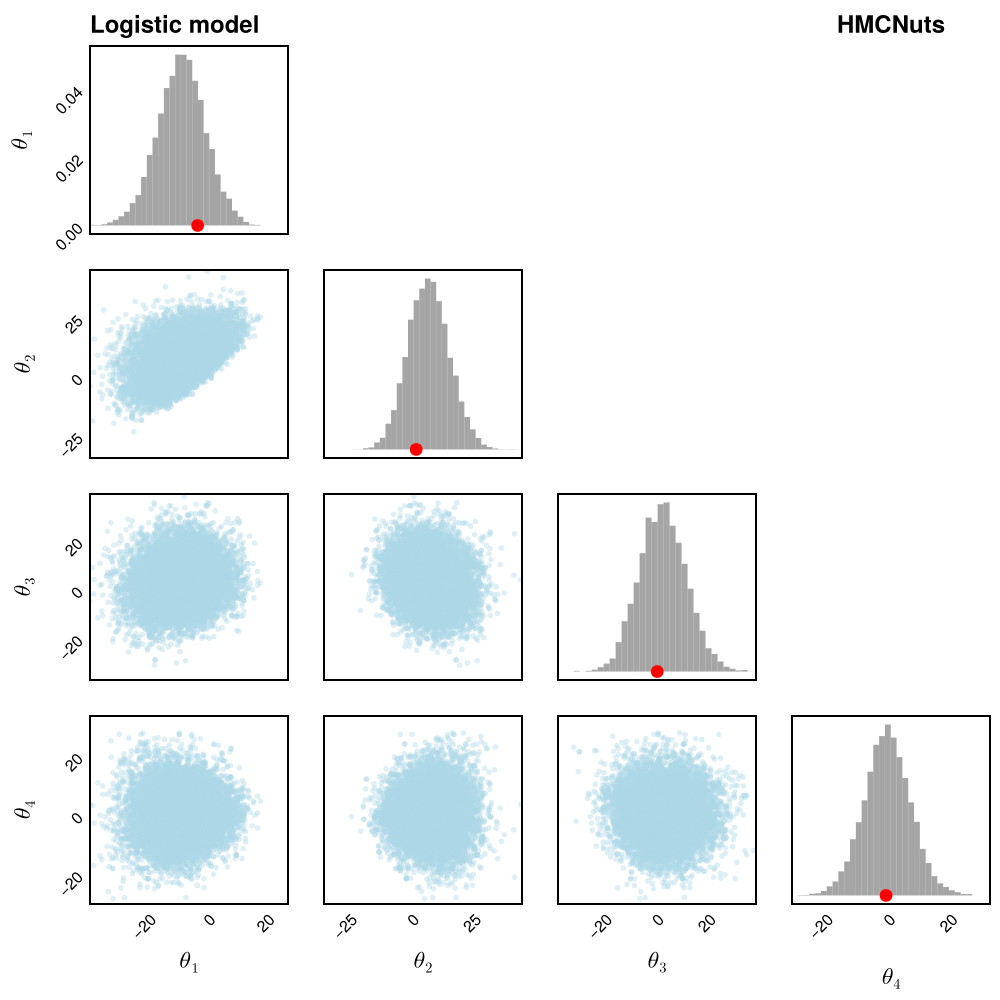}
        \\
        \includegraphics[scale = 0.5]{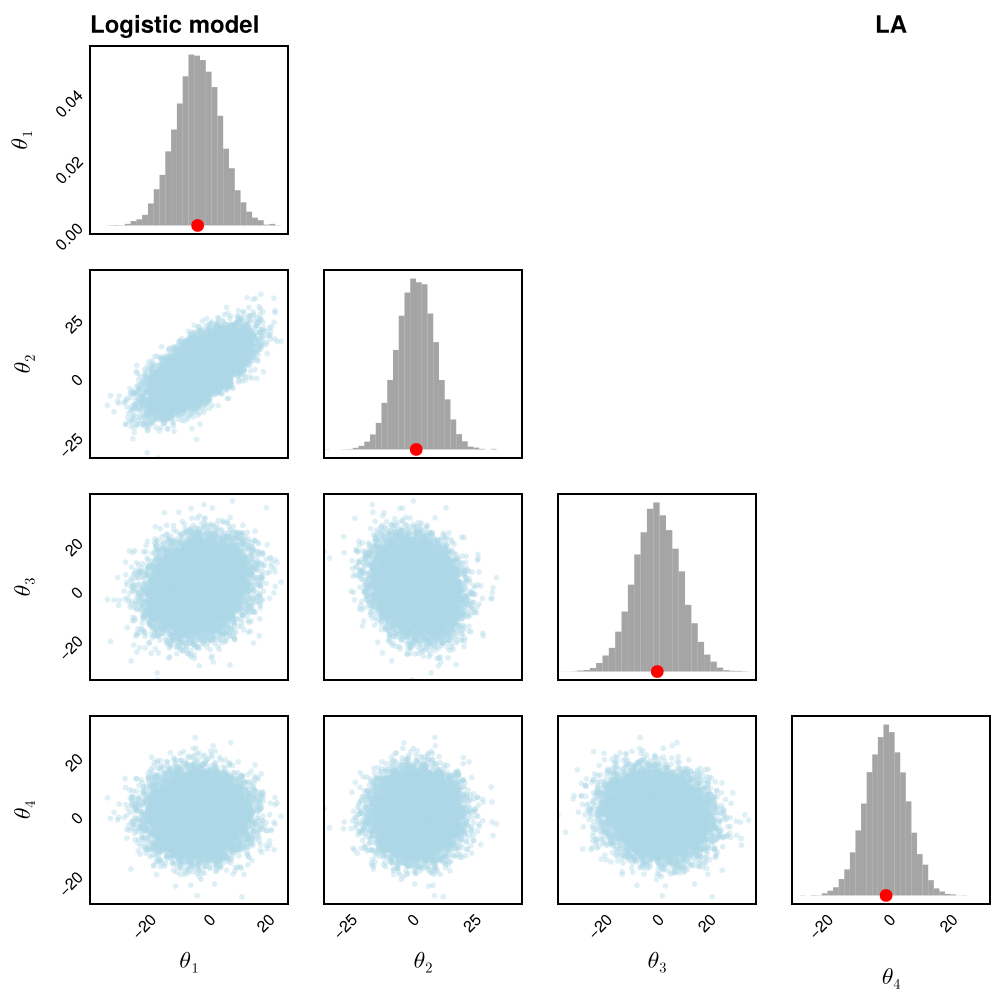}
        \hspace{-0.37cm}
        &
        \includegraphics[scale = 0.5]{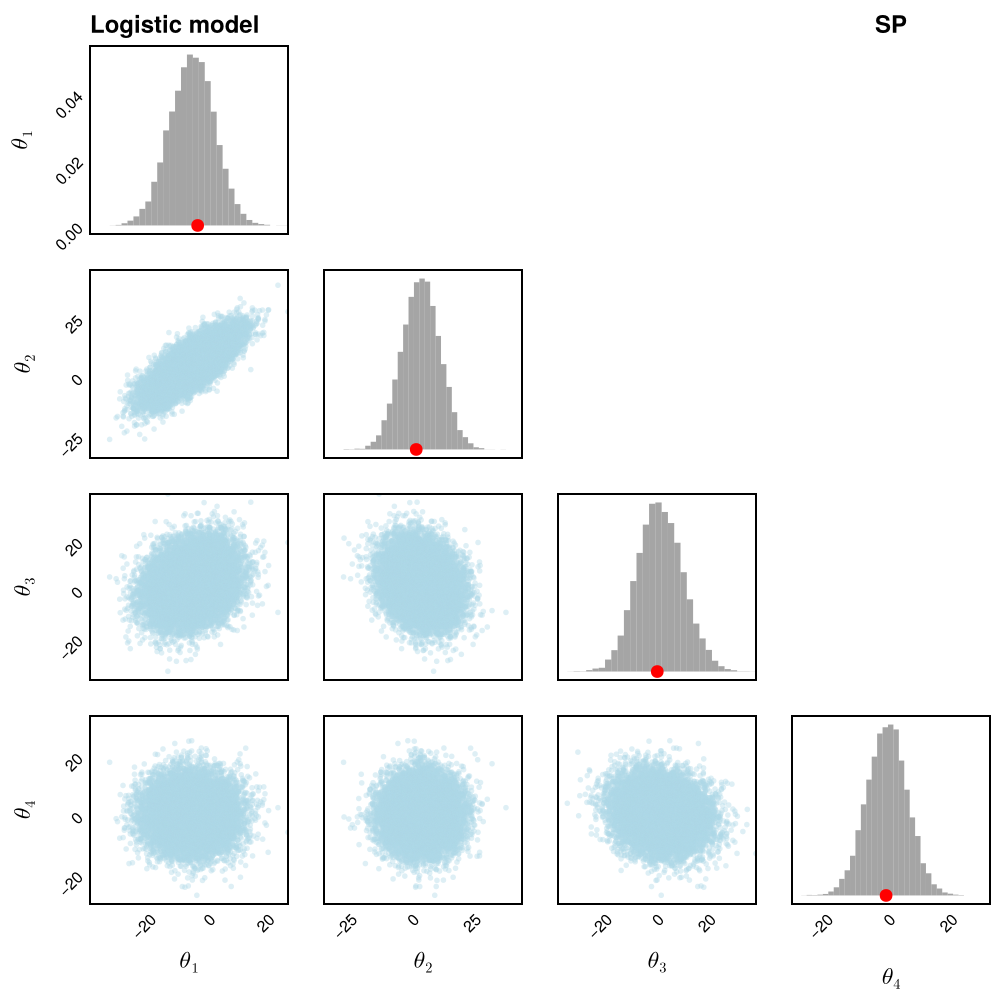}
    \end{tabular*}
        \caption{
            Illustration of the resulting scatter plot for the approximate methods \bWG\,, \bHMC\, (reference), \bLA\, and \bSP\, for the microassay dataset with Bernoulli regression model. In this case all the whole set of parameters $\bth$ with all 12000 samples is shown for visualization. In the top-left figure we show the scatter plots and histograms as estimators of the joint and marginal densities of the top-right distributions. On the bottom-left and bottom-right we respectively reproduce the same figures for the \bLA\, and \bSP\, cases
        } 
        \label{fig:reg-bernoulli-5}
    \end{center}
\end{figure}

The following additional scatter plots in Figure \eqref{fig:reg-poisson-1} and Figure \eqref{fig:reg-bernoulli-5} show the behavior of different approximation methods for the australian and microassay data set approached with Poisson and Bernoulli regression models, respectively. Visually, we observe similar scatter patterns between \bWG\, and \bHMC, compared to those scatter patterns of \bLA\, and \bSP. Indeed, the evaluation metrics such as $W_2$, are in closer agreement for both \bWG\, vs \bHMC\, and \bSP\, vs \bHMC\,. This indicates that the approximate distributions \bWG\, and \bSP\, provide similar error to the \bHMC\, approximation.

\section{Appendix : Extra experiments with neural-networks}

In this section, we additionally test the robustness of the wrapped Gaussian approximation in overparameterized NNs where different initializations of the optimisation routine typically lead to different optima of the posterior distribution. We reconsider the regression example with a Gaussian noise as well as a binary classification task subject to Bernoulli probabilistic model. For both experiments we show that the predictive distributions of $f_{\btheta}(\bx)$ is robust to different MAP estimates of the posterior distribution, whereas the predictive distribution with Laplace approximation does not seem to be. 

\begin{figure}[!ht]
    \vskip 0.0in
        \begin{center}
            \centerline{
                \includegraphics[scale = 0.9]{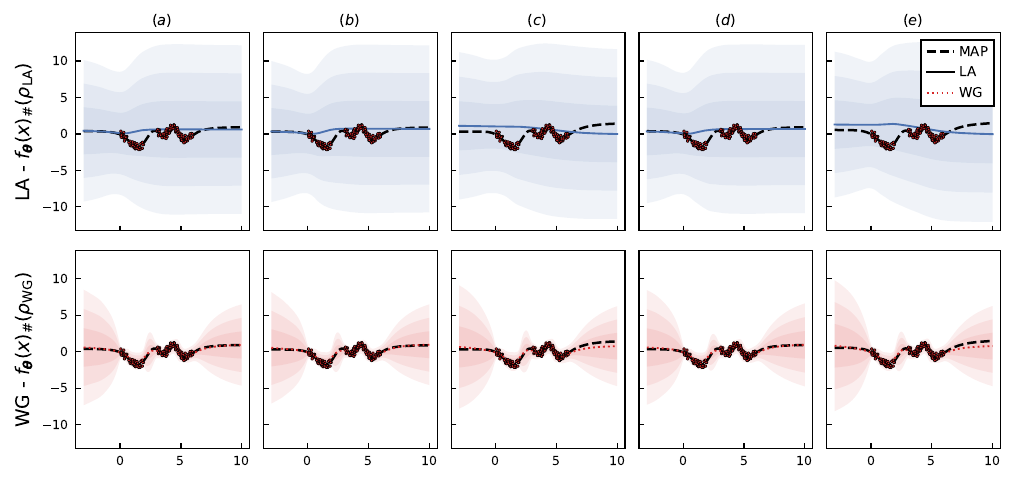}
            }
            \caption{The first row shows the predictive distributions for the \bLA\, approximation, where each column corresponds to a different initialisation of the optimiser and hence a different MAP estimate, giving rise to a different distribution $f_{\bth}(x_\ast)_{\#}(\rho_{\bLA})$. The second row shows the analogous predictive distributions $f_{\bth}(x_\ast)_{\#}(\rho_{\bWG})$ based on the \bWG\, approximation. In this case, the predictive distributions are consistently similar across different MAP estimates in both mean and variance: the predictive mean consistently follows each MAP estimate, with smaller variance in the regions where data are observed. This behaviour is not observed with \bLA.
            } 
            \label{fig:snelson-regression-figure}
        \end{center}
    \vskip -0.1in
\end{figure}

\begin{figure}[!hb]
    \vskip 0.0in
        \begin{center}
            \centerline{
                \includegraphics[scale = 0.89]{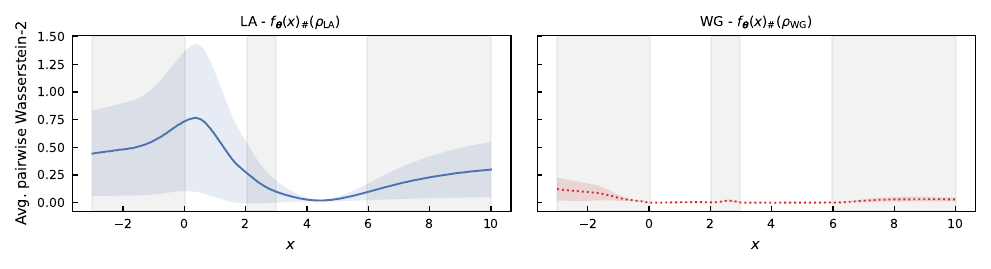}
            }
            \caption{This plot shows the average $W_2$ distance across the various predictive distributions. For each $x$, we obtain five predictive distributions $f_{\bth}(x)_{\#}(\rho_{\bLA})$ corresponding to the five different MAP estimates, compute all pairwise $W_2$ distances between them, and average the result. This is repeated for every $x$ in the given interval, and the resulting mean and variance are plotted. The same procedure is applied to the \bWG\, approximations. The \bWG\, distances show near-zero mean values with low variance throughout, whereas those of \bLA\, vary widely in both mean and variance. This confirms that the predictive distributions of the \bWG\ approximation are consistently insensitive to the choice of MAP estimate.} 
            \label{fig:wasserstein-figure}
        \end{center}
    \vskip -0.1in
\end{figure}

We approach both settings as follows: we randomly initialise 5 sets of parameters (a)--(e) (NN weights) and train them until convergence. At each optimum, we draw $S = 5 \times 10^3$ samples from the respective approximate \bWG\, posteriors and $K = 20$ samples from each predictive distribution, that is, from $f_{\bth}(x)_{\#}(\rho_{\bWG})$ for fixed values of $x \in [-3, 10]$. At every choice of $x$, we compute the pairwise $W_2$ distances between the samples of $f_{\bth}(x)_{\#}(\rho_{\bWG})$ across the different MAP cases (a)--(e) and average over these. For the NN regression case, we summarise these by the mean distance and variance per point $x$. We repeat the same procedure for the predictive distribution under the \bLA, that is, for $f_{\bth}(x)_{\#}(\rho_{\bLA})$. These results are shown in Figures~$\ref{fig:snelson-regression-figure}$ and~\ref{fig:wasserstein-figure}. For the classification case, we compute the spectra of the Fisher matrix 
$\mathcal{I}(\hat{\bth})$ for each MAP estimate.
%

Figure~\ref{fig:snelson-regression-figure} shows the resulting predictive distributions under each approximation across the five runs. The \bWG\, approximation appears to adapt to the local geometry of each mode, yielding nearly identical predictive distributions: its mean follows the MAP on $f_{\hat\bth}(x)$, the variance is tighter within the data support and grows smoothly outside the region of observed covariates, and this behaviour is stable from run to run. In contrast, \bLA\ performs poorly and is uncontrollable. This is supported by Figure~\ref{fig:wasserstein-figure}, where the $W_2$ distances between the predictive distributions obtained from different MAP estimates are computed and averaged over all pairs, and this is repeated for every $x$ in the given interval. For the \bWG\, cases, the distances are essentially zero within the observed covariate region and only slightly larger outside it, confirming that the predictive distributions based on \bWG\, are insensitive to which mode the optimiser reaches --- a property of utmost importance in downstream ML tasks with overparametrised models. For the \bLA\ cases, the distances remain large throughout, showing that the standard Laplace approximation is far more sensitive to the choice of MAP estimate.

\begin{figure}[!ht]
    \vskip 0.0in
        \begin{center}
            \centerline{
                \includegraphics[scale = 0.68]{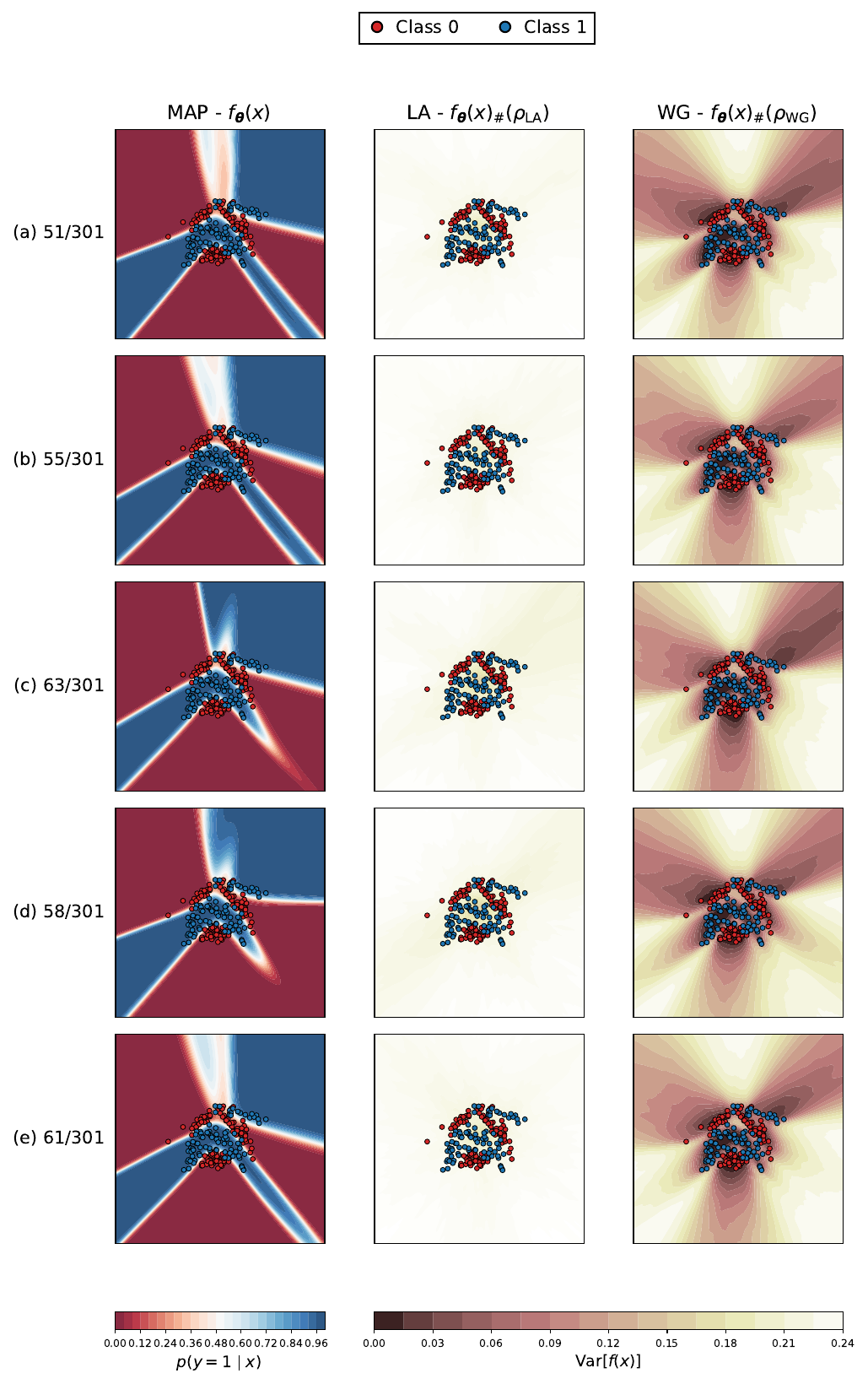}
            }
            \caption{The first column shows the decision boundary at each MAP estimate. The middle and right columns show the predictive variance for the \bLA\, and \bWG\, approximations, respectively. Once again, the \bWG\, approximation yields predictive variances that are both more accurate and insensitive to the choice of MAP estimate, compared to those of \bLA.} 
            \label{fig:classification-visualization}
        \end{center}
    \vskip -0.1in
\end{figure}

\begin{figure}[!ht]
    \vskip 0.0in
        \begin{center}
            \centerline{
                \includegraphics[scale = 0.9]{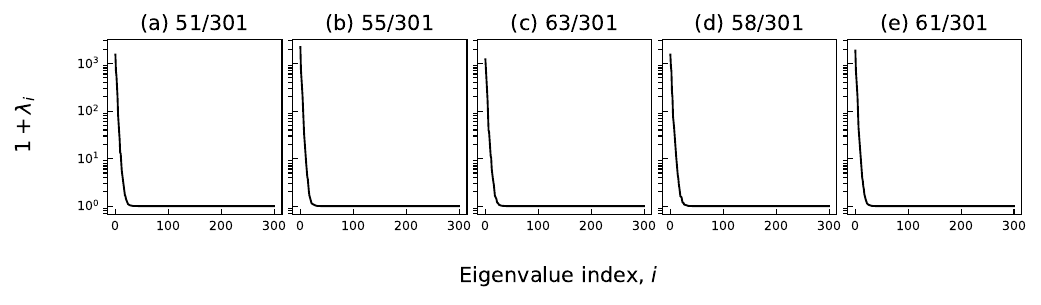}
            }
            \caption{This figure shows the spectra of the Fisher information matrix $\mathcal{I}(\hat{\bth})$ at each MAP estimate, with the eigenvalues on the $y$-axis and each column corresponding to a different MAP. All spectra share the same overall shape, with a sharp drop after roughly 50 directions. Thus, even though the effective dimension of the tangent space changes across MAP estimates, the \bWG\, approximation is flexible enough to adapt to the local geometry of each mode: it produces classifiers that agree within the data support and share the same uncertainty behaviour outside it, yielding well-calibrated and MAP-independent predictive uncertainties.} 
            \label{fig:ggn-spectrum}
        \end{center}
    \vskip -0.1in
\end{figure}

For the classification task, the covariate space is $\mathbb{R}^2$. We set $\phi_1 = \phi_2$ to the sigmoid linear unit activation function, $L = 2$, $d^i = 2$, $d^o = 1$, and $k_1 = k_2 = 15$, giving $d = 301$ parameters in total, and fit the network on $n = 200$ observations across five different initialisations of the optimiser. Note that this setting does not include an additional noise parameter $\sigma^2$ in the parametrisation. We repeat the same experiment as before, comparing the variance of the predictive distributions obtained from the \bWG\ and \bLA\ approximations at each of the five optima (a)--(e). We also show the spectrum of the Fisher matrix $\mathcal{I}(\hat{\bth})$ in Figure~\ref{fig:ggn-spectrum}, which reveals the number of numerically zero and non-zero eigenvalues at each optimum (a)--(e). These results are presented in Figures~\ref{fig:classification-visualization} and~\ref{fig:ggn-spectrum}.

Figure~\ref{fig:classification-visualization} shows the MAP decision boundary (left) together with the predictive variance $\mathbb{V}(f(x)) := \mathbb{V}_{\bth \sim \rho}(f_{\bth}(x))$ for $\rho = \rho_{\bLA}$ (middle) and $\rho = \rho_{\bWG}$ (right). The \bWG\, approximation quantifies the predictive variance well and consistently across runs -- it is confident within the data support and grows smoothly away from it -- whereas \bLA\ once again performs poorly. Figure~\ref{fig:ggn-spectrum} shows the corresponding spectra of $\mathcal{I}(\hat{\bth})$. These differ across the five optima but share the same overall shape, with a sharp drop after roughly 50 directions. Thus, even though the effective directions on the tangent space change from optimum to optimum due to the different MAP estimates, the \bWG\, approximation is flexible enough to adapt to the local geometry of each mode: it produces classifiers that agree within the data support and share the same variance behaviour outside it, yielding predictive distributions that are, again, insensitive to the choice of MAP estimate.

\subsection{Code for the neural-networks experiments}

{\tt code} : https://github.com/albertkjoller/beyond-laplace-torch


\end{appendices}


\end{document}